\documentclass[a4paper,11pt]{article}
\usepackage{amssymb,amsfonts,amsmath,amsbsy}
\usepackage{exscale,graphics,psfrag,overpic}
\usepackage{graphicx}
\usepackage{amsfonts}
\usepackage{caption}
\usepackage{subcaption}
\usepackage{authblk}

\newcommand{\etal}{\textit{et al.}\ }

\newcommand{\cE}{{\cal{E}}}
\newcommand{\cI}{{\cal{I}}}
\newcommand{\JE}{J_{\textrm{E}}}
\newcommand{\JI}{J_{\textrm{I}}}
\newcommand{\FE}{F_{\textrm{E}}}
\newcommand{\FI}{F_{\textrm{I}}}
\newcommand\thetamax{\theta_{\textrm{max}}}
\newcommand\thetamin{\theta_{\textrm{min}}}
\newcommand\phidist{\phi_\textrm{dist}}

\newcommand\Mdist{M_{\textrm{dist}}}
\newcommand\Mring{M_{\textrm{ring}}}
\newcommand\Mdrop{M_{\textrm{drop}}}
\newcommand\Mrings{M_{\textrm{rings}}}
\newcommand\Mringn{M_{\textrm{ring},n}}
\newcommand\tCR{t_{\textrm{CR}}}
\newcommand\tCA{t_{\textrm{CA}}}
\newcommand\tSS{t_{\textrm{SS}}}
\newcommand\tSJ{t_{\textrm{SJ}}}
\newcommand\tdist{t_{\textrm{dist}}}
\newcommand\tring{t_{\textrm{ring}}}
\newcommand\tringn{t_{\textrm{ring},n}}
\newcommand\thetastarcrit{\theta^*_{\textrm{crit}}}
\newcommand\thetastarmin{\theta^*_{\textrm{min}}}
\newcommand\thetamincrit{\theta_{\textrm{min,crit}}}

\DeclareUnicodeCharacter{2212}{-}
\DeclareUnicodeCharacter{2009}{-}

\title{Simultaneous evaporation and imbibition of a droplet on a flooded porous substrate}

\author[1]{David Craig\thanks{Email: david.craig@strath.ac.uk, ORCID: 0009-0004-2777-8717}}
\author[1]{Alexander W.\ Wray\thanks{alexander.wray@strath.ac.uk, ORCID: 0000-0002-3219-8272}}
\author[2]{Khellil Sefiane\thanks{k.sefiane@ed.ac.uk, ORCID: 0000-0003-3300-0210}}
\author[3,1]{Stephen K.\ Wilson\thanks{Author for correspondence, Email: sw3197@bath.ac.uk, ORCID: 0000-0001-7841-9643}}

\affil[1]{Department of Mathematics and Statistics, University of Strathclyde, Livingstone Tower, 26 Richmond Street, Glasgow G1 1XH, United Kingdom}
\affil[2]{School of Engineering, Institute for Multiscale Thermofluids, University of Edinburgh, The King’s Buildings, Mayfield Road, Edinburgh EH9 3JL, United Kingdom}
\affil[3]{Department of Mathematical Sciences, University of Bath, Claverton Down, Bath BA2 7AY, United Kingdom}

\date{4th November 2025}

\begin{document}

\maketitle

\begin{abstract}
A mathematical model for the evolution of, and deposition from, a thin particle-laden droplet on an infinitely thick, isotropic, flooded, porous substrate with interconnected pores undergoing simultaneous evaporation and imbibition is formulated and analysed.
In particular, analytical expressions for the evolution of the droplet, as well as for the flow within the droplet and the substrate, and for the transport and deposition onto the substrate of the particles are obtained for droplets evolving in four different modes.
While the physical mechanisms driving evaporation (namely, the diffusion-driven flow of vapour in the atmosphere away from the free surface of the droplet) and imbibition (namely, the pressure-driven flow of liquid in the substrate away from the base of the droplet) are rather different, perhaps rather unexpectedly, it is found that there are a number of qualitative and quantitative similarities as well as differences in the resulting behaviour of the droplet as it loses mass to its environment.
For example, it is shown that a droplet undergoing pure imbibition in the constant contact radius mode never completely imbibes (i.e., it has an infinitely long lifetime), but if it evaporates (either with or without imbibition also occurring) then it has a finite lifetime, and increasing the strength of evaporation and/or imbibition shortens its lifetime.
It is also shown that in the regime in which diffusion of particles is faster than axial advection but slower than radial advection of particles, the final deposit of particles left behind on the substrate after the droplet has completely evaporated and/or imbibed is independent of both the nature and the strength of the physical mechanism(s) driving the mass loss from the droplet.
Not only are these results of theoretical interest, but they are also relevant to a wide variety of practical applications, including ink-jet printing, DNA chip manufacturing, and disease diagnostics, that would benefit from an improved ability to predict and/or control the final deposit pattern from a droplet undergoing simultaneous evaporation and imbibition.
\end{abstract}

\section{Introduction}
\label{sec:introduction}

The behaviour of a sessile droplet on a porous substrate is not only of fundamental scientific interest, but is also of importance in a wide variety of practical applications, such as printing onto paper and textiles (see, for example, Chen \etal \cite{chen2023material}), paintwork restoration (see, for example, Zenit \cite{zenit2019painting}), and microfabrication (see, for example, Tseng \etal \cite{tseng2014research}).
Consequently, there have been extensive experimental, numerical, and theoretical investigations of the evaporation of a droplet on a solid substrate (see, for example, the review articles by Brutin \& Starov \cite{brutin2018recent}, Gelderblom \etal \cite{gelderblom2022evaporation}, and Wilson \& D'Ambrosio \cite{WilsonDAmbrosioReview2023}) and of the imbibition of a droplet on a porous substrate (see, for example, the review articles by Gambaryan-Roisman \cite{roisman2014} and Johnson \etal \cite{johnson2019kinetics}) in a variety of different contexts. 
However, thus far there have been relatively few experimental, numerical, or theoretical investigations of the situation studied in the present work, namely a droplet on a porous substrate undergoing simultaneous evaporation and imbibition.
Specifically, we use an approach based on lubrication theory (see, for example, Oron \etal \cite{oron1997long}, Craster \& Matar \cite{craster2009dynamics}, and Davis \cite{davis2017importance}) to formulate and analyse a mathematical model for the simultaneous evaporation and imbibition of a droplet.
In particular, we investigate the evolution of, and deposition from, a thin axisymmetric droplet on an infinitely thick, isotropic, flooded, porous substrate with interconnected pores undergoing simultaneous evaporation and imbibition.
The isotropy of the substrate means that there is flow within it in both radial and axial directions. This contrasts with the situation considered by some previous authors (see, for example, Davis \& Hocking \cite{DavisHocking1999,DavisHocking2000}) in which the pores are assumed to be non-connected and vertically aligned, so that the porosity and permeability of the substrate are anisotropic, and hence there is only flow within it in the vertical direction.

The theoretical study of the diffusion-limited evaporation of a droplet on a solid substrate was pioneered by Mangel \& Baer \cite{mangel1962evaporation}, Picknett \& Bexon \cite{Picknett1977}, Birdi \etal \cite{Birdi1989}, and Bourges-Monnier \& Shanahan \cite{Bourges1995}. The diffusion-limited evaporation of a droplet is driven by the difference between the saturated concentration of vapour at the free surface of the droplet and the ambient concentration of vapour in the atmosphere. In practice, the timescales for diffusion and the adjustment of the free surface are both short compared to the typical lifetime of a droplet (see, for example, Larson \cite{larson2014transport}), and so both processes are often assumed to be quasi-steady.
In their pioneering study, Picknett \& Bexon \cite{Picknett1977} determined the evolution, and hence the lifetime, of evaporating droplets with pinned (fixed) and unpinned (receding) contact lines.
Many subsequent authors, such as 
Birdi \etal \cite{Birdi1989}, 
Bourges-Monnier \& Shanahan \cite{Bourges1995}, 
Deegan \cite{deegan1997capillary}, 
Deegan \etal \cite{deegan2000contact,deegan2000pattern}, 
Hu \& Larson \cite{hu2002evaporation,hu2005microfluid,hu2005marangoni,hu2006marangoni}, 
Sultan \etal \cite{sultan2004diffusion,sultan2005evaporation}, 
McHale \etal \cite{mchale2005analysis}, 
Popov \cite{Popov2005}, 
Dunn \etal \cite{Dunn2008,Dunn2009substrate}, 
Zheng \cite{zheng2009study}, 
Freed-Brown \cite{freed2014evaporative}, 
Stauber \etal \cite{stauber2014lifetimes,stauber2015evaporation,stauber2015lifetimes}, 
Wray \etal \cite{wray2014electrostatic,wray2020competitive,wray2021contact}, 
Du \& Deegan \cite{du2015ring}, 
Kang \etal \cite{kang2016alternative}, 
Moore \etal \cite{Moore2016SessileAP},
Boulogne \etal \cite{boulogne2017coffee}, 
Boulogne \& Dollet \cite{boulogne2018convective}, 
Schofield \etal \cite{schofield2018lifetimes,schofield2020shielding}, 
D'Ambrosio \etal \cite{DAmbrosio2021well,DAmbrosio2023effect,DAmbrosio2025gravity,DAmbrosio2025movingcontactline},
Masoud \etal \cite{masoud2021evaporation}, 
Moore \etal \cite{moore2021nascent,moore2022nascent},
Gelderblom \etal \cite{gelderblom2022evaporation}, 
Seyfert \etal \cite{seyfert2022stability},
Moore \& Wray \cite{wray2023noncircular,wray2024high,wray2024novel},
Yariv \cite{yariv2023lifetime},
Erdem \etal \cite{erdem2024numerical},
Jeon \etal \cite{jeon2025acceleration},
Mei \& Zhou \cite{mei2025evaporation}, and
O'Brien \etal \cite{obrien2025quantifying},
have built on the work of Picknett \& Bexon \cite{Picknett1977} to obtain exact, asymptotic, approximate, and/or numerical descriptions of the evolution, and hence the lifetime, of droplets undergoing diffusion-limited evaporation in a variety of different situations. 

The theoretical study of the imbibition of a droplet on a porous substrate was pioneered by Davis \& Hocking \cite{DavisHocking1999,DavisHocking2000}. The imbibition of a droplet is driven by the difference between the Laplace pressure within the droplet and the pressure within the substrate. When the porous substrate is flooded there are no liquid-air interfaces, and therefore no capillary pressure, within the substrate, and hence the imbibition is driven by the difference between the pressure within the droplet and atmospheric pressure. In practice, the timescales for imbibition and the adjustment of the free surface are both short compared to the typical lifetime of a droplet (see, for example, Starov \etal \cite{starov2002spreadingdry}), and so both processes are often assumed to be quasi-steady. Davis \& Hocking \cite{DavisHocking1999} considered several situations involving the imbibition of a thin two-dimensional droplet. Specifically, they considered the evolution of such droplets on both horizontal and inclined porous substrates, where the pores within the substrate were either interconnected or non-connected and vertically aligned. In a companion paper, Davis \& Hocking \cite{DavisHocking2000} considered several situations involving (and related to) the imbibition of a thin two-dimensional droplet on an initially dry porous substrate. In contrast to the large body of work on the evaporation of droplets on non-porous substrates, relatively few subsequent authors (with the exception of Alleborn \& Raszillier \cite{alleborn2004spreading}, Esp\'{i}n \& Kumar \cite{espin2015}, Pham \& Kumar \cite{Pham2019}, Zheng \cite{zheng2022growth}, and Hartmann \& Thiele \cite{hartmann2025gradient}) have built on the work of Davis \& Hocking \cite{DavisHocking1999,DavisHocking2000} to obtain exact, asymptotic, approximate, and numerical descriptions of the evolution of a droplet on a flooded porous substrate undergoing imbibition. In particular, we highlight the work of Pham \& Kumar \cite{Pham2019} who investigated the simultaneous spreading, evaporation, and imbibition of a thin particle-laden droplet on a finitely thick, flooded, porous substrate with non-connected and vertically aligned pores. In this work the evaporation was described using the one-sided model (see, for example, Burelbach \etal \cite{burelbach1988nonlinear}, Murisic \& Kondic \cite{murisic2011evaporation}, and Larsson \& Kumar \cite{larsson2023comparison}) rather than the diffusion-limited model. Specifically, Pham \& Kumar \cite{Pham2019} described the evolution of, and deposition from, a droplet in the presence of intermolecular forces, a precursor layer, disjoining pressure, and surface roughness. Pham \& Kumar \cite{Pham2019} identified that evaporation and imbibition have qualitatively similar effects on the contact line dynamics of a droplet, and as a result, have qualitatively similar effects on particle deposition.

In a wide variety of practical applications, the spatial distribution of the final deposit pattern of particles left behind on the substrate after a particle-laden droplet has completely evaporated and/or imbibed is of considerable importance. For instance, in practical applications concerning droplets in ink-jet printing (see, for example, He \& Derby \cite{he2017controlling}) and DNA chip manufacturing (see, for example, Dugas \etal \cite{dugas2005droplet}) the desired outcome is usually a spatially uniform deposit. In contrast, in practical applications concerning droplets in disease diagnostics (see, for example, Carre{\'o}n \etal \cite{carreon2018texture}) and conductive coatings (see, for example, Layani \etal \cite{layani2009transparent}) the desired outcome is usually either a single ring deposit or multiple ring deposits. These and other practical applications would benefit from an improved ability to predict and/or control the final deposit pattern from a droplet undergoing simultaneous evaporation and imbibition (see, for example, the review articles by Larson \cite{larson2014transport}, Sefiane \cite{sefiane2014patterns}, Mampallil \& Eral \cite{mampallil2018review}, Parsa \etal \cite{parsa2018mechanisms}, Lohse \cite{LohseAnnualReview2022}, and Thampi \& Basavaraj \cite{thampi2023drying}). 

The theoretical and experimental studies by Deegan \etal \cite{deegan1997capillary,deegan2000contact} and Deegan \cite{deegan2000pattern} are credited with providing the first detailed explanation of the coffee-ring effect, in which a ring-shaped deposit is left behind at the position of the contact line when a particle-laden droplet with a pinned contact line evaporates. Deegan \etal \cite{deegan1997capillary,deegan2000contact} and Deegan \cite{deegan2000pattern} identified that as a droplet evaporates an outward flow carries the suspended particles towards the contact line, leading to the formation of a characteristic coffee-ring deposit. Many subsequent authors, such as
Popov \cite{Popov2005}, 
Hu \& Larson \cite{hu2006marangoni}, 
Zheng \cite{zheng2009study},
Mar{\'\i}n \etal \cite{marin2011order,marin2011rush}, 
Nguyen \etal \cite{nguyen2012theoretical},
Siregar \etal \cite{siregar2013numerical}, 
Wray \etal \cite{wray2014electrostatic,wray2020competitive,wray2021contact}, 
Freed-Brown \cite{freed2014evaporative}, 
Man \& Doi \cite{man2016ring}, 
Zigelman \& Manor \cite{zigelman2016model,zigelman2018deposition}, 
Pham \& Kumar \cite{pham2017drying}, 
Yang \etal \cite{yang2021deposition}, 
Matav{\v{z}} \etal \cite{matavvz2022coffee}, 
Coombs \etal \cite{coombs2024colloidal1,coombs2024colloidal2}, and 
D'Ambrosio \etal \cite{DAmbrosio2025adsorption,DAmbrosio2025movingcontactline}
have built on the work of Deegan \etal \cite{deegan1997capillary,deegan2000contact} and Deegan \cite{deegan2000pattern} to investigate the final deposit pattern formed when a particle-laden droplet on a solid substrate undergoes evaporation.

Dou \etal \cite{dou2011ink} and Dou \& Derby \cite{douderby2012} pioneered the investigation of the formation of a coffee-ring deposit when a droplet imbibes into a porous substrate. Dou \& Derby \cite{douderby2012} identified that (in a similar manner to diffusion-limited evaporation) as a droplet imbibes an outward flow carries the suspended particles towards the contact line, leading to the formation of a coffee-ring deposit. Several subsequent authors, such as Boulogne \etal \cite{boulogne2015homogeneous, boulogne2016tuning}, Nilghaz \etal \cite{nilghaz2015coffee}, Al-Milaji \etal \cite{al2019inkjet}, and Hwang \etal \cite{hwang2023monte} have built on the works of Dou \etal \cite{dou2011ink} and Dou \& Derby \cite{douderby2012} to investigate the final deposit pattern formed when a droplet on a porous substrate undergoes imbibition. However, relatively few subsequent authors (with the exception of Pack \etal \cite{pack2015colloidal}, Pham \& Kumar \cite{Pham2019}, Li \etal \cite{li2020absorption}, Kumar \etal \cite{kumar2022pattern}, Wijburg \etal \cite{wijburg2023transport}, Wang \& Darhuber \cite{wang2024numerical}, and Niu \etal \cite{niu2024revisiting}) have investigated the final deposit pattern formed when a droplet undergoes simultaneous evaporation and imbibition.

Motivated by the wide variety of practical applications which would benefit from theoretical insight into the evolution of a droplet on a porous substrate and the formation of various final deposit patterns, in the present work we investigate the evolution of, and deposition from, a droplet undergoing simultaneous evaporation and imbibition.
Specifically, in Sections \ref{sec:model} and \ref{sec:evolution} we formulate and analyse a mathematical model for the evolution of a thin particle-laden droplet on an infinitely thick, isotropic, flooded, porous substrate with interconnected pores undergoing simultaneous evaporation and imbibition.
In Section \ref{sec:flow} we determine the flow within the droplet and the substrate, in Section \ref{sec:particles} we describe the transport and deposition onto the substrate of the particles, and in Section \ref{sec:paths} we calculate the paths of the particles in the regime in which diffusion of particles is slower than both radial and axial advection of particles. 
Finally, in Section \ref{sec:conclusions} we summarise the conclusions of the present work and discuss some potential directions for future work. 

\section{Mathematical Model}
\label{sec:model}

\subsection{Problem Formulation}
\label{sec:model_formulation}

\begin{figure}[tp]
\centering
\includegraphics[width=0.75\textwidth]{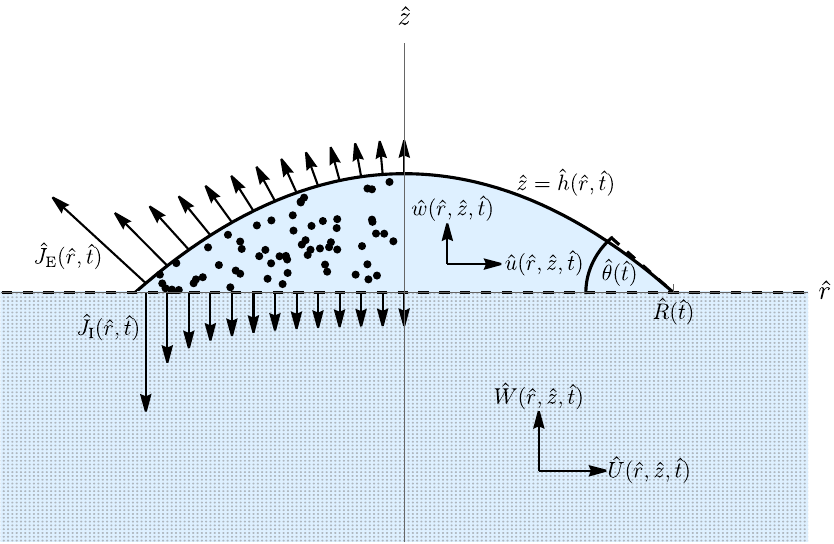}
\caption{
Sketch of a thin, axisymmetric, particle-laden droplet on an infinitely thick, isotropic, flooded, porous substrate with interconnected pores undergoing simultaneous evaporation and imbibition.
}
\label{fig:figure1}
\end{figure}

We consider a thin axisymmetric droplet of liquid on an infinitely thick, isotropic, flooded, porous substrate with interconnected pores undergoing simultaneous evaporation and imbibition. We refer the description to polar coordinates $\hat{r}$ and $\hat{z}$ with $O\hat{z}$ perpendicular to the surface of the substrate at $\hat{z}=0$, as sketched in Figure \ref{fig:figure1}. We denote the free surface of the droplet by $\hat{z}=\hat{h}(\hat{r},\hat{t})$, the volume of the droplet by $\hat{V}=\hat{V}(\hat{t})$, the radial and axial components of the velocity within the droplet by $\hat{u}=\hat{u}(\hat{r},\hat{z},\hat{t})$ and $\hat{w}=\hat{w}(\hat{r},\hat{z},\hat{t})$, respectively, and the pressure within the droplet by $\hat{p}=\hat{p}(\hat{r},\hat{z},\hat{t})$, where $\hat{t}$ denotes time. The local evaporation and imbibition mass fluxes are denoted by $\hat{J}_{\textrm{E}}=\hat{J}_{\textrm{E}}(\hat{r},\hat{t})$ and $\hat{J}_{\textrm{I}}=\hat{J}_{\textrm{I}}(\hat{r},\hat{t})$, and the corresponding global evaporation and imbibition mass fluxes are denoted by $\hat{F}_{\textrm{E}}=\hat{F}_{\textrm{E}}(\hat{t})$ and $\hat{F}_{\textrm{I}}=\hat{F}_{\textrm{I}}(\hat{t})$, while the contact radius and contact angle are denoted by $\hat{R}=\hat{R}(\hat{t})$ and $\hat{\theta}=\hat{\theta}(\hat{t})$, with their initial values denoted by $\hat{R}_0$ and $\hat{\theta}_0$, respectively. The droplet is deposited onto the surface of the substrate at $\hat{t}=0$, and thereafter undergoes mass loss due to simultaneous evaporation and imbibition. We do not consider the relatively short initial period during which the droplet rapidly adjusts to its quasi-static shape after it is deposited.

We non-dimensionalise and scale the variables appropriately for a thin droplet according to
\begin{gather}
r = \frac{\hat{r}}{\hat{R}_0}, \quad 
z = \frac{\hat{z}}{\hat{\theta}_0\hat{R}_0}, \quad 
t =\frac{\hat{U}_{\textrm{ref}} \hat{t}}{\hat{R}_0}, \quad
R = \frac{\hat{R}}{\hat{R}_0} \quad 
\theta = \frac{\hat{\theta}}{\hat{\theta}_0}, \quad
V = \frac{\hat{V}}{\hat{\theta}_0 \hat{R}_0^3}, \quad 
u = \frac{\hat{u}}{\hat{U}_{\textrm{ref}}}, \quad
w = \frac{\hat{w}}{\hat{U}_{\textrm{ref}} \hat{\theta}_0}, \nonumber \\
p = \frac{\hat{R}_0 (\hat{p}-\hat{p}_\infty)}{\hat{\gamma} \hat{\theta}_0}, \quad 
\JE = \frac{\hat{J}_{\textrm{E}}}{\hat{J}_{\textrm{E,ref}}}, \quad 
\JI = \frac{\hat{J}_{\textrm{I}}}{\hat{J}_{\textrm{I,ref}}}, \quad 
\FE = \frac{\hat{F}_{\textrm{E}}}{\hat{R}_0^2 \hat{J}_{\textrm{E,ref}}}, \quad 
\FI = \frac{\hat{F}_{\textrm{I}}}{\hat{R}_0^2 \hat{J}_{\textrm{I,ref}}},
\label{eq:nondim}
\end{gather} 
where 
$\hat{U}_{\textrm{ref}}$ is an appropriate characteristic radial velocity scale, 
$\hat{p}_\infty$ is the constant atmospheric pressure,
$\hat{\gamma}$ is the constant coefficient of surface tension at the liquid--gas interface,
$\hat{J}_{\textrm{E,ref}}$ is an appropriate characteristic local evaporation mass flux scale for the evaporation of liquid from the free surface of the droplet into the atmosphere, and
$\hat{J}_{\textrm{I,ref}}$ is an appropriate characteristic local imbibition mass flux scale for the imbibition of liquid from the base of the droplet into the substrate. 
The dimensionless ratios that characterise evaporation and imbibition, denoted by $\cE$ and $\cI$, hereafter referred to as the evaporation number and the imbibition number, are defined by
\begin{equation}
\label{eq:cEandcI}
\cE = \frac{\hat{J}_{\textrm{E,ref}}}{\hat{\rho} \hat{\theta}_0 \hat{U}_{\textrm{ref}}}
\quad \hbox{and} \quad
\cI = \frac{\hat{J}_{\textrm{I,ref}}}{\hat{\rho} \hat{\theta}_0 \hat{U}_{\textrm{ref}}},
\end{equation}
respectively, where $\hat{\rho}$ is the constant density of the liquid.
Before defining $\hat{U}_{\textrm{ref}}$, $\hat{J}_{{\textrm{E}}_{\textrm{ref}}}$, $\hat{J}_{{\textrm{I}}_{\textrm{ref}}}$, and appropriate characteristic timescales for evaporation and imbibition, we first state the key assumptions and derive the global mass balance equation for a droplet undergoing simultaneous evaporation and imbibition.

\subsection{The Hydrodynamic Problem}
\label{sec:model_evolution}

In this subsection, we describe the evolution of, and the flow within, a thin droplet undergoing simultaneous evaporation and imbibition. Specifically, we consider the situation in which the droplet is sufficiently small that the effect of gravity is negligible and surface tension is sufficiently strong that the free surface of the droplet evolves quasi-statically. More specifically, we consider situations in which the appropriately defined Bond number $\textrm{Bo}$ and capillary number $\textrm{Ca}$, namely
\begin{equation}
\label{eq:BoandCa}
\textrm{Bo} = \frac{\hat{\rho}\hat{g}\hat{R}_0^2}{\hat{\gamma}} 
\quad \textrm{and} \quad 
\textrm{Ca} = \frac{\hat{\mu} \hat{U}_{\textrm{ref}}}{\hat{\gamma} \hat{\theta}_0^3},
\end{equation}
respectively, are both small and satisfy $\hat{\theta}_0^2, \textrm{Bo} \ll \textrm{Ca} \ll 1$, where
$\hat{g}$ is the magnitude of acceleration due to gravity and
$\hat{\mu}$ is the constant dynamic viscosity of the liquid.

We seek an asymptotic solution for the pressure $p$ in the form 
\begin{equation}
\label{eq:p_expansion}
p = p^{(0)}(r,z,t) + \textrm{Ca} \ p^{(1)}(r,z,t) + \textrm{O} \left(\hat{\theta}_0^2, \textrm{Bo}, \textrm{Ca}^2\right),
\end{equation}
with corresponding expressions for the other variables. However, since the pressure $p$ and the concentration of particles $\phi$ introduced subsequently in Section \ref{sec:model_transport} are the only quantities that we require beyond leading order, for brevity, we omit the superscript ``$(0)$'' on all of the other leading-order variables.

At leading order in \textrm{Ca}, the Stokes equations yield $\partial p^{(0)}/\partial r = \partial p^{(0)}/\partial z=0$, and hence the leading-order pressure $p^{(0)}$ is independent of $r$ and $z$, i.e., $p^{(0)}=p^{(0)}(t)$, and is given by the normal stress condition at the free surface to be
\begin{equation}
\label{eq:p0_equation}
p^{(0)} = - \frac{1}{r} \frac{\partial}{\partial r} \left(r \frac{\partial h}{\partial r}\right).
\end{equation} 
Hence, the free surface of the droplet $h$ satisfies
\begin{equation}
\label{eq:h_equation}
\frac{\partial}{\partial r} \left[\frac{1}{r}\frac{\partial}{\partial r}\left(r \frac{\partial h}{\partial r}\right)\right] = 0 
\end{equation}
subject to $h=0$ and $\partial h / \partial r = - \theta$ at $r=R$.
An additional requirement is that $h$ must be finite at $r=0$, and hence the shape of the free surface takes the familiar paraboloidal form 
\begin{equation}
\label{eq:h_solution}
h = \frac{\theta (R^2-r^2)}{2 R}.
\end{equation}
Evaluating the leading-order pressure \eqref{eq:p0_equation} using \eqref{eq:h_solution} yields
\begin{equation}
\label{eq:p_0}
p^{(0)} = \frac{2 \theta}{R}.
\end{equation} 
We note that the leading-order pressure given by \eqref{eq:p_0} is spatially uniform. The volume of the droplet is given by 
\begin{equation}
\label{eq:V}
V = 2 \pi \int_{0}^{R} h (r,t) r \ \textrm{d}r = \frac{\pi \theta R^3}{4}. 
\end{equation} 

At leading order in the thin-film limit, the continuity and Stokes equations reduce to the familiar axisymmetric lubrication equations, namely
\begin{equation}
\label{eq:lubrication_equations}
\frac{1}{r} \frac{\partial(r u)}{\partial r}+\frac{\partial w}{\partial z} = 0, \quad 
\frac{\partial^2 u}{\partial z^2} = \frac{\partial p^{(1)}}{\partial r}, \quad 
\frac{\partial p^{(1)}}{\partial z} = 0,
\end{equation}
subject to the tangential stress condition,
\begin{equation}
\label{eq:tangential_stress}
\frac{\partial u}{\partial z}=0 \quad \textrm{on} \quad z=h \quad \textrm{for} \quad 0 \le r \le R,
\end{equation}
and the kinematic condition,
\begin{equation}
\label{eq:kinematic}
\frac{\partial h}{\partial t} + u \frac{\partial h}{\partial r} - w = - \cE \JE \quad \textrm{on} \quad z=h \quad \textrm{for} \quad 0 \le r \le R,
\end{equation}
on the free surface, 
together with a no-slip condition,
\begin{equation}
\label{eq:noslip}
u = 0 \quad \textrm{on} \quad z=0 \quad \textrm{for} \quad 0 \le r \le R,
\end{equation}
and a penetration condition,
\begin{equation}
\label{eq:penetration}
w = - \cI \JI \quad \textrm{on} \quad z=0 \quad \textrm{for} \quad 0 \le r \le R,
\end{equation}
on the solid-liquid interface (i.e., on the surface of the substrate within the footprint of the droplet). The penetration condition \eqref{eq:penetration} is a key feature of the present model and couples the flow within the droplet to the flow within the substrate.
Applying conditions \eqref{eq:tangential_stress}, \eqref{eq:noslip}, and \eqref{eq:penetration} to the solutions of \eqref{eq:lubrication_equations}, the radial and axial components of the velocity within the droplet are given by
\begin{equation}
\label{eq:uandw}
u = - \frac{(2h-z)z}{2}\frac{\partial p^{(1)}}{\partial r}, \quad 
w = \frac{z^2}{6r}\frac{\partial}{\partial r}\left((3h-z)r\frac{\partial p^{(1)}}{\partial r}\right) - \cI \JI,
\end{equation}
respectively. The kinematic condition \eqref{eq:kinematic} can be rewritten as
\begin{equation}
\label{eq:kinematic_rewritten}
\frac{\partial h}{\partial t} + \frac{1}{r}\frac{\partial}{\partial r} \left(r Q\right) = - \left( \cE \JE + \cI \JI \right),
\end{equation}
where $Q=Q(r,t)$ is the local radial volume flux of liquid given by
\begin{equation}
\label{eq:Q}
Q = \int_{0}^{h} u \ \textrm{d}z = -\frac{h^3}{3} \frac{\partial p^{(1)}}{\partial r}.
\end{equation}
The depth-averaged radial velocity $\bar{u}=\bar{u}(r,t)$ is given by
\begin{equation}
\label{eq:ubar}
\bar{u}	= \frac{1}{h}\int_{0}^{h} u \ \textrm{d}z=\frac{Q}{h},
\end{equation}
and in Section \ref{sec:flow} it will be convenient to rearrange the kinematic condition \eqref{eq:kinematic_rewritten} to express $Q$ as
\begin{equation}
\label{eq:kinematic_integrated}
Q = - \frac{1}{r} \int_{0}^{r} \left(\frac{\partial h}{\partial t} + \cE \JE + \cI \JI\right) r \ \textrm{d}r.
\end{equation}
Using \eqref{eq:Q} to eliminate the first-order pressure $p^{(1)}$ from the expression for $u$ in \eqref{eq:uandw} and then evaluating the integral in the expression for $w$ in \eqref{eq:uandw} yields
\begin{equation}
\label{eq:uandw_rewritten}
u = \frac{3 Q (2h-z) z}{2 h^3}, \quad
w = - \frac{z^2}{2 r} \frac{\partial}{\partial r} \left[\frac{r Q (3h-z)}{h^3} \right] - \cI \JI.
\end{equation}

Integrating the kinematic condition \eqref{eq:kinematic_rewritten} from $r=0$ to $r=R$ leads to the global mass balance equation
\begin{equation}
\label{eq:global_1}
\frac{\textrm{d}V}{\textrm{d}t} = - 2 \pi \int_{0}^{R} \left( \cE \JE+\cI \JI\right) r \ \textrm{d}r,
\end{equation} 
and substituting the expression for the droplet volume \eqref{eq:V} into \eqref{eq:global_1} yields an evolution equation for $R$ and $\theta$, namely
\begin{equation}
\label{eq:evolution_1}
\frac{\textrm{d}\left(\theta R^3\right)}{\textrm{d}t} = - 8 \int_{0}^{R} \left( \cE \JE+ \cI \JI\right) r \ \textrm{d}r.
\end{equation}
However, \eqref{eq:evolution_1} is only one equation for two unknowns, and so in order to determine the evolution, and hence the lifetime, of a droplet we must either determine or (as we do here) specify the manner (i.e., the mode) in which the droplet evolves as it undergoes mass loss into the atmosphere and/or the substrate.

\begin{figure}[tp]
\centering
\begin{subfigure}[b]{0.3\textwidth}
\centering
\includegraphics[width=\textwidth]{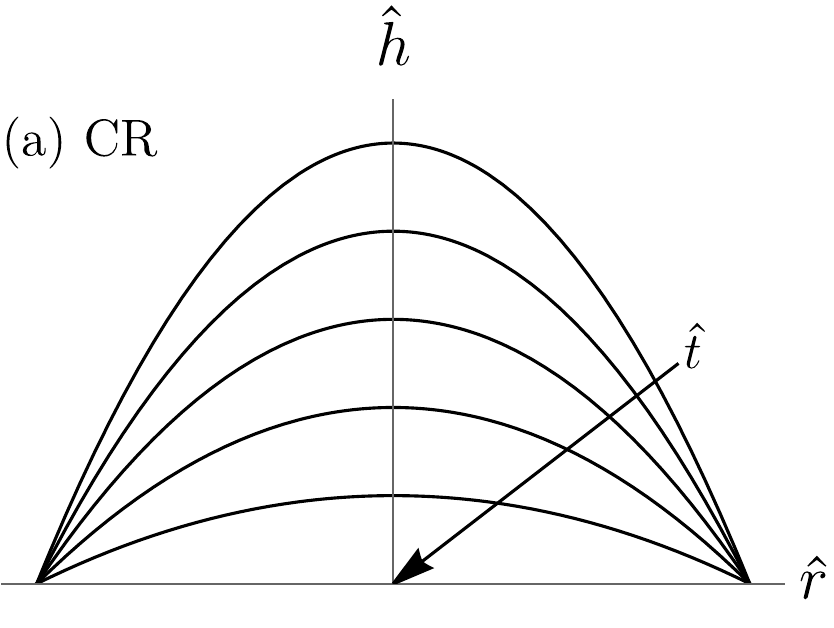}
\end{subfigure}
\hfill
\begin{subfigure}[b]{0.3\textwidth}
\centering
\includegraphics[width=\textwidth]{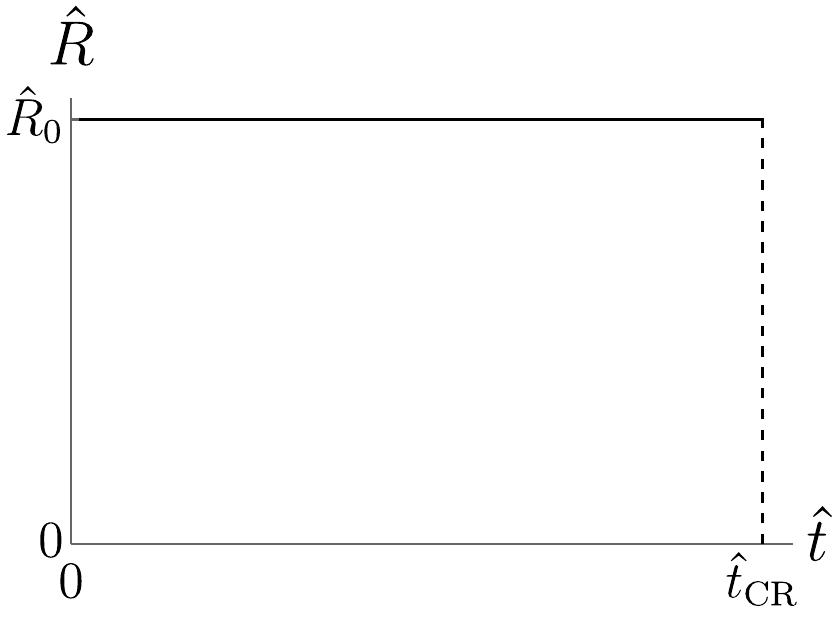}
\end{subfigure}
\hfill
\begin{subfigure}[b]{0.3\textwidth}
\centering
\includegraphics[width=\textwidth]{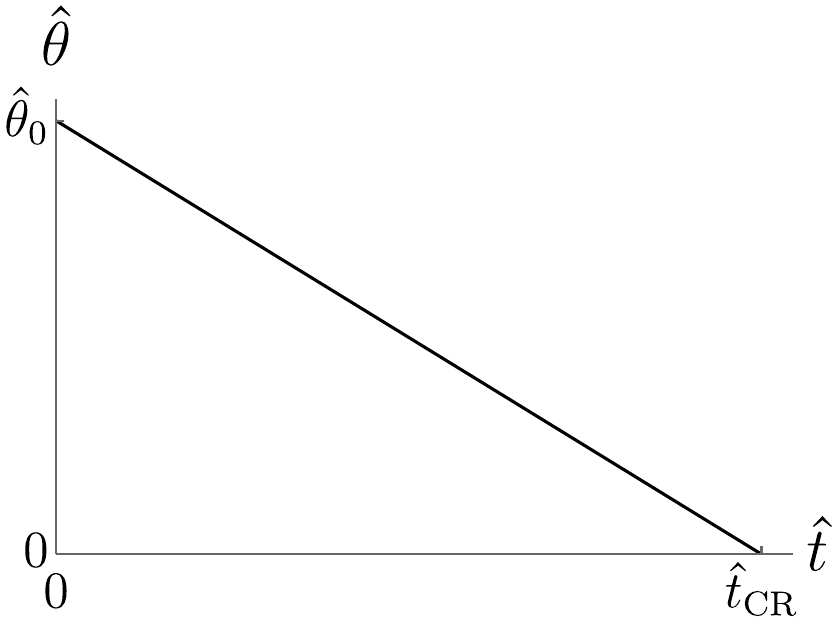}
\end{subfigure}
\begin{subfigure}[b]{0.3\textwidth}
\centering
\includegraphics[width=\textwidth]{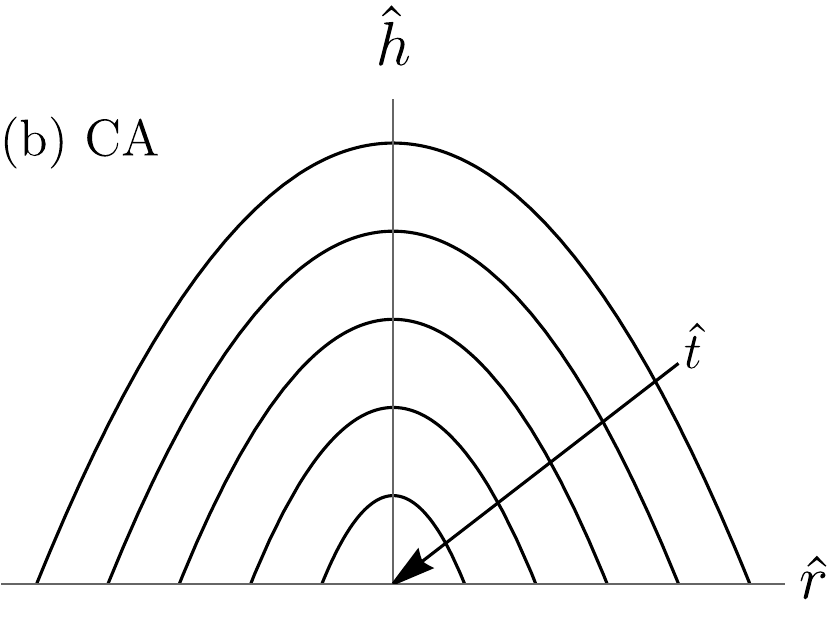}
\end{subfigure}
\hfill
\begin{subfigure}[b]{0.3\textwidth}
\centering
\includegraphics[width=\textwidth]{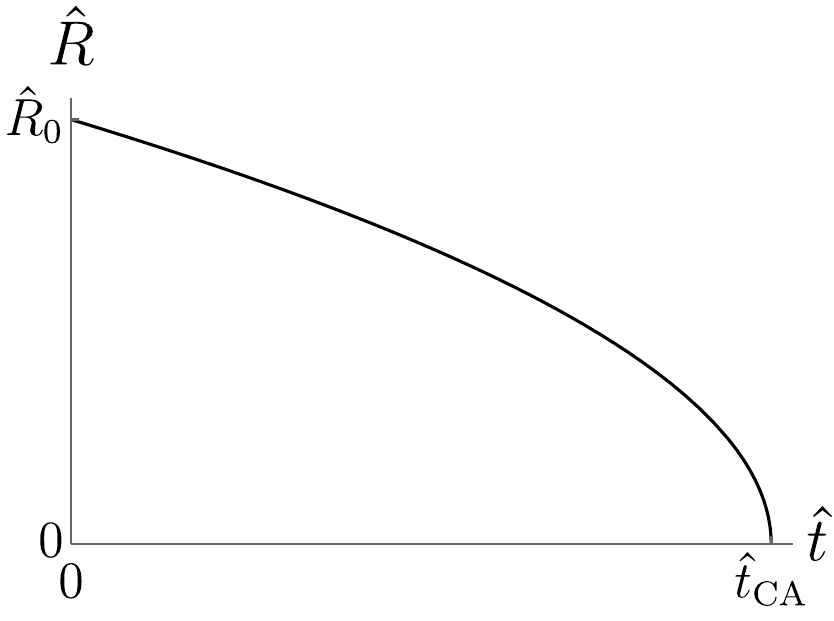}
\end{subfigure}
\hfill
\begin{subfigure}[b]{0.3\textwidth}
\centering
\includegraphics[width=\textwidth]{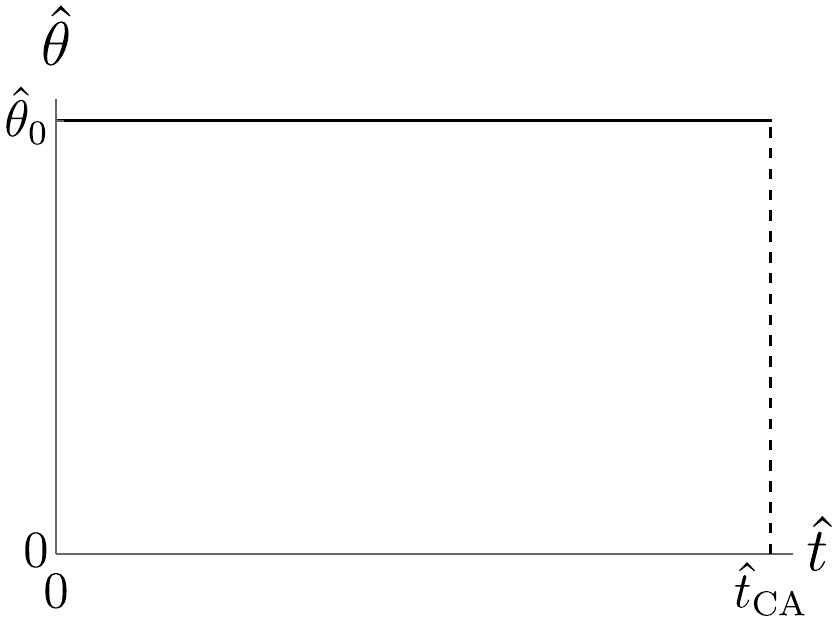}
\end{subfigure}
\begin{subfigure}[b]{0.3\textwidth}
\centering
\includegraphics[width=\textwidth]{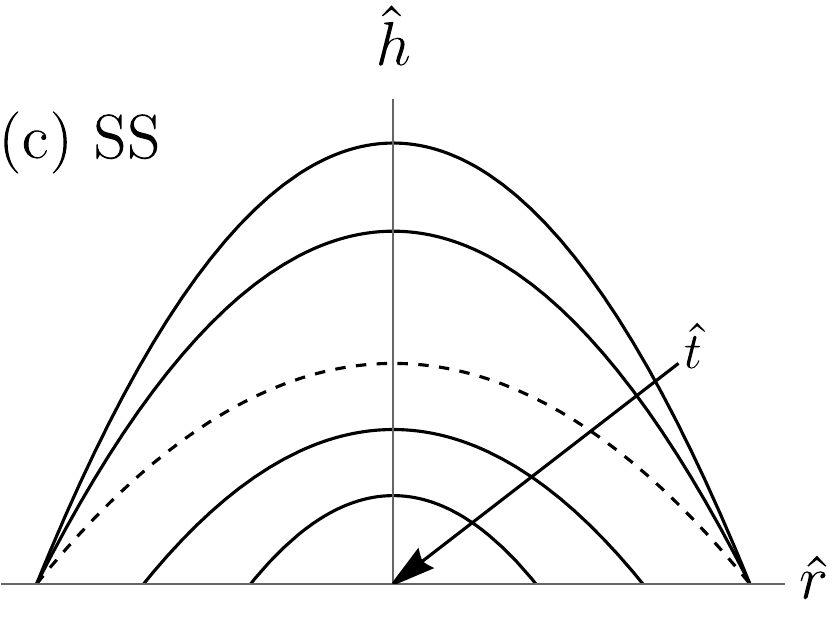}
\end{subfigure}
\hfill
\begin{subfigure}[b]{0.3\textwidth}
\centering
\includegraphics[width=\textwidth]{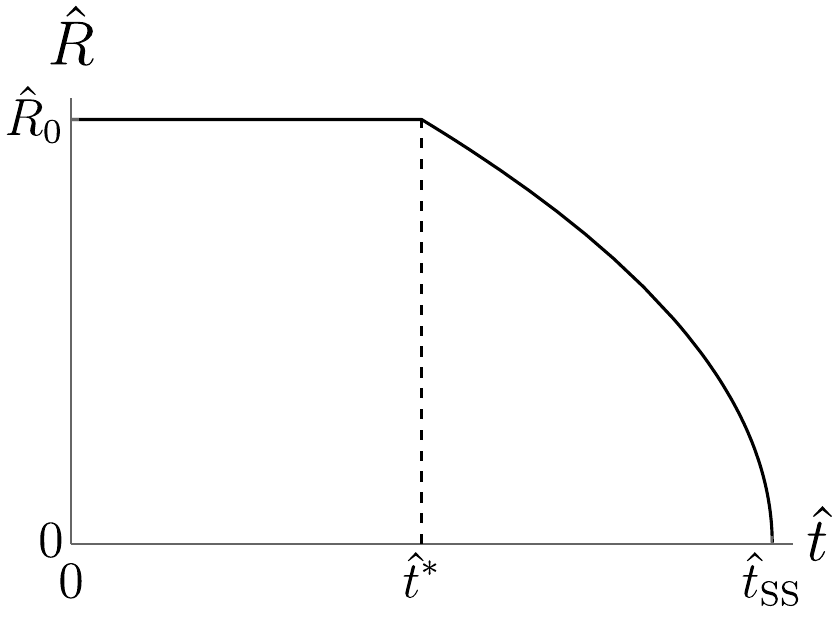}
\end{subfigure}
\hfill
\begin{subfigure}[b]{0.3\textwidth}
\centering
\includegraphics[width=\textwidth]{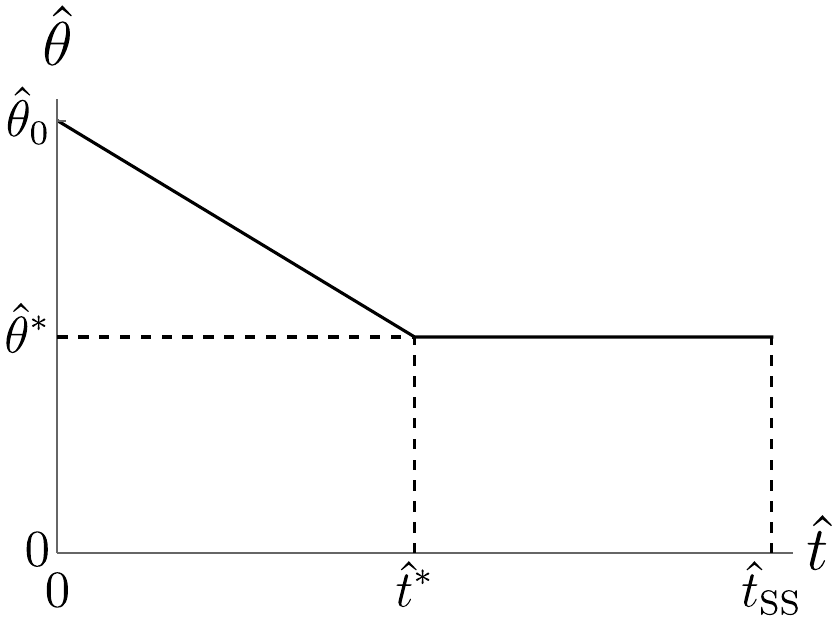}
\end{subfigure}
\begin{subfigure}[b]{0.3\textwidth}
\centering
\includegraphics[width=\textwidth]{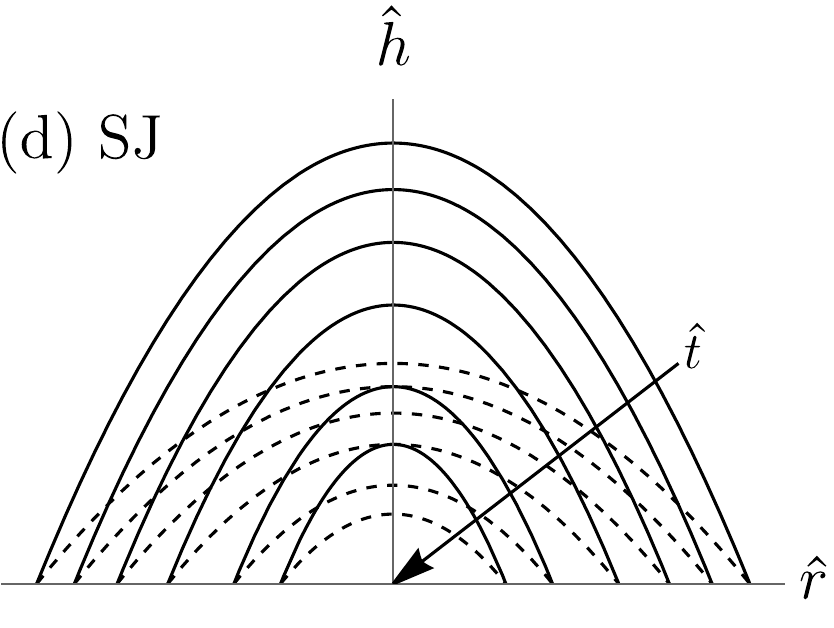}
\end{subfigure}
\hfill
\begin{subfigure}[b]{0.3\textwidth}
\centering
\includegraphics[width=\textwidth]{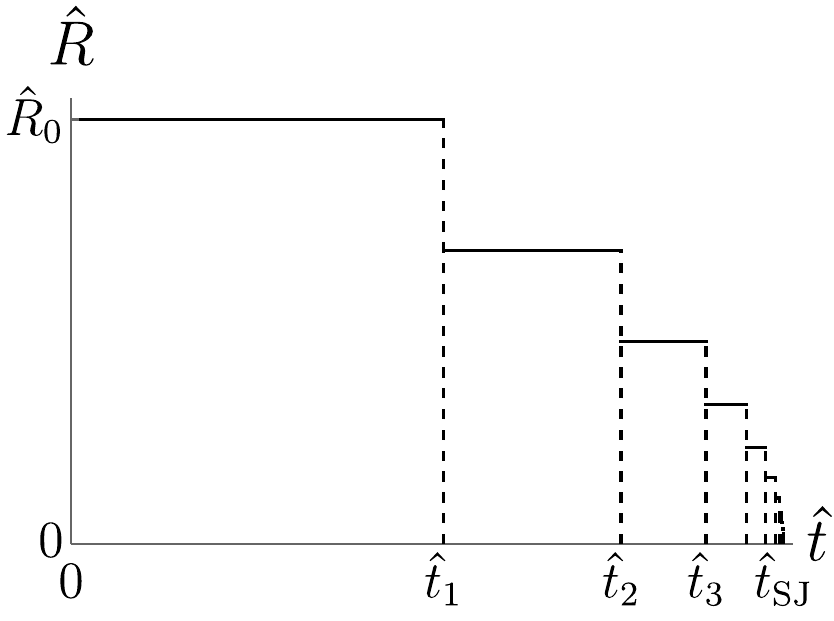}
\end{subfigure}
\hfill
\begin{subfigure}[b]{0.3\textwidth}
\centering
\hspace*{-0.6cm}
\includegraphics[width=\textwidth]{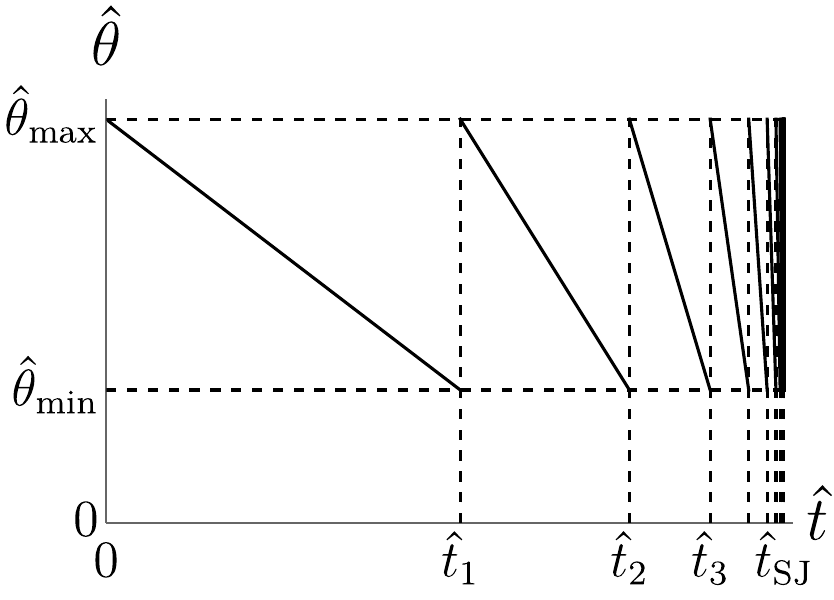}
\end{subfigure}
\caption{
Sketches of the evolutions of the free surface $\hat{h}(\hat{r},\hat{t})$, the contact radius $\hat{R}(\hat{t})$, and the contact angle $\hat{\theta}(\hat{t})$, respectively. Each row of the figure corresponds to a different mode of evolution: (a) the CR mode, (b) the CA mode, (c) the SS mode, and (d) the SJ mode. The vertical dashed lines correspond to the lifetime of the droplet, while the additional vertical dashed lines in (c) correspond to the instant $\hat{t}=\hat{t}^*$ at which the droplet contact line de-pins, and the additional vertical dashed lines in (d) correspond to the instants $\hat{t}=\hat{t}_n$ ($n=1,2,3,\ldots$) at which the droplet contact line de-pins and jumps inwards from $\hat{R}_n$ to $\hat{R}_{n+1} (<\hat{R}_n)$. The arrows indicate the directions of increasing $\hat{t}$.
}
\label{fig:figure2}
\end{figure}

The modes of evolution which have received the most attention within the literature are the so-called extreme modes, namely the constant contact radius (CR) mode and the constant contact angle (CA) mode. In the CR mode the contact line remains pinned at the initial value $\hat{R}\equiv \hat{R}_0$ and the contact angle decreases to zero. In contrast, in the CA mode the contact angle remains constant, and equal to its initial value $\hat{\theta}\equiv \hat{\theta}_0$, and the contact radius decreases to zero. 

There are also a variety of mixed modes in which both the contact radius and the contact angle vary. One combination of the extreme modes that frequently occurs in practice is the Stick--Slide (SS) mode, as described by Picknett and Bexon \cite{Picknett1977} (and many other authors, such as Bourges \etal \cite{Bourges1995}, Gelderblom \etal \cite{gelderblom2013}, Nguyen \etal \cite{nguyen2013}, Trybala \etal \cite{trybala2013}, Boulogne \etal \cite{boulogne2015homogeneous,boulogne2016tuning}, and Johnson \etal \cite{johnson2019kinetics}) in which the droplet first evolves in a CR phase with a pinned contact line $\hat{R} \equiv \hat{R}_0$ until the contact angle reaches a critical (receding) value, denoted by $\hat{\theta}=\hat{\theta}^*$, where $0 \le \hat{\theta}^* \le \hat{\theta}_0$. The time at which the contact line de-pins is denoted by $\hat{t}=\hat{t}^*$. Once the critical contact angle has been reached, the contact line de-pins, and thereafter the droplet evolves in a CA phase with a constant contact angle $\hat{\theta} \equiv \hat{\theta}^*$ for the remainder of its evolution. The SS mode reduces to the CR mode when $\hat{\theta}^*=0$, in which case the contact line never de-pins, and reduces to the CA mode when $\hat{\theta}^*=\hat{\theta}_0$, in which case the contact line immediately de-pins. 
	
Another mode that occurs in practice is the Stick--Jump (SJ) mode, as described by Deegan \etal \cite{deegan2000contact} (and many other authors, such as Moffat \etal \cite{Moffat2009}, Shanahan \& Sefiane \cite{Shananhan2009}, and Orejon \etal \cite{Orejon2011}), in which the droplet initially evolves in a CR phase with a pinned contact line. However, when the contact angle reaches a critical (minimum) value, denoted by $\hat{\theta}=\hat{\theta}_{\textrm{min}}$, the contact line de-pins and jumps instantaneously inwards to a new position with a smaller contact radius, while the contact angle jumps instantaneously up to a critical (maximum) value, denoted by $\hat{\theta}=\hat{\theta}_{\textrm{max}}$, where $0 \le \hat{\theta}_{\textrm{min}} \le \hat{\theta}_{\textrm{max}} = \hat{\theta}_0$. This pattern then repeats itself, with a (theoretically infinite) sequence of CR phases separated by a (theoretically infinite) sequence of instantaneous jump phases. The values of $\hat{R}$ and $\hat{\theta}$ in the $n$th CR phase ($n=1,2,3\ldots$), which lasts from $\hat{t}=\hat{t}_{n-1}$ to $\hat{t}=\hat{t}_n$, where $\hat{t}_0=0$ and $\hat{t}=\hat{t}_n$ ($n=1,2,3,\ldots$) is the time at which the $n$th jump phase occurs, are denoted by $\hat{R}_n$ and $\hat{\theta}_n=\hat{\theta}_n(\hat{t})$, respectively, with $\hat{R}_1 = \hat{R}_0$ and $\hat{\theta}_1 = \hat{\theta}_0 = \hat{\theta}_{\textrm{max}}$. The SJ mode incorporates the CR and CA modes as special cases. Specifically, the SJ mode reduces to the CR mode when $\hat{\theta}_{\textrm{min}}=0$, and reduces to the CA mode in the limit $\hat{\theta}_{\textrm{min}}\to\hat{\theta}_\textrm{max}^- = \hat{\theta}_0^-$.
	
Figure \ref{fig:figure2} shows sketches of the evolutions of the free surface, contact radius, and contact angle in the four modes discussed here. The lifetimes of a droplet evolving in the CR, CA, SS and SJ modes are denoted by $\hat{t}_\textrm{CR}$, $\hat{t}_\textrm{CA}$, $\hat{t}_\textrm{SS}$, and $\hat{t}_\textrm{SJ}$, respectively. 

\subsection{The Evaporative Problem}
\label{sec:model_evaporation}
 
In this subsection, we describe the evaporation of liquid from the free surface of a thin droplet into the atmosphere according to the basic version of the well-known diffusion-limited model with constant diffusion coefficient $\hat{D}$. For simplicity, in the present work we follow Pham and Kumar \cite{Pham2019} and assume that there is no evaporation from the ``unwetted'' surface of the substrate. The concentration of vapour in the atmosphere is denoted by $\hat{c}=\hat{c}(\hat{r},\hat{z},\hat{t})$. The constant concentration of vapour at the free surface of the droplet $\hat{z}=\hat{h}(\hat{r},\hat{t})$ is denoted by $\hat{c}=\hat{c}_{\mathrm{sat}}$, and the constant far-field concentration is denoted by $\hat{c}=\hat{c}_\infty$.
Before proceeding further, we introduce the additional non-dimensional variables involved in the evaporative problem, namely
\begin{equation}
\label{eq:variables_evaporation}
c = \frac{\hat{c} - \hat{c}_\infty}{\hat{c}_{\textrm{sat}}-\hat{c}_\infty}
\quad \textrm{and} \quad 
z^{\textrm{a}} = \frac{\hat{z}}{\hat{R}_0},
\end{equation} 
where $z^{\textrm{a}}$ denotes the non-dimensional $z$ coordinate in the atmosphere.
Appropriate choices of $\hat{U}_{\textrm{E,ref}}$ and $\hat{J}_{\textrm{E,ref}}$ for the diffusion-limited evaporation problem are
\begin{equation}
\label{eq:UrefandJref_evaporation}
\hat{U}_{\textrm{E,ref}} = \frac{\hat{D} (\hat{c}_{\textrm{sat}}-\hat{c}_\infty)}{\hat{\rho} \hat{\theta}_0 \hat{R}_0}
\quad \textrm{and} \quad 
\hat{J}_{\textrm{E,ref}} = \hat{\rho} \hat{\theta}_0 \hat{U}_{\textrm{E,ref}} = \frac{\hat{D} (\hat{c}_{\textrm{sat}}-\hat{c}_\infty)}{\hat{R}_0}
\end{equation}
(see, for example, Wilson \& D'Ambrosio \cite{WilsonDAmbrosioReview2023}).

Since the droplet is thin, the concentration of vapour $c=c(r,z^{\textrm{a}},t)$ satisfies Laplace's equation in the upper half-space,
\begin{equation}
\label{eq:{eq:laplace_atmosphere}}	
\frac{1}{r} \frac{\partial}{\partial r} \left(r \frac{\partial c}{\partial r}\right) + \frac{\partial^2 c}{\partial z^{\textrm{a}2}} = 0 \quad \textrm{in} \quad z^{\textrm{a}}>0, 
\end{equation}
subject to the appropriate boundary and far-field conditions, namely
\begin{equation}	
\label{eq:BC_evaporation_1}
c = 1 \quad \textrm{on} \quad z^{\textrm{a}} = 0 \quad \textrm{for} \quad 0 \le r \le R,
\end{equation}
\begin{equation}
\label{eq:BC_evaporation_2}	
\frac{\partial c}{\partial z^{\textrm{a}}} = 0 \quad \textrm{on} \quad z^{\textrm{a}} = 0 \quad \textrm{for} \quad r > R,
\end{equation}
\begin{equation}
\label{eq:BC_evaporation_3}	
c \rightarrow 0 \quad \textrm{as} \quad (r^2 + z^{\textrm{a}2})^{1/2} \rightarrow \infty.
\end{equation}
Note that, in general, the boundary condition \eqref{eq:BC_evaporation_1} applies on the free surface $z=h(r,t)$, but for a thin droplet, such as that considered in the present work, it can be transferred to the plane $z^{\textrm{a}}=0$ by Taylor expanding about $z^{\textrm{a}}=0$. Equations \eqref{eq:{eq:laplace_atmosphere}}--\eqref{eq:BC_evaporation_3} may be solved to obtain the well-known solution for the concentration of vapour in the atmosphere, namely
\begin{equation}
\label{eq:c_solution}
c = \frac{2}{\pi}\arcsin \frac{2}{\left[(R+r)^2 + z^{\textrm{a2}}\right]^{1/2} + \left[(R-r)^2 + z^{\textrm{a2}}\right]^{1/2}}.
\end{equation}
The local evaporation mass flux is then given by
\begin{equation}
\label{eq:JE}
\JE = - \frac{\partial c}{\partial z^{\textrm{a}}} = \frac{2}{\pi\left(R^2-r^2 \right)^{1/2}} \quad \textrm{on} \quad z^{\textrm{a}}=0 \quad \textrm{for} \quad 0 \le r \le R,
\end{equation}
and hence the global evaporation mass flux is
\begin{equation}
\label{eq:FE}
\FE = 2 \pi \int_0^{R} \JE \, r \ \textrm{d}r = 4R.
\end{equation}
An appropriate characteristic timescale for the diffusion-limited evaporation of a thin droplet is
\begin{equation}
\label{eq:tEref}
\hat{t}_\textrm{E,ref} = \frac{\hat{\rho} \hat{\theta}_0 \hat{R}_0^2}{\hat{D} (\hat{c}_{\textrm{sat}}-\hat{c}_\infty)}
\end{equation}
(see, for example, Wilson \& D'Ambrosio \cite{WilsonDAmbrosioReview2023}).

\subsection{The Imbibition Problem}
\label{sec:model_imbibition}

In this subsection, we describe the imbibition of liquid from the base of a thin droplet into an infinitely thick, isotropic, flooded, porous substrate with interconnected pores. The radial and axial components of the macroscopic velocity within the substrate are denoted by $\hat{U}=\hat{U}(\hat{r},\hat{z},\hat{t})$ and $\hat{W}=\hat{W}(\hat{r},\hat{z},\hat{t})$, respectively, and the pressure within the substrate is denoted by $\hat{P}=\hat{P}(\hat{r},\hat{z},\hat{t})$. The constant porosity of the substrate, denoted by $\varphi$ ($0 < \varphi < 1$), is a (dimensionless) measure of the volume-averaged fraction of the interconnected pores within the substrate, and the constant permeability of the substrate, denoted by $\hat{k}$, is an intrinsic property of the substrate that describes the viscous resistance to the flow through the pores. In general, a larger internal pore surface area leads to a larger friction within the substrate and, hence, a lower permeability. The pressure at the surface of the substrate within the footprint of the droplet is equal to the (spatially-uniform) pressure at the base of the droplet, and the constant far-field pressure (assumed to be equal to the atmospheric pressure) is denoted by $\hat{P}_\infty \, (=\hat{p}_\infty)$.
Before proceeding further, we introduce the additional non-dimensional variables involved in the imbibition problem, namely
\begin{equation}
\label{eq:variables_imbibition}
U = \frac{\hat{\rho}\hat{U}}{\hat{J}_{\textrm{I,ref}}}, \quad
W = \frac{\hat{\rho}\hat{W}}{\hat{J}_{\textrm{I,ref}}}, \quad
P = \frac{\hat{R}_0(\hat{P}-\hat{P}_\infty)}{\hat{\gamma} \hat{\theta}_0},
\quad \textrm{and} \quad
z^{\textrm{s}} = \frac{\hat{z}}{\hat{R}_0},
\end{equation} 
where $z^{\textrm{s}} \, (=z^{\textrm{a}})$ denotes the non-dimensional $z$ coordinate in the substrate.
Appropriate choices of $\hat{U}_{\textrm{I,ref}}$ and $\hat{J}_{\textrm{I,ref}}$ for the imbibition problem are given by
\begin{equation}
\label{eq:UrefandJref_imbibition}
\hat{U}_{\textrm{I,ref}} = \frac{\hat{\gamma} \hat{k}}{\hat{\mu} \hat{R}_0^2} 
\quad \textrm{and} \quad 
\hat{J}_{\textrm{I,ref}} = \hat{\rho} \hat{\theta}_0 \hat{U}_{\textrm{I,ref}} = \frac{\hat{\rho} \hat{\gamma} \hat{k} \hat{\theta}_0}{\hat{\mu} \hat{R}_0^2}. 
\end{equation}

In the present work, Darcy's law (see, for example, Whitaker \cite{whitaker1986flow}) is used to describe the flow of liquid within the substrate, and so the radial and axial components of the macroscopic velocity are related to the pressure according to
\begin{equation}
\label{eq:UandW}
U = - \frac{1}{\varphi}\frac{\partial P}{\partial r} \quad \textrm{and} \quad
W = - \frac{1}{\varphi}\frac{\partial P}{\partial z^{\textrm{s}}}.
\end{equation}
{Local conservation of mass of liquid within the substrate leads to the appropriate} continuity equation,
\begin{equation}
\label{eq:conservation_substrate}
\frac{1}{r}\frac{\partial (r U)}{\partial r} + \frac{\partial W}{\partial z^{\textrm{s}}} = 0,
\end{equation}
and substitution of \eqref{eq:UandW} into \eqref{eq:conservation_substrate} shows that the pressure $P=P(r,z^{\textrm{s}},t)$ satisfies Laplace's equation in the lower half-space, 
\begin{equation}
\label{eq:laplace_substrate}
\frac{1}{r} \frac{\partial}{\partial r} \left(r \frac{\partial P}{\partial r}\right) + \frac{\partial^2 P}{\partial z^{\textrm{s}2}} = 0 \quad \textrm{in} \quad z^{\textrm{s}} < 0,
\end{equation} 
subject to the appropriate boundary and far-field conditions, namely
\begin{equation}
\label{eq:BC_imbibition_1}
P = p^{(0)} \quad \textrm{on} \quad z^{\textrm{s}} = 0 \quad \textrm{for} \quad 0 \le r \le R,
\end{equation}
\begin{equation}
\label{eq:BC_imbibition_2}
\frac{\partial P}{\partial z^{\textrm{s}}} = 0 \quad \textrm{on} \quad z^{\textrm{s}}=0 \quad \textrm{for} \quad r \geq R,
\end{equation}
\begin{equation}
\label{eq:BC_imbibition_3}
P \rightarrow 0 \quad \textrm{as} \quad (r^2 + z^{\textrm{s}2})^{1/2} \rightarrow \infty.
\end{equation}
Analogous to solving equations \eqref{eq:{eq:laplace_atmosphere}}--\eqref{eq:BC_evaporation_3} to obtain the solution for the concentration of vapour in the atmosphere given by \eqref{eq:c_solution}, equations \eqref{eq:laplace_substrate}--\eqref{eq:BC_imbibition_3} may be solved to obtain the corresponding solution for the pressure within the substrate, namely
\begin{equation}
\label{eq:P_solution}
P = \frac{4 \theta}{\pi R} \arcsin \frac{2 R}{\left[(R + r)^2 + z^{\textrm{s}2}\right]^{1/2} + \left[(R-r)^2 + z^{\textrm{s}2}\right]^{1/2}}.
\end{equation}
The local imbibition mass flux is then given by
\begin{equation}
\label{eq:JI}
\JI = \frac{\partial P}{\partial z^{\textrm{s}}} = \frac{4 \theta}{\pi R \left(R^2-r^2 \right)^{1/2}} \quad \textrm{on} \quad z^{\textrm{s}} = 0 \quad \textrm{for} \quad 0 \le r \le R,
\end{equation}
and hence the global imbibition mass flux is
\begin{equation}
\label{eq:FI}
\FI = 2 \pi \int_0^{R} \JI \, r \ \textrm{d}r = 8\theta.
\end{equation}
Note that, in contrast to the global evaporation mass flux $\FE$ given by \eqref{eq:FE}, which depends on $R$ but not $\theta$, the global imbibition mass flux $\FI$ given by \eqref{eq:FI} depends on $\theta$ but not $R$. However, both depend on $t$ parametrically via either $R$ or $\theta$.
An appropriate characteristic timescale for the imbibition of a thin droplet on an infinitely thick, isotropic, flooded, porous substrate is 
\begin{equation}
\label{eq:tIref}
\hat{t}_\textrm{I,ref} = \frac{\hat{\mu} \hat{R}_0^3}{\hat{\gamma}\hat{k}}.
\end{equation}

\subsection{The Particle Transport and Deposition Problem}
\label{sec:model_transport}

In this subsection, we describe the transport and deposition onto the substrate of a dilute suspension of neutrally-buoyant and hydrodynamically passive particles from a particle-laden droplet (see, for example, Kang \etal \cite{kang2016alternative}, Moore \etal \cite{moore2021nascent}, Wray \etal \cite{wray2021contact}, and Wilson \& D'Ambrosio \cite{WilsonDAmbrosioReview2023}).

The concentration of particles per unit volume within the bulk of the droplet $\hat{\phi}=\hat{\phi}(\hat{r},\hat{z},\hat{t})$ satisfies an advection-diffusion equation subject to a penetration condition on the surface of the substrate and a no-flux condition on the free surface of the droplet. The concentration of particles per unit volume within the bulk of the droplet is non-dimensionalised according to $\phi=\hat{\phi}/ \hat{\phi}_{\textrm{ref}}$, where $\hat{\phi}_{\textrm{ref}}$ is the spatial average of the initial concentration of particles. Hence $\phi=\phi(r,z,t)$ satisfies 
\begin{equation}
\label{eq:phi_equation}
\textrm{Pe}^* \left(\frac{\partial \phi}{\partial t} + u \frac{\partial \phi}{\partial r} + w \frac{\partial \phi}{\partial z}\right)
= \frac{\hat{\theta}_0^2}{r} \frac{\partial}{\partial r} \left(r \frac{\partial \phi}{\partial r}\right) + \frac{\partial^2 \phi}{\partial z^2},
\end{equation}
where $\textrm{Pe}^*$ is the appropriately defined reduced P\'{e}clet number that is the ratio of the characteristic advective and diffusive particle transport timescales, namely
\begin{equation}
\label{eq:Pestar}
\textrm{Pe}^* = \hat{\theta}_0^2 \textrm{Pe} = \frac{\hat{\theta}_0^2 \hat{R}_0 \hat{U}_{\textrm{ref}}}{\hat{D}_\textrm{p}},
\end{equation}
where $\hat{D}_\textrm{p}$ is the constant diffusivity of the particles within the droplet. 
For simplicity, in the present work we assume that particles do not pass into the substrate and that the accumulation of particles on the surface of the substrate does not inhibit the flow of liquid into the substrate, but both of these assumptions could be re-visited in future work.
Hence \eqref{eq:phi_equation} is subject to the boundary conditions
\begin{equation}
\label{eq:phi_equation_BC_1} 
\frac{\partial \phi}{\partial z} = - \textrm{Pe}^* \cI \JI \, \phi \quad \textrm{on} \quad z=0
\end{equation}
and
\begin{equation}
\label{eq:phi_equation_BC_2}
\frac{1}{\sqrt{1 + \hat{\theta}_0^2 \left(\partial h / \partial r\right)^2}} \left(\frac{\partial \phi}{\partial z} - \hat{\theta}_0^2 \frac{\partial h}{\partial r} \frac{\partial \phi}{\partial r}\right) = \textrm{Pe}^* \cE \JE \, \phi \quad \textrm{on} \quad z=h. 
\end{equation}

We consider the regime in which diffusion of particles is faster than axial advection but slower than radial advection of particles. In other words, we consider the regime in which the reduced P\'{e}clet number satisfies $\hat{\theta}_0^2 \ll \textrm{Pe}^* \ll 1$ (see, for example, Jensen \& Grotberg \cite{jensen1993spreading} and D'Ambrosio \etal \cite{DAmbrosio2023effect}). 
(In Section \ref{sec:paths} we will consider the paths of the particles within the droplet in the alternative regime $\textrm{Pe}^* \gg 1$ in which diffusion of particles is slower than both radial and axial advection of particles, and hence at leading order diffusion of particles plays no role and the transport of particles is purely advective.)
Hence, we seek an asymptotic solution for the concentration of particles $\phi$ in the form
\begin{equation}
\label{eq:phi_expansion}
\phi = \phi^{(0)}(r,z,t) + \textrm{Pe}^* \phi^{(1)}(r,z,t) + \textrm{O} \left(\hat{\theta}_0^2, \textrm{Pe}^{*2}\right).
\end{equation}

At leading order in $\textrm{Pe}^*$, equations \eqref{eq:phi_equation}, \eqref{eq:phi_equation_BC_1}, and \eqref{eq:phi_equation_BC_2} reduce to 
\begin{equation}
\label{eq:phi0_equation}
\frac{\partial^2 \phi^{(0)}}{\partial z^2} = 0
\end{equation}
subject to
\begin{equation}
\label{eq:phi0_BC}
\frac{\partial \phi^{(0)}}{\partial z} = 0 \quad \textrm{on} \quad z =0 \quad \textrm{and} \quad z=h,
\end{equation}
and hence the leading-order concentration $\phi^{(0)}$ is independent of $z$, i.e., $\phi^{(0)}=\phi^{(0)}(r,t)$.

At first order in $\textrm{Pe}^*$, equations \eqref{eq:phi_equation}, \eqref{eq:phi_equation_BC_1}, and \eqref{eq:phi_equation_BC_2} reduce to 
\begin{equation}
\label{eq:phi1_equation}
\frac{\partial^2 \phi^{(1)}}{\partial z^2} = \frac{\partial \phi^{(0)}}{\partial t} + u \frac{\partial \phi^{(0)}}{\partial r}
\end{equation}
subject to
\begin{equation}
\label{eq:phi1_BC_1}
\frac{\partial \phi^{(1)}}{\partial z} = - \phi \cI \JI \quad \textrm{on} \quad z=0 
\end{equation}
and
\begin{equation}
\label{eq:phi1_BC_2}
\frac{\partial \phi^{(1)}}{\partial z} = \phi \cE \JE \quad \textrm{on} \quad z=h. 
\end{equation}
Integrating \eqref{eq:phi1_equation} with respect to $z$ subject to \eqref{eq:phi1_BC_1} and \eqref{eq:phi1_BC_2}, and omitting the superscript ``$(0)$'' for clarity, we arrive at the equation governing the evolution of the leading-order concentration of particles within the droplet, namely
\begin{equation}
\label{eq:phi_equation_redux}
\frac{\partial \phi}{\partial t} + \bar{u} \frac{\partial \phi}{\partial r} = \frac{\cE \JE + \cI \JI}{h} \phi.
\end{equation}
Equation \eqref{eq:phi_equation_redux} can be written in characteristic form as
\begin{equation}
\label{eq:characteristic_form}
\frac{\textrm{d}\phi}{\textrm{d}t} = \frac{\cE \JE + \cI \JI}{h} \phi \quad \textrm{on the characteristics determined by} \quad \frac{\textrm{d}r}{\textrm{d}t} = \bar{u},
\end{equation}
subject to the initial condition $r=r_0$ ($0 \le r_0 \le 1$) and $\phi=\phi(r,0)=\phi_0(r)$ at $t=0$, where $r_0=r_0(r,t)$ denotes the initial radial position of the particle that is at the radial position $r$ at time $t$ (see, for example, Boulogne \etal \cite{boulogne2017coffee} and D'Ambrosio \etal \cite{DAmbrosio2023effect}).

\begin{figure}[tp]
\centering
\begin{subfigure}[b]{0.45\textwidth}
\centering
\includegraphics[keepaspectratio,width=1\textwidth,height=0.175\textheight]{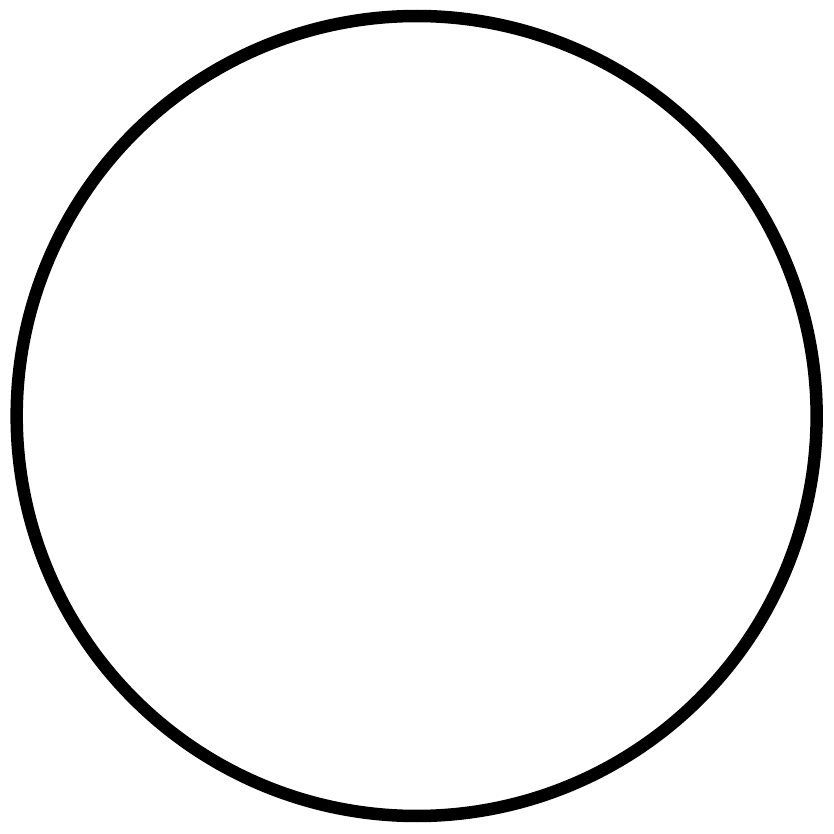}
\caption{}
\vspace{0.5cm}
\end{subfigure}
\hfill
\begin{subfigure}[b]{0.45\textwidth}
\centering
\includegraphics[keepaspectratio,width=1\textwidth,height=0.175\textheight]{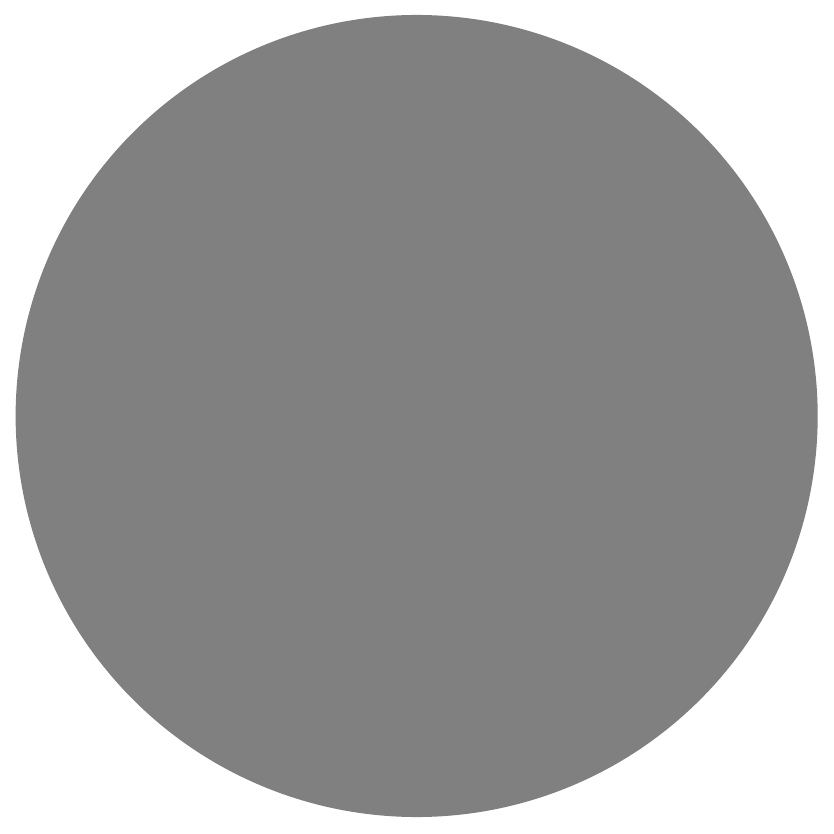}
\caption{}
\vspace{0.5cm}
\end{subfigure}
\hfill
\begin{subfigure}[b]{0.45\textwidth}
\centering
\includegraphics[keepaspectratio,width=1\textwidth,height=0.175\textheight]{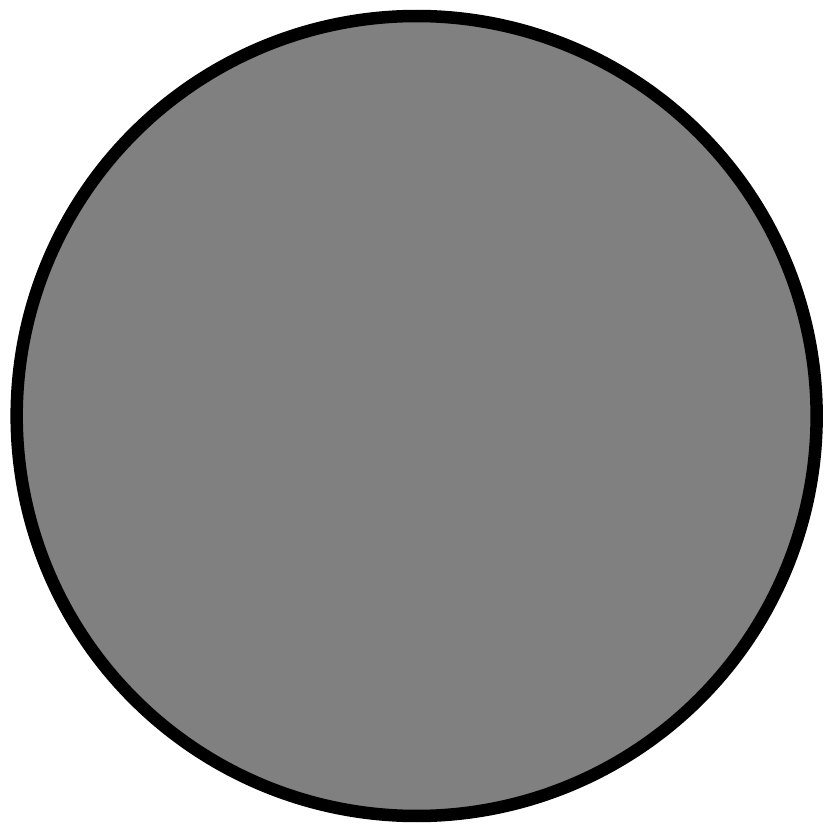}
\caption{}
\end{subfigure}
\hfill
\begin{subfigure}[b]{0.45\textwidth}
\centering
\includegraphics[keepaspectratio,width=1\textwidth,height=0.175\textheight]{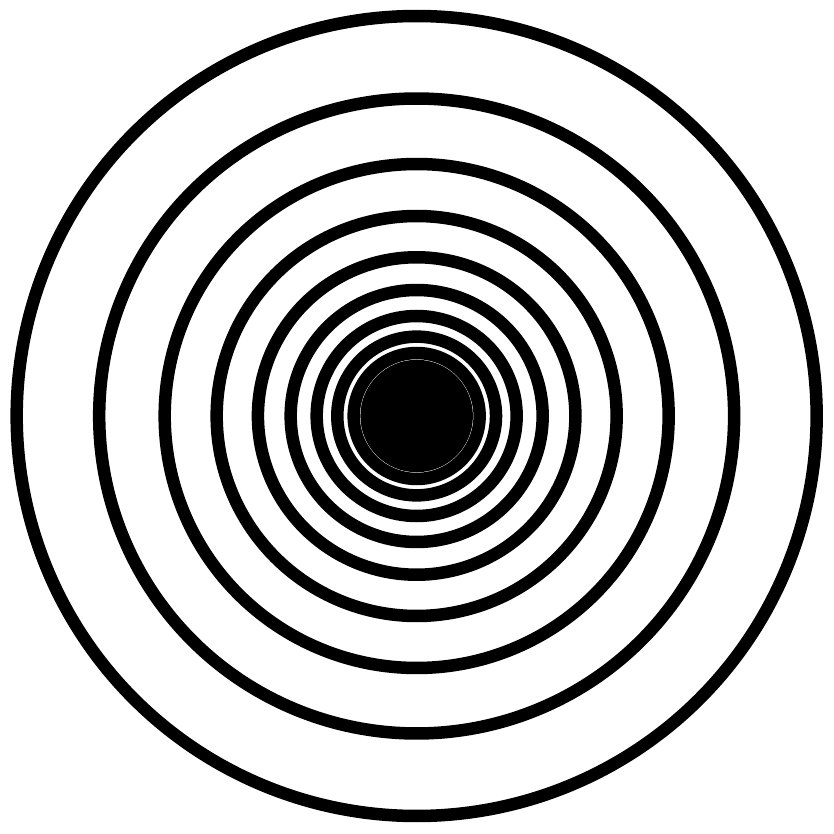}
\caption{}
\end{subfigure}
\caption{
Sketches of the final deposit patterns arising from particle-laden droplets evolving in the CR, CA, SS and SJ modes, respectively, namely (a) a ring deposit, (b) a distributed deposit, (c) a combined ring and distributed deposit, and (d) multiple ring deposits.
}
\label{fig:figure3}
\end{figure}

The mass of particles per unit area within the bulk of the droplet is given by $\phi h$, and hence the mass of particles in the bulk of the droplet, denoted by $\Mdrop=\Mdrop(t)$ and non-dimensionalised by $\hat{\theta}_0 \hat{R}_0^3 \hat{\phi}_{\textrm{ref}}$, is given by
\begin{equation}
\label{eq:Mdrop_1}
\Mdrop = 2 \pi \int_{0}^{R(t)} \phi(r,t) h(r,t) r \ \textrm{d}r,
\end{equation}
which can alternatively be expressed as
\begin{equation}
\label{eq:Mdrop_2}
\Mdrop = 2 \pi \int_{0}^{r_0(R(t),t)} \phi(r,0) h(r,0) r \ \textrm{d}r.
\end{equation}

The initial mass of particles in the bulk of the droplet is denoted by $M_0=\Mdrop(0)$. 
For simplicity, in the remainder of the present work we take the initial concentration of particles in the bulk of the droplet to be spatially uniform, i.e., $\phi_0(r) \equiv 1$, in which case $M_0$ is given by
\begin{equation}
\label{eq:M0}
M_0 = 2 \pi \int_{0}^{1} \phi_0(r) h(r,0) r \ \textrm{d}r = 2 \pi \int_{0}^{1} h(r,0) r \ \textrm{d}r = \frac{\pi}{4}.
\end{equation}

The final deposit pattern (i.e., the pattern of particles left behind on the substrate after the droplet has completely evaporated and/or imbibed) depends on the manner (i.e., the mode) in which the droplet evolved. The four modes of evolution considered in the present work give rise to the four qualitatively different final deposit patterns sketched in Figure \ref{fig:figure3}, namely a ring deposit, a distributed deposit, a combined ring and distributed deposit, and multiple ring deposits. In the next four subsections we describe how these four final deposit patterns arise from droplets evolving in the CR, CA, SS and SJ modes, respectively.

\subsubsection{Particle Deposition in the CR Mode}
\label{sec:model_transport_CR}

In the CR mode, the particles are deposited onto the substrate via the pinned contact line, resulting in the formation of a ring deposit at $r=1$ (i.e., the well-known coffee-ring effect), as sketched in Figure \ref{fig:figure3}(a).
The mass flux of particles from the bulk of the droplet into the contact line is given by
\begin{equation}
\textrm{lim}_{r \to 1^-} \, 2 \pi \phi h \bar{u} r
\end{equation}
(see, for example, Deegan \etal \cite{deegan2000pattern} and D'Ambrosio \etal \cite{DAmbrosio2023effect}), 
and so the mass of particles in the ring deposit at time $t$, denoted by $\Mring=\Mring(t)$ and non-dimensionalised by $\hat{\theta}_0 \hat{R}_0^3 \hat{\phi}_{\textrm{ref}}$, is given by
\begin{equation}
\label{eq:Mring_1}
\Mring = 2 \pi \int_{0}^{t} \lim_{r \to 1^-} \phi(r,t) h(r,t) \bar{u}(r,t) r \ \textrm{d}t,
\end{equation}
which can alternatively be expressed as
\begin{equation}
\label{eq:Mring_2}
\Mring = 2 \pi \int_{r_0(1,t)}^{1} \phi(r, 0) h(r,0) r \ \textrm{d}r.
\end{equation}
Note that, since all of the particles are initially within the bulk of the droplet, but are all eventually transferred to the ring deposit, $\Mring(0)=0$ and $\Mring(\tCR)=M_0$.

\subsubsection{Particle Deposition in the CA Mode}
\label{sec:model_transport_CA}

In the CA mode, the particles are again deposited onto the substrate via the contact line, but, since the contact line is continuously receding, this results in the formation of a distributed deposit within the initial footprint of the droplet (i.e., in $0 \le r < 1$), as sketched in Figure \ref{fig:figure3}(b).
The mass flux of particles from the bulk of the droplet into the contact line is given by
\begin{equation}
\textrm{lim}_{r \to R(t)^-} \, 2 \pi \phi h \left(\bar{u}-\frac{\textrm{d}R}{\textrm{d}t}\right) r
\end{equation}
(see D'Ambrosio \etal \cite{DAmbrosio2025movingcontactline}), 
and so the mass of particles in the distributed deposit that occupies the annular region $R(t) \le r < 1$ at time $t$, denoted by $\Mdist=\Mdist(t)$ and non-dimensionalised by $\hat{\theta}_0 \hat{R}_0^3 \hat{\phi}_{\textrm{ref}} $, is given by
\begin{equation}
\label{eq:Mdist_1}
\Mdist = 2 \pi \int_{0}^{t} \lim_{r \to R(t)^-} \phi(r,t) h(r,t) \left(\bar{u}(r,t)-\frac{\textrm{d}R}{\textrm{d}t}\right) r \ \textrm{d}t,
\end{equation}
which can alternatively be expressed as
\begin{equation}
\label{eq:Mdist_2}
\Mdist = 2 \pi \int_{r_0(R(t),t)}^{1} \phi(r,0) h(r,0) r \ \textrm{d}r.
\end{equation}
Note that, since all of the particles are initially within the bulk of the droplet, but are all eventually transferred to the distributed deposit, $\Mdist(0)=0$ and $\Mdist(\tCA)=M_0$.

Following the approach of Freed-Brown \cite{freed2014evaporative} and D'Ambrosio \etal \cite{DAmbrosio2025movingcontactline}, the final density per unit area of the distributed deposit of particles on the substrate, denoted by $\phidist = \phidist(r)$ and non-dimensionalised by $\hat{\theta}_0 \hat{R}_0 \hat{\phi}_\textrm{ref}$, is given by
\begin{equation}
\label{eq:phidist_CA_equation}
\phidist(R) = \phi(r_0(R,t),0) h(r_0(R,t),0) \frac{r_0(R,t)}{R} \frac{\partial (r_0(R,t))}{\partial R}
\quad \hbox{for} \quad
0 \le r < 1,
\end{equation}
and $\Mdist$ can be expressed in terms of $\phidist$ as
\begin{equation}
\label{eq:Mdist_CA_equation}
\Mdist = 2 \pi \int_{R(t)}^{1} \phidist(r) r \ \textrm{d} r.
\end{equation}

\subsubsection{Particle Deposition in the SS Mode}
\label{sec:model_transport_SS}

In the SS mode, the droplet evolves in a CR phase with $R \equiv 1$ for $0 \le t \le t^*$ and then in a CA phase with $0 \le R < 1$ for $t^* < t \le \tSS$, resulting in the formation of a combined ring and distributed deposit comprising a ring deposit at $r=1$ and a distributed deposit in $0 \le r < 1$, as sketched in Figure \ref{fig:figure3}(c).
The mass of particles in the ring deposit is given by \eqref{eq:Mring_1} and \eqref{eq:Mring_2} for $0 \le t \le t^*$ and by $\Mring \equiv \Mring(t^*)$ for $t^* < t \le \tSS$, and
the mass of particles in the distributed deposit is given by $\Mdist \equiv 0$ for $0 \le t \le t^*$ and by
\begin{equation}
\label{eq:Mdist_3}
\Mdist = 2 \pi \int_{t^*}^{t} \lim_{r \to R(t)^-}\phi(r,t) h(r,t) \left(\bar{u}(r,t)-\frac{\textrm{d}R}{\textrm{d}t}\right) r \ \textrm{d}t
\end{equation}
for $t^* < t \le \tSS$, the latter of which can alternatively be expressed as either
\begin{equation}
\label{eq:Mdist_4}
\Mdist = 2 \pi \int_{r_0(R(t),t)}^{1} \phi(r,t^*) h(r,t^*) r \ \textrm{d}r
\end{equation}
or \eqref{eq:Mdist_CA_equation}, where $\phidist$ is given by
\begin{equation}
\label{eq:phidist_SS_equation}
\phidist(R) = \phi(r_0(R,t),t^*) h(r_0(R,t),t^*) \frac{r_0(R,t)}{R} \frac{\partial (r_0(R,t))}{\partial R}
\quad \hbox{for} \quad
0 \le r < 1.
\end{equation}

\subsubsection{Particle Deposition in the SJ Mode}
\label{sec:model_transport_SJ}

In the SJ mode, the droplet evolves in a (theoretically infinite) sequence of CR phases with $R \equiv R_n$ for $t_{n-1} < t < t_n$ separated by a (theoretically infinite) sequence of instantaneous jump phases at $t=t_n$ ($n=1,2,3,\ldots$), resulting in the formation of a (theoretically infinite number of) ring deposits, as sketched in Figure \ref{fig:figure3}(d).
The mass of particles in the $n$th ($n=1,2,3,\ldots$) ring deposit, denoted by $\Mringn=\Mringn(t)$ and non-dimensionalised by $\hat{\theta}_0 \hat{R}_0^3 \phi_\textrm{ref}$, is given by $\Mringn\equiv0$ for $0 \le t < t_{n-1}$, by
\begin{equation}
\label{eq:Mringn_1}
\Mringn = 2 \pi \int_{t_{n-1}}^{t} \lim_{r \to R_n^-} \phi(r,t) h(r,t) \bar{u}(r,t) r \ \textrm{d}t,
\end{equation}
which can alternatively be expressed as
\begin{equation}
\label{eq:Mringn_2}
\Mringn = 2 \pi \int_{r_{0,n}(R_n,t)}^{R_n} \phi(r, t_{n-1}) h(r,t_{n-1}) r \ \textrm{d}r,
\end{equation}
for $t_{n-1} < t < t_n$, and by $\Mringn \equiv \Mringn(t_n)$ for $t_n < t \le \tSJ$.
The total mass of particles in the multiple ring deposits, denoted by $\Mrings=\Mrings(t)$, is then determined by summing the masses of the particles in the ring deposits that have either been previously created or are still being created at time $t$.

\section{Evolution and Lifetime of the Droplet}
\label{sec:evolution}

In this section, we determine the evolution, and hence the lifetime, of the droplet in the CR, CA, SS and SJ modes. 

Substituting the expressions for the local evaporation mass flux $\JE$ given by \eqref{eq:JE} and the local imbibition mass flux $\JI$ given by \eqref{eq:JI} into the evolution equation \eqref{eq:evolution_1} and performing the integration, yields
\begin{equation}
\label{eq:evolution_2}
\frac{\textrm{d}\left(\theta R^3\right)}{\textrm{d}t} = - \frac{16}{\pi} \left(\cE R + 2 \cI\theta\right).
\end{equation}

In situations in which evaporation dominates, it is most appropriate to choose $\hat{t}_\textrm{ref}=\hat{t}_\textrm{E,ref}$ given by \eqref{eq:tEref} and $\hat{U}_{\textrm{ref}}=\hat{U}_{\textrm{E,ref}}$ and $\hat{J}_{\textrm{ref}}=\hat{J}_{\textrm{E,ref}}$ given by equation \eqref{eq:UrefandJref_evaporation} in the non-dimensionalisation \eqref{eq:nondim}, and hence
\begin{equation}
\label{eq:cEandcI_evaporation}
\cE = 1 
\quad \textrm{and} \quad
\cI = \frac{\hat{t}_\textrm{E,ref}}{\hat{t}_\textrm{I,ref}}
= \frac{\hat{U}_{\textrm{I,ref}}}{\hat{U}_{\textrm{E,ref}}}
= \frac{\hat{J}_{\textrm{I,ref}}}{\hat{J}_{\textrm{E,ref}}}
= \frac{\hat{\rho} \hat{\gamma} \hat{k} \hat{\theta}_0}{\hat{D} (\hat{c}_{\textrm{sat}}-\hat{c}_\infty) \hat{\mu}\hat{R}_0}.
\end{equation}
On the other hand, in situations in which imbibition dominates, it is most appropriate to choose $\hat{t}_\textrm{ref}=\hat{t}_\textrm{I,ref}$ given by \eqref{eq:tIref} and $\hat{U}_{\textrm{ref}}=\hat{U}_{\textrm{I,ref}}$ and $\hat{J}_{\textrm{ref}}=\hat{J}_{\textrm{I,ref}}$ given by equation \eqref{eq:UrefandJref_imbibition} in the non-dimensionalisation \eqref{eq:nondim}, and hence
\begin{equation}
\label{eq:cEandcI_imbibition}
\cE = \frac{\hat{t}_\textrm{I,ref}}{\hat{t}_\textrm{E,ref}}
= \frac{\hat{U}_{\textrm{E,ref}}}{\hat{U}_{\textrm{I,ref}}}
= \frac{\hat{J}_{\textrm{E,ref}}}{\hat{J}_{\textrm{I,ref}}}
= \frac{\hat{D} (\hat{c}_{\textrm{sat}}-\hat{c}_\infty) \hat{\mu}\hat{R}_0}{\hat{\rho} \hat{\gamma} \hat{k} \hat{\theta}_0}
\quad \textrm{and} \quad
\cI = 1.
\end{equation}
For clarity of presentation, we leave $\hat{t}_\textrm{ref}$, $\hat{U}_{\textrm{ref}}$, and $\hat{J}_{\textrm{ref}}$ unspecified and retain both $\cE$ and $\cI$ in the general expressions given in the remainder of the present work. In particular, taking this approach means that the results in the special case of a droplet undergoing pure evaporation expressed using the evaporation scaling leading to \eqref{eq:cEandcI_evaporation} can be obtained by setting $\cE=1$ and taking the limit $\cI \to 0$, while the results in the special case of a droplet undergoing pure imbibition expressed using the imbibition scaling leading to \eqref{eq:cEandcI_imbibition} can be obtained by setting $\cI=1$ and taking the limit $\cE \to 0$.

\subsection{Evolution and Lifetime in the CR Mode}
\label{sec:CR_evolution}

\begin{figure}[tp]
\centering
\begin{subfigure}[b]{0.45\textwidth}
\centering
\includegraphics[keepaspectratio,width=1\textwidth,height=0.2\textheight]{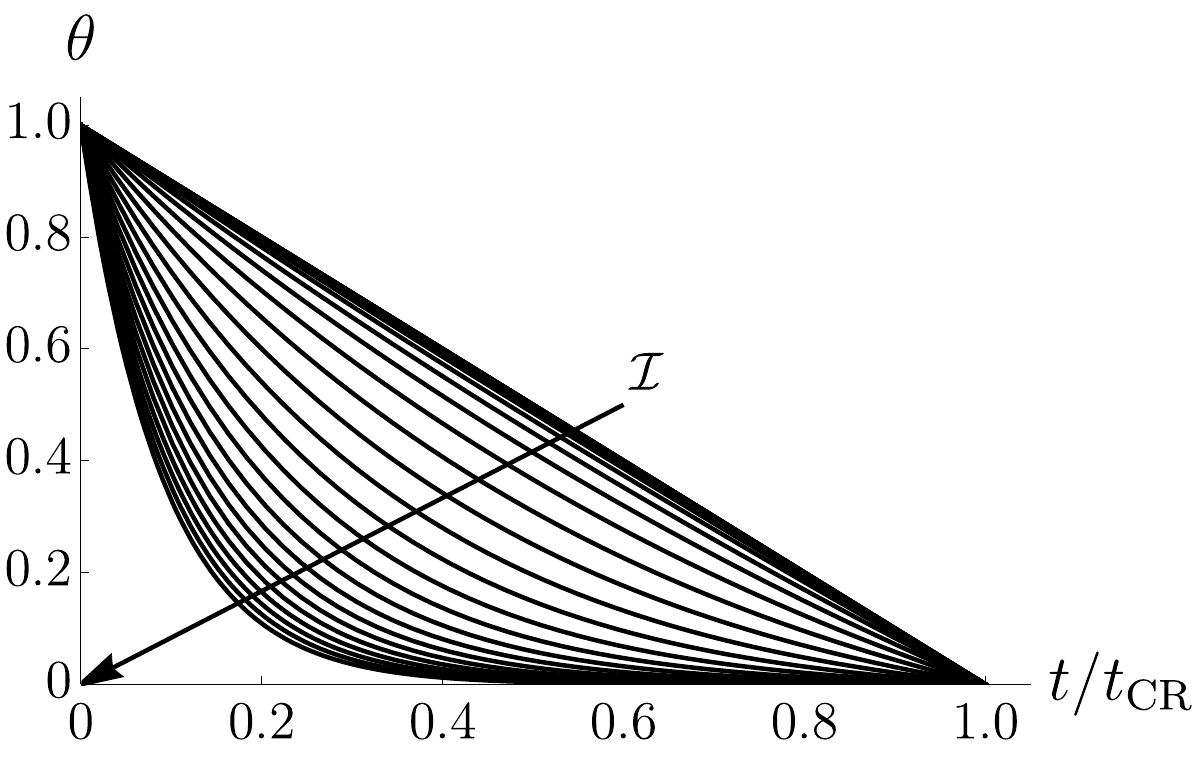}
\caption{}
\end{subfigure}
\centering
\begin{subfigure}[b]{0.45\textwidth}
\centering
\includegraphics[keepaspectratio,width=1\textwidth,height=0.2\textheight]{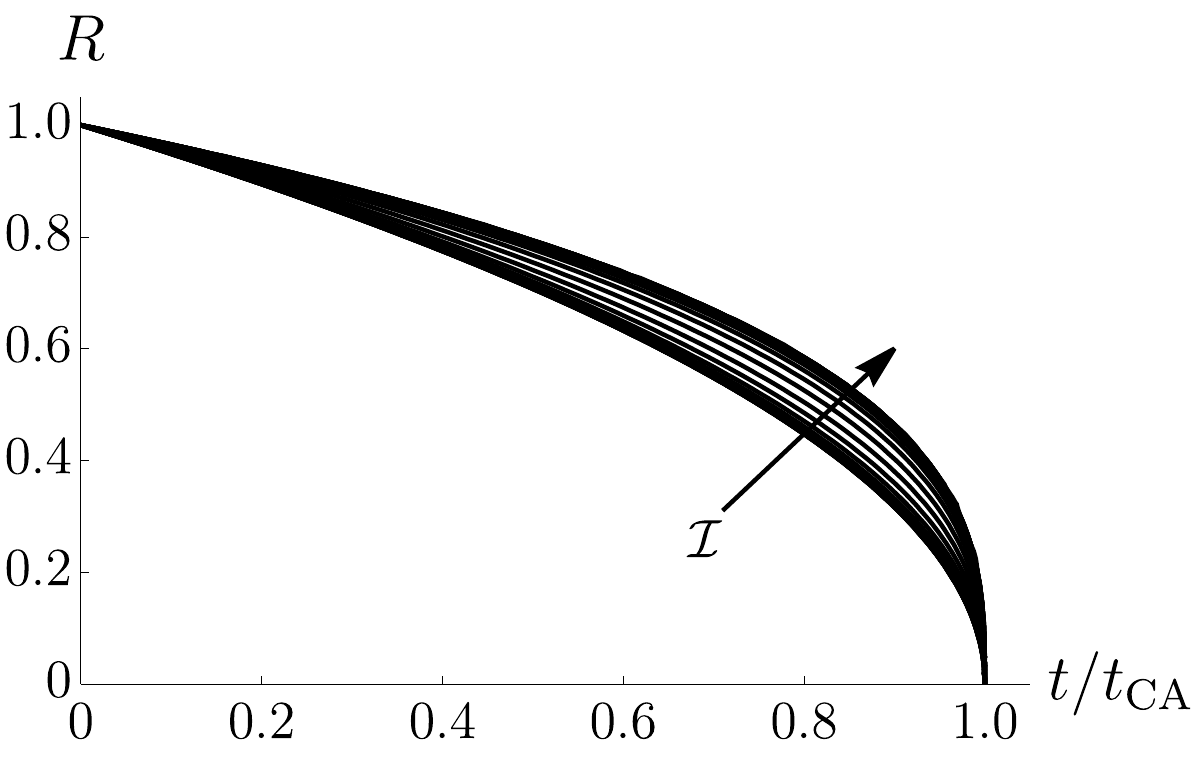}
\caption{}
\end{subfigure}
\caption{
The evolutions of 
(a) $\theta$ for a droplet evolving in the CR mode given by \eqref{eq:CR_evolution} 
plotted as a function of $t/\tCR$
for $\cE=1$ and $\cI=2^n$ ($n=-15,-14,\ldots,15$) 
and 
(b) $R$ for a droplet evolving in the CA mode given by \eqref{eq:CA_evolution} 
plotted as a function of $t/\tCA$
also for $\cE=1$ and $\cI=2^n$ ($n=-15,-14,\ldots,15$).
The arrows indicate the directions of increasing $\cI$.
}
\label{fig:figure4}
\end{figure}

Setting $R \equiv 1$ in \eqref{eq:evolution_2} and solving for $\theta$ shows that a droplet undergoing simultaneous evaporation and imbibition in the CR mode evolves according to
\begin{equation}
\label{eq:CR_evolution}
R \equiv 1, \quad 
\theta = \frac{1}{2\cI}\left[(\cE+2\cI)\exp\left(- \frac{32 \cI}{\pi} t \right) - \cE\right],
\end{equation}
and so has lifetime
\begin{equation}
\label{eq:tCR}
\tCR = \frac{\pi}{32 \cI} \log\left(\frac{\cE+2\cI}{\cE}\right).
\end{equation}
Figure \ref{fig:figure4}(a) shows the evolution of $\theta$ given by \eqref{eq:CR_evolution} for a range of values of $\cI$ plotted as a function of $t/\tCR$. In particular, \eqref{eq:tCR} shows that, as expected, increasing the strength of evaporation and/or imbibition (i.e., increasing $\cE$ and/or $\cI$) shortens the lifetime of the droplet.
In particular, a droplet undergoing pure evaporation evolves according to
\begin{equation}
\label{eq:CR_evolution_evaporation}
R \equiv 1, \quad 
\theta = 1 - \frac{16}{\pi}t,
\end{equation}
and so has lifetime $\tCR=t_\textrm{CR,E}$ given by
\begin{equation}
\label{eq:tCR_evaporation}
t_\textrm{CR,E} = \frac{\pi}{16}
\end{equation}
(see, for example, Wilson \& Duffy \cite{WilsonDuffyChapter2022}), while a droplet undergoing pure imbibition evolves according to
\begin{equation}
\label{eq:CR_evolution_imbibition}
R \equiv 1, \quad 
\theta = \exp\left(- \frac{32}{\pi}t \right),
\end{equation}
and so never completely imbibes, i.e., it has an infinitely long lifetime $\tCR=t_\textrm{CR,I}=\infty$.
Physically the qualitative difference in the behaviour of a droplet undergoing pure evaporation and pure imbibition in the CR mode arises because the global evaporation mass flux $\FE\equiv4$ given by \eqref{eq:FE} remains constant as the droplet evolves, leading to the finite lifetime \eqref{eq:tCR_evaporation} in the case of pure evaporation, whereas, since the leading-order pressure within the droplet $p^{(0)}=2\theta$ given by \eqref{eq:p_0} tends to zero as the droplet evolves, the global imbibition mass flux $\FI=8\theta$ given by \eqref{eq:FI} also tends to zero, leading to an infinitely long lifetime in the case of pure imbibition.

\subsection{Evolution and Lifetime in the CA Mode}
\label{sec:CA_evolution}

Setting $\theta \equiv 1$ in \eqref{eq:evolution_2} and obtaining an implicit equation for $R$ shows that a droplet undergoing simultaneous evaporation and imbibition in the CA mode evolves according to
\begin{equation}
\label{eq:CA_evolution}
\frac{3 \pi}{32 \cE^3} \left[8 \cI^2 \log\left(\frac{\cE + 2 \cI}{\cE R + 2 \cI}\right) + \cE (1-R) \left\{\cE (1+R)-4 \cI\right\}\right] = t, \quad 
\theta \equiv 1,
\end{equation}
and so has lifetime
\begin{equation}
\label{eq:tCA}
\tCA = \frac{3 \pi}{32 \cE^3} \left[8 \cI^2 \log\left(\frac{\cE+2 \cI}{2 \cI}\right) + \cE (\cE-4 \cI)\right].
\end{equation}
Figure \ref{fig:figure4}(b) shows the evolution of $R$ given by \eqref{eq:CA_evolution} for a range of values of $\cI$ plotted as a function of $t/\tCA$. In particular, \eqref{eq:tCA} shows that, as in the CR mode, increasing $\cE$ and/or $\cI$ shortens the lifetime of the droplet.
In particular, a droplet undergoing pure evaporation evolves according to
\begin{equation}
\label{eq:CA_evaporation}
R = \left(1-\frac{32}{3 \pi}t\right)^{1/2}, \quad 
\theta \equiv 1, 
\end{equation}
and so has lifetime $\tCA=t_\textrm{CA,E}$ given by
\begin{equation}
\label{eq:tCA_evaporation}
t_\textrm{CA,E} = \frac{3 \pi}{32} = \frac{3}{2}t_\textrm{CR,E}
\end{equation}
(see, for example, Wilson \& Duffy \cite{WilsonDuffyChapter2022}), while a droplet undergoing pure imbibition evolves according to
\begin{equation}
\label{eq:CA_imbibition}
R = \left(1-\frac{32}{\pi}t\right)^{1/3}, \quad 
\theta \equiv 1, 
\end{equation}
and so has lifetime $\tCA=t_\textrm{CA,I}$ given by
\begin{equation}
\label{eq:tCA_imbibition}
t_\textrm{CA,I} = \frac{\pi}{32}. 
\end{equation}
According to \eqref{eq:CA_evaporation} $R^2$ and $V^{2/3}$ are both linear in $t$, results that are often referred to in the literature as the ``$d^2$ law'' and the ``$2/3$ law'' of pure evaporation, respectively. Similarly, according to \eqref{eq:CA_imbibition} $R^3$ and $V$ are both linear in $t$, and so by analogy we could term these results the ``$d^3$ law'' and the ``$1$ law'' of pure imbibition, respectively.
Physically the qualitative difference in the behaviour of a droplet undergoing pure imbibition in the CR and CA modes arises because, as discussed in Section \ref{sec:CR_evolution}, in the CR mode the global imbibition mass flux $\FI=8\theta$ given by \eqref{eq:FI} tends to zero as the droplet imbibes, leading to an infinitely long lifetime, whereas in the CA mode, since the leading-order pressure within the droplet $p^{(0)}=2/R$ given by \eqref{eq:p_0} tends to infinity as the droplet imbibes, $\FI \equiv 8$ remains constant, leading to the finite lifetime \eqref{eq:tCA_imbibition}.

\subsection{Evolution and Lifetime in the SS Mode}
\label{sec:SS_evolution}

\begin{figure}[tp]
\centering
\begin{subfigure}[b]{0.45\textwidth}
\centering
\includegraphics[keepaspectratio,width=1\textwidth,height=0.2\textheight]{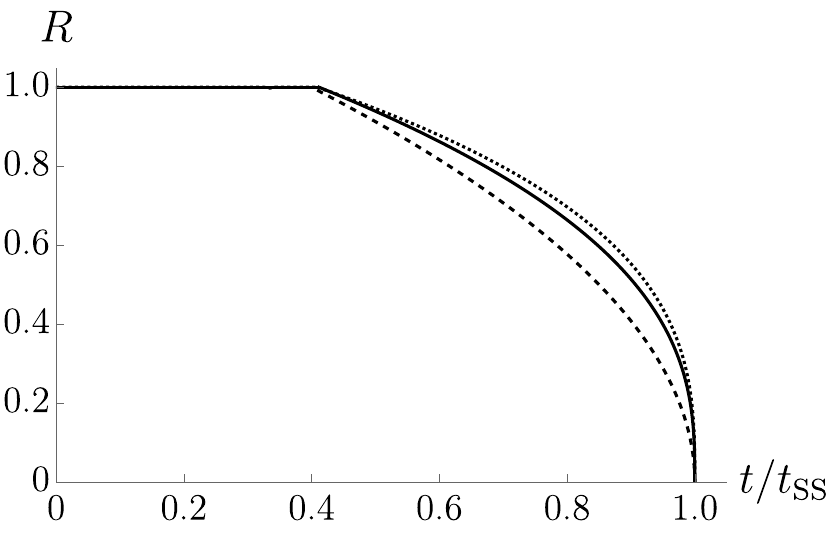}
\caption{}
\end{subfigure}
\hfill
\begin{subfigure}[b]{0.45\textwidth}
\centering
\includegraphics[keepaspectratio,width=1\textwidth,height=0.2\textheight]{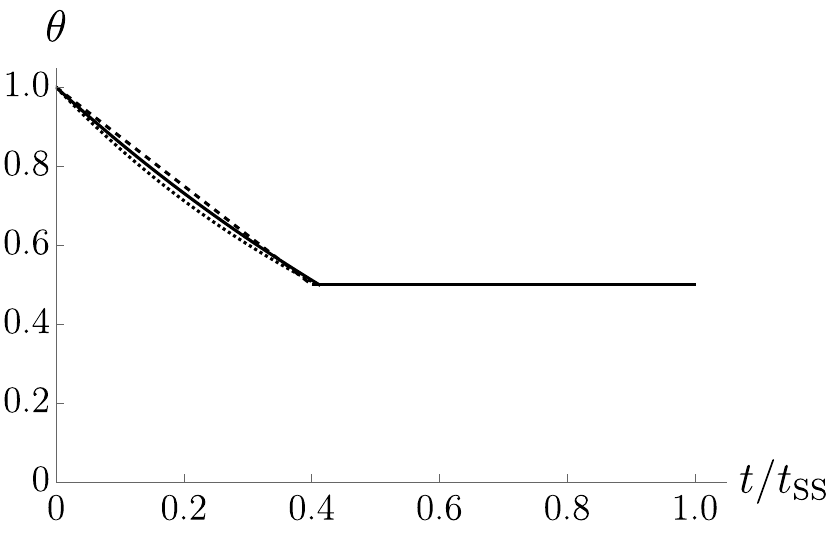}
\caption{}
\end{subfigure}
\caption{
The evolutions of 
(a) $R$ and (b) $\theta$ 
for a droplet evolving in the SS mode 
given by \eqref{eq:SS_evolution_CA_phase} and \eqref{eq:CR_evolution} with \eqref{eq:tstar}
for pure evaporation (dashed line), 
simultaneous evaporation and imbibition with $\cE=\cI=1$ (solid line), 
and 
pure imbibition (dotted line) 
plotted as functions of $t/\tSS$ when $\theta^*=1/2$.
}
\label{fig:figure5}
\end{figure}

\begin{figure}[tp]
\centering
\begin{subfigure}[b]{0.45\textwidth}
\centering
\includegraphics[keepaspectratio,width=1\textwidth,height=0.2\textheight]{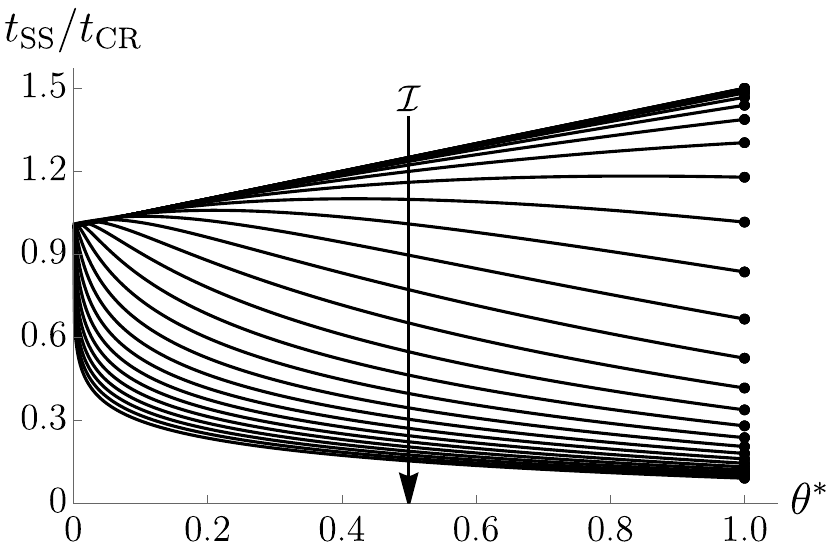}
\caption*{}
\end{subfigure}
\caption{
$\tSS/\tCR$ given by \eqref{eq:tSS} 
plotted as a function of $\theta^*$
for $\cE=1$ and $\cI=2^n$ ($n=-15,-14,\ldots,15$).
The dots ($\bullet$) indicate the values of $\tCA/\tCR$ when $\theta^*=1$, and the arrow indicates the direction of increasing $\cI$.
}
\label{fig:figure6}
\end{figure}

A droplet undergoing simultaneous evaporation and imbibition in the SS mode evolves according to \eqref{eq:CR_evolution} in the CR phase for $0 \le t \le t^*$, where
\begin{equation}
\label{eq:tstar}
t^* = \frac{\pi}{32 \cI}\log\left(\frac{\cE + 2\cI}{\cE + 2\cI\theta^*}\right),
\end{equation}
and according to 
\begin{equation}
\label{eq:SS_evolution_CA_phase}
\frac{3 \pi \theta^*}{32 \cE^3}\left[8 \cI^2 \theta^{*2} \log\left(\frac{\cE + 2 \cI \theta^*}{\cE R + 2 \cI \theta^*}\right) + \cE (1-R) \left\{\cE(1+R)-4 \cI \theta^*\right\}\right] = t - t^*, \quad
\theta \equiv \theta^*
\end{equation}
in the CA phase for $t^* < t \le \tSS$, and so has lifetime 
\begin{equation}
\label{eq:tSS}
\tSS = \frac{\pi}{32 \cE^3 \cI}\left[\cE^3\log\left(\frac{\cE + 2\cI}{\cE + 2\cI\theta^*}\right) + 3 \cI \theta^* \left\{8 \cI^2 \theta^{*2} \log\left(\frac{\cE + 2 \cI \theta^*}{2 \cI \theta^*}\right) + \cE (\cE-4 \cI \theta^*) \right\}\right].
\end{equation}
Figure \ref{fig:figure5} shows the evolutions of $R$ and $\theta$ given by \eqref{eq:SS_evolution_CA_phase} and \eqref{eq:CR_evolution} with \eqref{eq:tstar} for pure evaporation (dashed line), simultaneous evaporation and imbibition (solid line), and pure imbibition (dotted line) plotted as functions of $t/\tSS$ when $\theta^*=1/2$.
Figure \ref{fig:figure6} shows $\tSS/\tCR$ given by \eqref{eq:tSS} for a range of values of $\cI$ plotted as a function of $\theta^*$. In particular, Figure \ref{fig:figure6} illustrates that $\tSS$ is, in general, a non-monotonic function of $\theta^*$, and that $\tSS$ does not, as might have naively been expected, always lie between $\tCR$ and $\tCA$.
In particular, a droplet undergoing pure evaporation evolves according to \eqref{eq:CR_evolution_evaporation} in the CR phase for $0 \le t \le t^*$, where $t^*=t^*_\textrm{E}$ is given by
\begin{equation}
\label{eq:tstar_evaporation}
t^*_\textrm{E} = \frac{\pi (1-\theta^*)}{16},
\end{equation}
and according to 
\begin{equation}
\label{eq:SS_evolution_CA_phase_evaporation}
R = \left[1 - \frac{32}{3 \pi \theta^*} (t-t^*_\textrm{E})\right]^{1/2}, \quad 
\theta \equiv \theta^*
\end{equation}
in the CA phase for $t^* < t \le \tSS$, and so has lifetime $\tSS=t_\textrm{SS,E}$ given by
\begin{equation}
\label{eq:tSS_evaporation}
t_\textrm{SS,E} = \frac{\pi (2 + \theta^*)}{32}
\end{equation}
(see, for example, Wilson \& Duffy \cite{WilsonDuffyChapter2022}).
On the other hand, a droplet undergoing pure imbibition evolves according to \eqref{eq:CR_evolution_imbibition} in the CR phase for $0 \le t \le t^*$, where $t^*=t^*_\textrm{I}$ is given by
\begin{equation}
\label{eq:tstar_imbibition}
t^*_\textrm{I} = \frac{\pi}{32} (-\log\theta^*),
\end{equation}
and according to
\begin{equation}
\label{eq:SS_evolution_CA_phase_imbibition}
R = \left[1 - \frac{32}{\pi} (t-t^*_\textrm{I})\right]^{1/3}, \quad 
\theta \equiv \theta^*
\end{equation}
in the CA phase for $t^* < t \le \tSS$, and so has lifetime $\tSS=t_\textrm{SS,I}$ given by
\begin{equation}
t_\textrm{SS,I} = \frac{\pi}{32} \left(1 - \log\theta^*\right).
\label{eq:tSS_imbibition}
\end{equation}

\subsection{Evolution and Lifetime in the SJ Mode}
\label{sec:SJ_evolution}

\begin{figure}[tp]
\centering
\begin{subfigure}[b]{0.45\textwidth}
\centering
\includegraphics[keepaspectratio,width=1\textwidth,height=0.2\textheight]{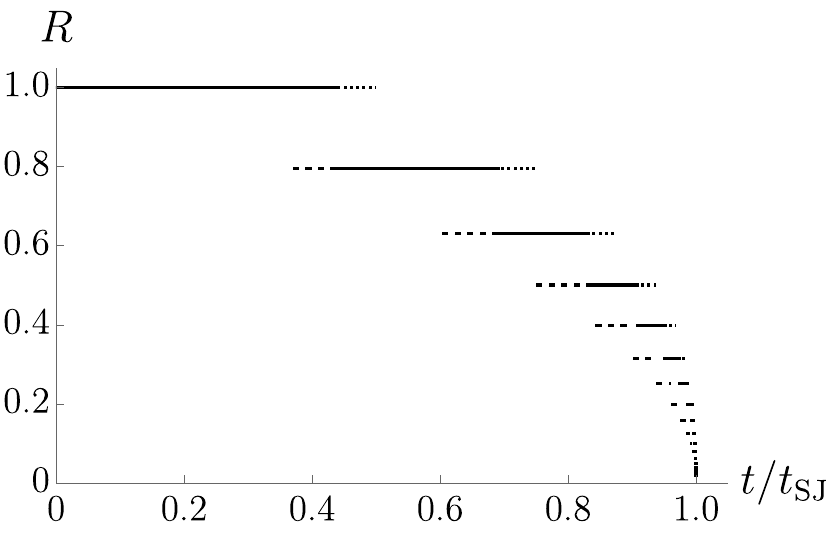}
\caption{}
\end{subfigure}
\hfill
\begin{subfigure}[b]{0.45\textwidth}
\centering
\includegraphics[keepaspectratio,width=1\textwidth,height=0.2\textheight]{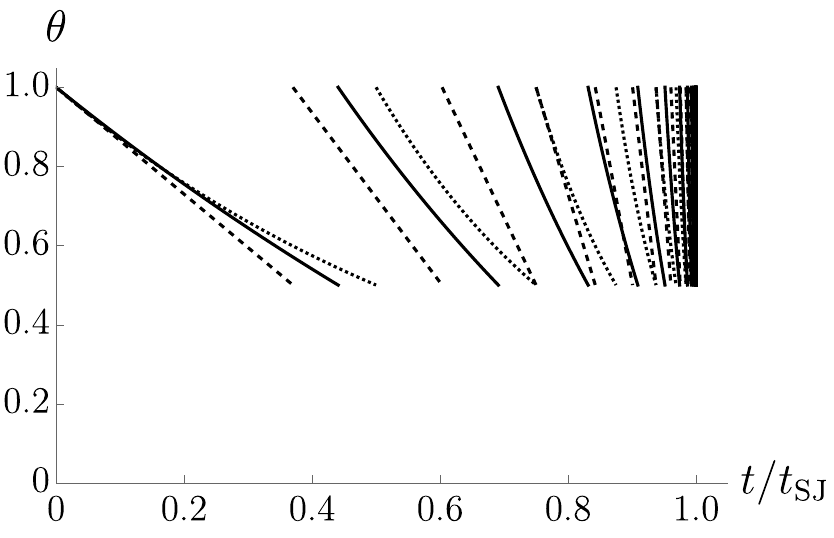}
\caption{}
\end{subfigure}
\caption{
The evolutions of 
(a) $R$ and (b) $\theta$ 
for a droplet evolving in the SJ mode 
given by \eqref{eq:SJ_evolution} with \eqref{eq:tn}
for pure evaporation (dashed line), 
simultaneous evaporation and imbibition with $\cE=\cI=1$ (solid line), and 
pure imbibition (dotted line) 
plotted as functions of $t/\tSJ$ when $\thetamin=1/2$.
}
\label{fig:figure7}
\end{figure}

\begin{figure}[tp]
\centering
\begin{subfigure}[b]{0.45\textwidth}
\centering
\includegraphics[keepaspectratio,width=1\textwidth,height=0.2\textheight]{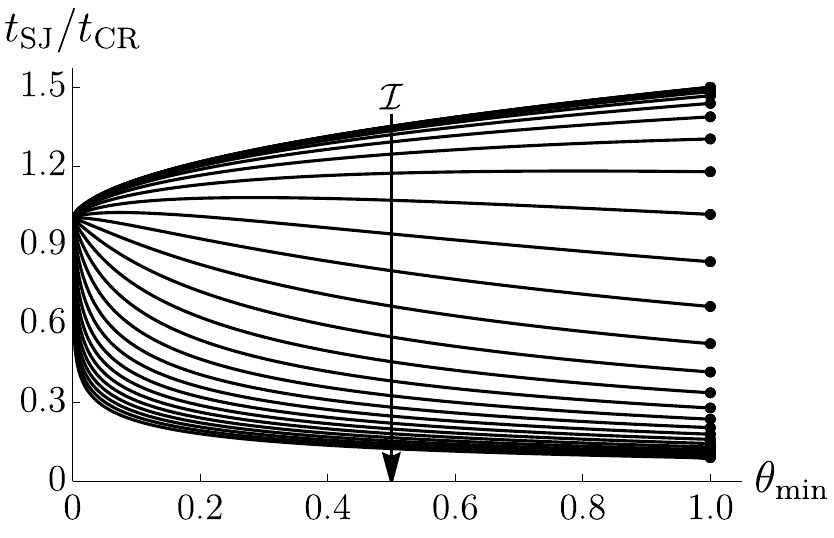}
\caption*{}
\end{subfigure}
\caption{
$\tSJ/\tCR$ given by \eqref{eq:tSJ}
plotted as a function of $\thetamin$
for $\cE=1$ and $\cI=2^n$ ($n=-15,-14,\ldots,15$). 
The dots ($\bullet$) indicate the values of $\tCA/\tCR$ in the limit $\thetamin \to \thetamax^- = 1^-$, and the arrow indicates the direction of increasing $\cI$.
}
\label{fig:figure8}
\end{figure}

A droplet undergoing simultaneous evaporation and imbibition in the SJ mode evolves according to
\begin{equation}
\label{eq:SJ_evolution}
R \equiv R_n = \left( \frac{\thetamin}{\thetamax} \right)^{(n-1)/3}, \quad 
\theta = \left(\thetamax + \frac{\cE R_n}{2 \cI} \right) \exp \left[-\frac{32 \cI}{\pi R_n^3} (t-t_{n-1}) \right] - \frac{\cE R_n}{2 \cI}
\end{equation}
in the $n$th CR phase for $t_{n-1} < t < t_n$ ($n=1,2,3,\ldots$), the duration of the $n$th CR phase, denoted by $\delta t_n=t_n - t_{n-1}$, is given by
\begin{equation}
\label{eq:deltatn}
\delta t_n = \frac{\pi}{32 \cI} \left(\frac{\thetamin}{\thetamax}\right)^{n-1} \log\left[\frac{\cE \left(\frac{\thetamin}{\thetamax}\right)^{(n-1)/3} + 2 \cI \thetamax}{\cE \left(\frac{\thetamin}{\thetamax}\right)^{(n-1)/3} + 2 \cI \thetamin}\right],
\end{equation}
the time at which the $n$th CR phase ends, namely $t=t_n$, is given by 
\begin{equation}
\label{eq:tn}
t_n = \sum_{i=1}^{n} \delta t_{i},
\end{equation}
and so has lifetime
\begin{equation}
\label{eq:tSJ}
\tSJ=\lim_{n \to \infty} t_n,
\end{equation}
which must, in general, be evaluated numerically, which we implemented using Mathematica \cite{Mathematica}.
Figure \ref{fig:figure7} shows the evolutions of $R$ and $\theta$ given by \eqref{eq:SJ_evolution} with \eqref{eq:tn} for pure evaporation (dashed line), simultaneous evaporation and imbibition (solid line), and pure imbibition (dotted line) plotted as functions of $t/\tSJ$ when $\thetamin=1/2$.
Figure \ref{fig:figure8} shows $\tSJ/\tCR$ given by \eqref{eq:tSJ} for a range of values of $\cI$ plotted as a function of $\thetamin$. In particular, Figure \ref{fig:figure8} illustrates, similarly to the variation of $\tSS$ with $\theta^*$ in the SS mode shown in Figure \ref{fig:figure6}, that $\tSJ$ is, in general, a non-monotonic function of $\thetamin$, and that $\tSJ$ does not, as might have naively been expected, always lie between $\tCR$ and $\tCA$.
In particular, a droplet undergoing pure evaporation evolves according to
\begin{equation}
\label{eq:SJ_evolution_evaporation}
R \equiv R_n = \left( \frac{\thetamin}{\thetamax} \right)^{(n-1)/3}, \quad 
\theta = \thetamax - \frac{16}{\pi R_n^2} (t-t_{n-1})
\end{equation}
in the $n$th CR phase for $t_{n-1} < t < t_n$ ($n=1,2,3,\ldots$), with
\begin{equation}
\label{eq:deltatn_evaporation}
\delta t_{n,\textrm{E}} = \frac{\pi}{16} \left( \thetamax-\thetamin\right) \left( \frac{\thetamin}{\thetamax} \right)^{2(n-1)/3}
\end{equation}
and
\begin{equation}
\label{eq:tn_evaporation}
t_{n,\textrm{E}} = \frac{\pi \left(\thetamax-\thetamin\right)}{16} \left[\frac{1-\left(\frac{\thetamin}{\thetamax}\right)^{{2n}/{3}}}{1-\left(\frac{\thetamin}{\thetamax}\right)^{2/3}}\right],
\end{equation}
and so has lifetime
\begin{equation}
\label{eq:tSJ_evaporation}
t_\textrm{SJ,E} = \frac{\pi\left(\thetamax-\thetamin\right)}{16} \frac{\thetamax^{2/3}}{\thetamax^{2/3}-\thetamin^{2/3}}
\end{equation}
(see, for example, Wilson \& Duffy \cite{WilsonDuffyChapter2022}).
On the other hand, a droplet undergoing pure imbibition evolves according to
\begin{equation}
\label{eq:SJ_evolution_imbibition}
R \equiv R_n = \left( \frac{\thetamin}{\thetamax} \right)^{(n-1)/3}, \quad 
\theta =\thetamax \, \exp\left[-\frac{32}{\pi R_n^3}(t-t_{n-1})\right]
\end{equation}
in the $n$th CR phase for $t_{n-1} < t < t_n$ ($n=1,2,3,\ldots$), with
\begin{equation}
\label{eq:deltatn_imbibition}
\delta t_{n,\textrm{I}} = \frac{\pi}{32} \left( \frac{\thetamin}{\thetamax} \right)^{n-1} \log\left(\frac{\thetamax}{\thetamin}\right)
\end{equation}
and
\begin{equation}
\label{eq:tn_imbibition}
t_{n,\textrm{I}} = \frac{\pi}{32} \left[\frac{1-\left(\frac{\thetamin}{\thetamax}\right)^{n}}{1-\left(\frac{\thetamin}{\thetamax}\right)}\right] \log\left(\frac{\thetamax}{\thetamin}\right),
\end{equation}
and so has lifetime
\begin{equation}
\label{eq:tSJ_imbibition}
t_{\textrm{SJ,I}} = \frac{\pi\thetamax}{32(\thetamax-\thetamin)} \log\left(\frac{\thetamax}{\thetamin}\right).
\end{equation}

\section{Flow within the Droplet and the Substrate}
\label{sec:flow}

In this section, we determine the flow within the droplet and the substrate in the CR, CA, SS and SJ modes.

\subsection{Flow in the CR Mode}
\label{sec:flow_CR}

\begin{figure}[tp]
\centering
\begin{subfigure}[b]{0.45\textwidth}
\centering
\includegraphics[keepaspectratio,width=1\textwidth,height=0.2\textheight]{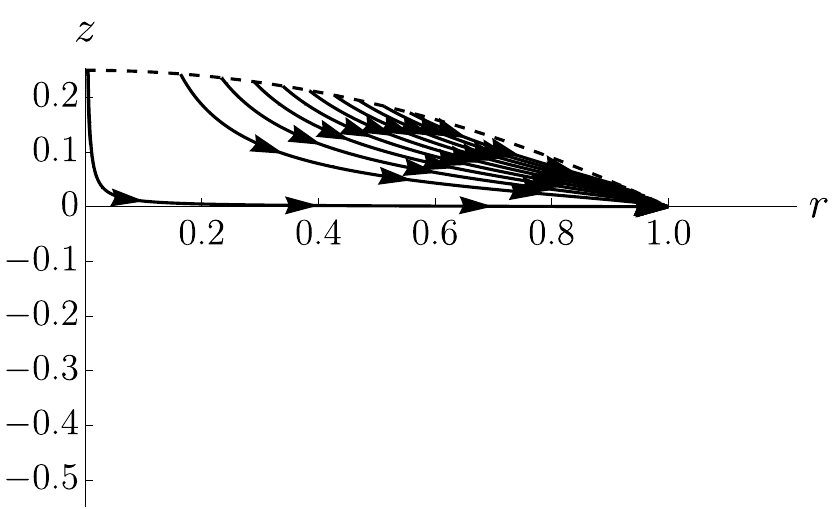}
\caption{}
\end{subfigure}
\hfill
\begin{subfigure}[b]{0.45\textwidth}
\centering
\includegraphics[keepaspectratio,width=1\textwidth,height=0.2\textheight]{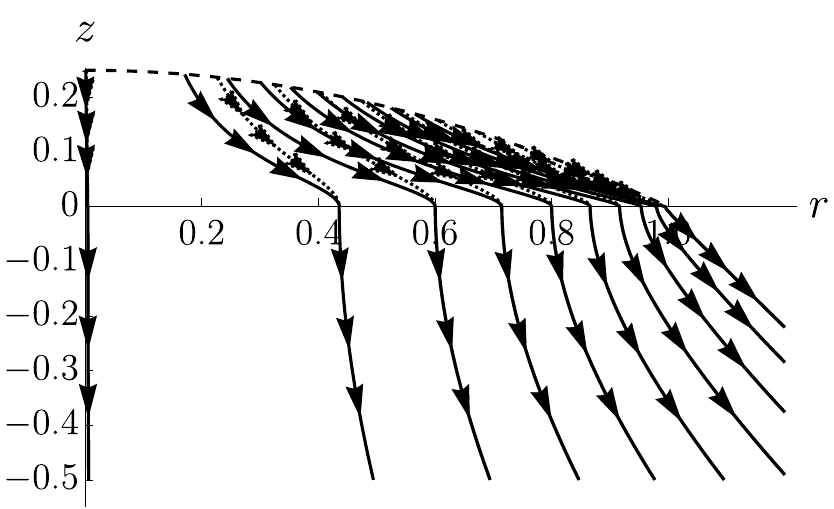}
\caption{}
\end{subfigure}
\caption{
Instantaneous streamlines of the flow within the droplet and the substrate for a droplet evolving in the CR mode calculated from \eqref{eq:u_CR}, \eqref{eq:w_CR}, \eqref{eq:U_CR}, and \eqref{eq:W_CR} for (a) pure evaporation and (b) simultaneous evaporation and imbibition with $\cE=\cI=1$ (solid lines) and pure imbibition (dotted lines) when $\theta=1/2$.
}
\label{fig:figure9}
\end{figure}

Substituting the expressions for $h$, $\JE$, $\JI$, $R$, and $\theta$ given by \eqref{eq:h_solution}, \eqref{eq:JE}, \eqref{eq:JI}, and \eqref{eq:CR_evolution}, respectively, into \eqref{eq:kinematic_integrated} and evaluating the integral, the local radial volume flux $Q$ in the CR mode is given by
\begin{equation}
\label{eq:Q_CR}
Q = \frac{2 (\cE + 2\cI\theta)}{\pi r} \left[ \left(1-r^2\right)^{1/2} - \left(1-r^2\right)^2 \right],
\end{equation}
and hence from \eqref{eq:ubar} the depth-averaged radial velocity $\bar{u}$ in the CR mode is given by
\begin{equation}
\label{eq:ubar_CR}
\bar{u} = \frac{4 (\cE + 2\cI\theta)}{\pi \theta r} \left[\left(1-r^2\right)^{-1/2}-\left(1-r^2\right)\right].
\end{equation} 
Evaluating \eqref{eq:uandw_rewritten} using \eqref{eq:Q_CR} the radial and axial components of the velocity within the droplet $u$ and $w$ are given by
\begin{equation}
\label{eq:u_CR}
u = \frac{24 (\cE + 2\cI\theta) z}{\pi \theta^3 r \left(1-r^2\right)^{5/2}} \left[ 1-\left(1-r^2\right)^{3/2} \right] \left[\theta\left(1-r^2\right)-z\right]
\end{equation}
and
\begin{gather}
w = \frac{4 (\cE + 2\cI\theta) z^2}{\pi \theta^3 \left(1-r^2\right)^{7/2}} \left[ \left\{10-4\left(1-r^2\right)^{3/2}\right\}z - 9 \theta \left(1-r^2\right) \right] - \frac{4 \cI \theta}{\pi \left(1-r^2\right)^{1/2}},
\label{eq:w_CR}
\end{gather}
while evaluating \eqref{eq:UandW} using \eqref{eq:P_solution} the radial and axial components of the velocity within the substrate $U$ and $W$ are given by
\begin{equation}
\label{eq:U_CR}
U = \frac{8 \cI \theta \left( \frac{1+r}{\left[ (1+r)^2+z^{\textrm{s}2} \right]^{1/2}} - \frac{1-r}{\left[ (1-r)^2+z^{\textrm{s}2}\right]^{1/2}}\right)}{\pi \varphi d(d^2-4)^{1/2}}
\end{equation}
and
\begin{equation}
\label{eq:W_CR}
W = \frac{8 \cI \theta z^{\textrm{s}} \left(\frac{1}{\left[ (1+r)^2+z^{\textrm{s}2} \right]^{1/2}} - \frac{1}{\left[ (1-r)^2+z^{\textrm{s}2} \right]^{1/2}}\right)}{\pi \varphi d(d^2-4)^{1/2}},
\end{equation}
where we have introduced the notation
\begin{equation}
\label{eq:d_CR}
d = \left[ (1+r)^2+z^{\textrm{s}2} \right]^{1/2} + \left[ (1-r)^2+z^{\textrm{s}2} \right]^{1/2}.
\end{equation}
Note that $Q$, $\bar{u}$, $u$ and $w+\cI\JI$ depend on $\cE$ and $\cI$ in the combination $\cE+2\cI\theta$, while $U$ and $W$ depend linearly on $\cI\theta$ but are independent of $\cE$.
For a droplet undergoing pure evaporation, \eqref{eq:Q_CR}, \eqref{eq:ubar_CR}, \eqref{eq:u_CR}, and \eqref{eq:w_CR} reduce to the corresponding expressions given by several previous authors (see, for example, Boulogne \etal \cite{boulogne2017coffee}, Gelderblom \etal \cite{gelderblom2022evaporation}\footnote{Note that Gelderblom \etal \cite{gelderblom2022evaporation} define $Q = h \bar{u} r$, and so their expression for $Q$ in the CR mode is equivalent to $Q r$ in the present work.}, and Wilson \& D'Ambrosio \cite{WilsonDAmbrosioReview2023}).

Inspection of \eqref{eq:Q_CR}, \eqref{eq:ubar_CR}, \eqref{eq:u_CR}, and \eqref{eq:w_CR} reveals that $Q \ge 0$, $\bar{u} \ge 0$, $u \ge 0$ and $w \le 0$, i.e., the flow is always outwards and downwards in both the droplet and the substrate. 
In particular, $Q=O(r) \to 0^+$ and $\bar{u}=O(r) \to 0^+$ as $r \to 0^+$ and $Q=O((1-r)^{1/2}) \to 0^+$ and $\bar{u}=O((1-r)^{-1/2}) \to \infty$ as $r \to 1^-$.
Figure \ref{fig:figure9} shows instantaneous streamlines of the flow for a droplet evolving in the CR mode calculated from \eqref{eq:u_CR}, \eqref{eq:w_CR}, \eqref{eq:U_CR}, and \eqref{eq:W_CR} for (a) pure evaporation and (b) simultaneous evaporation and imbibition (solid lines) and pure imbibition (dotted lines), all for the same value of $\theta$. Note that, because they are for the same value of $\theta$, the flow within the substrate is identical for the two situations shown in Figure \ref{fig:figure9}(b).

\subsection{Flow in the CA Mode} 
\label{sec:flow_CA}

Substituting the expressions for $h$, $\JE$, $\JI$, $R$, and $\theta$ given by \eqref{eq:h_solution}, \eqref{eq:JE}, \eqref{eq:JI}, and \eqref{eq:CA_evolution}, respectively, into \eqref{eq:kinematic_integrated} and evaluating the integral, $Q$ is given by
\begin{equation}
\label{eq:Q_CA}
Q = \frac{2 (\cE R + 2 \cI)}{\pi r} \left[\left(1-\frac{r^2}{R^2}\right)^{1/2}-\left(1+\frac{r^2}{3 R^2}\right) \left(1-\frac{r^2}{R^2}\right)\right],
\end{equation}
and hence from \eqref{eq:ubar} $\bar{u}$ is given by
\begin{equation}
\label{eq:ubar_CA}
\bar{u} = \frac{4 (\cE R + 2 \cI)}{\pi R r} \left[ \left(1-\frac{r^2}{R^2}\right)^{-1/2}-\left(1+\frac{r^2}{3 R^2}\right)\right].
\end{equation}
Evaluating \eqref{eq:uandw_rewritten} using \eqref{eq:Q_CA} $u$ and $w$ are given by
\begin{equation}
\label{eq:u_CA}
u = \frac{24 (\cE R + 2 \cI)} z{\pi R^3 r \left(1-\frac{r^2}{R^2}\right)^{5/2}} \left[1-\left(1+\frac{r^2}{3 R^2}\right) \left(1-\frac{r^2}{R^2}\right)^{1/2}\right] \left[R\left(1-\frac{r^2}{R^2}\right)-z\right]
\end{equation}
and
\begin{gather}
w = \frac{8 (\cE R + 2 \cI) z^2}{\pi R^5 \left(1-\frac{r^2}{R^2}\right)^{7/2}} \left[
\left\{5-\frac{2}{3}\left(1-\frac{r^2}{R^2}\right)^{1/2}\left(7+\frac{r^2}{R^2}\right)\right\}z \right. \nonumber \\
\left. \mbox{} + R\left(1-\frac{r^2}{R^2}\right)\left\{-\frac{9}{2}+4\left(1-\frac{r^2}{R^2}\right)^{1/2}\right\} \right] - \frac{4 \cI}{\pi R^2 \left(1-\frac{r^2}{R^2}\right)^{1/2}},
\label{eq:w_CA}
\end{gather}
and evaluating \eqref{eq:UandW} using \eqref{eq:P_solution} $U$ and $W$ are given by 
\begin{equation}
\label{eq:U_CA}
U = \frac{8 \cI \left(\frac{R+r}{\left[ (R+r)^2+z^{\textrm{s}2} \right]^{1/2}}-\frac{R-r}{\left[ (R-r)^2+z^{\textrm{s}2} \right]^{1/2}}\right)}{\pi \varphi d(d^2-4R^2)^{1/2}}
\end{equation}
and
\begin{equation}
\label{eq:W_CA}
W = \frac{8 \cI z^{\textrm{s}} \left(\frac{1}{\left[ (R+r)^2+z^{\textrm{s}2} \right]^{1/2}}-\frac{1}{\left[ (R-r)^2+z^{\textrm{s}2}\right]^{1/2}}\right)}{\pi \varphi d(d^2-4R^2)^{1/2}},
\end{equation}
where we have introduced the notation
\begin{equation}
\label{eq:d_CA}
d = \left[ (R+r)^2+z^{\textrm{s}2} \right]^{1/2} + \left[ (R-r)^2+z^{\textrm{s}2} \right]^{1/2}.
\end{equation}
Note that $Q$, $\bar{u}$, $u$ and $w + \cI\JI$ depend on $\cE$ and $\cI$ in the combination $\cE R+2\cI$, while $U$ and $W$ depend linearly on $\cI$ but are independent of $\cE$.
For a droplet undergoing pure evaporation, \eqref{eq:Q_CA}, \eqref{eq:ubar_CA}, \eqref{eq:u_CA}, and \eqref{eq:w_CA} reduce to the corresponding expressions given by Gelderblom \etal \cite{gelderblom2022evaporation} and D'Ambrosio \etal \cite{DAmbrosio2025movingcontactline}.

Inspection of \eqref{eq:Q_CA}, \eqref{eq:ubar_CA}, \eqref{eq:u_CA}, and \eqref{eq:w_CA} reveals that $Q \ge 0$, $\bar{u} \ge 0$, $u \ge 0$ and $w \le 0$, i.e., as in the CR mode, the flow is always outwards and downwards in both the droplet and the substrate. 
In particular, $Q=O(r) \to 0^+$ and $\bar{u}=O(r) \to 0^+$ as $r \to 0^+$ and $Q=O((R-r)^{1/2}) \to 0^+$ and $\bar{u}=O((R-r)^{-1/2}) \to \infty$ as $r \to R^-$.
The instantaneous streamlines of the flow for a droplet evolving in the CA mode are qualitatively similar to those in the CR mode shown in Figure \ref{fig:figure9}, and so are omitted here for brevity.

\subsection{Flow in the SS Mode}
\label{sec:flow_SS}

Substituting the expressions for $h$, $\JE$, $\JI$, $R$, and $\theta$ given by \eqref{eq:h_solution}, \eqref{eq:JE}, \eqref{eq:JI}, \eqref{eq:CR_evolution}, and \eqref{eq:SS_evolution_CA_phase}, respectively, into \eqref{eq:kinematic_integrated} and evaluating the integral, $Q$ is given by \eqref{eq:Q_CR} in the CR phase for $0 \le t \le t^*$ and by
\begin{equation}
\label{eq:Q_SS}
Q = \frac{2 (\cE R + 2 \cI \theta^*)}{\pi r} \left[\left(1-\frac{r^2}{R^2}\right)^{1/2}-\left(1+ \frac{r^2}{3 R^2}\right) \left(1-\frac{r^2}{R^2}\right)\right]
\end{equation}
in the CA phase for $t^* < t \le \tSS$, and hence from \eqref{eq:ubar} $\bar{u}$ is given by \eqref{eq:ubar_CR} in the CR phase for $0 \le t \le t^*$ and by
\begin{equation}
\label{eq:ubar_SS}
\bar{u} = \frac{4 (\cE R + 2 \cI \theta^*)}{\pi R \theta^* r} \left[\left( 1 - \frac{r^2}{R^2}\right)^{-1/2}-\left(1+\frac{r^2}{3 R^2}\right)\right]
\end{equation}
in the CA phase for $t^* < t \le \tSS$. 
The expressions for $u$, $w$, $U$ and $W$ are given by Craig \cite{craigthesis2025} but are omitted here for brevity, but inspection of \eqref{eq:Q_SS} and \eqref{eq:ubar_SS} reveals that $Q \ge 0$ and $\bar{u} \ge 0$, i.e., the mean radial flow within the droplet is always outwards.
As in the CA mode, the instantaneous streamlines of the flow for a droplet evolving in the SS mode are qualitatively similar to those in the CR mode shown in Figure \ref{fig:figure9}, and so are also omitted here for brevity.

\subsection{Flow in the SJ Mode}
\label{sec:flow_SJ}

Substituting the expressions for $h$, $\JE$, $\JI$, $R$, and $\theta$ given by \eqref{eq:h_solution}, \eqref{eq:JE}, \eqref{eq:JI}, and \eqref{eq:SJ_evolution}, respectively, into \eqref{eq:kinematic_integrated} and evaluating the integral, $Q$ in the $n$th CR phase is given by
\begin{equation}
\label{eq:Q_SJ}
Q = \frac{2 (\cE R_n + 2 \cI \theta)}{\pi r} \left[\left( 1-\frac{r^2}{R_n^2}\right)^{1/2}-\left(1-\frac{r^2}{R_n^2}\right)^2\right] \quad \textrm{for} \quad t_{n-1} < t < t_n \quad (n=1,2,3,\ldots),
\end{equation}
and hence from \eqref{eq:ubar} $\bar{u}$ in the $n$th CR phase is given by
\begin{equation}
\label{eq:ubar_SJ}
\bar{u} = \frac{4 (\cE R_n + 2 \cI \theta)}{\pi R_n \theta r}\left[ \left(1-\frac{r^2}{R_n^2} \right)^{-1/2}-\left(1-\frac{r^2}{R_n^2}\right) \right] \quad \textrm{for} \quad t_{n-1} < t < t_n \quad (n=1,2,3,\ldots).
\end{equation}
As in the SS mode, the expressions for $u$, $w$, $U$ and $W$ are given by Craig \cite{craigthesis2025} but are omitted here for brevity, but inspection of \eqref{eq:Q_SJ} and \eqref{eq:ubar_SJ} reveals that $Q \ge 0$ and $\bar{u} \ge 0$, i.e., the mean radial flow within the droplet is always outwards.
As in the CA and SS modes, the instantaneous streamlines of the flow for a droplet evolving in the SJ mode are qualitatively similar to those in the CR mode shown in Figure \ref{fig:figure9}, and so are also omitted here for brevity.

\section{Transport and Deposition of the Particles}
\label{sec:particles}

In this section, we describe the transport of, and deposition onto the substrate of, the particles. Specifically, we determine the evolution of the concentration of particles within the droplet, $\phi$, and the mass of particles within the droplet, $\Mdrop$, and deposited onto the substrate, $\Mring$ or $\Mrings$ and/or $\Mdist$, and (where appropriate) the final density per unit area of the distributed deposit of particles on the substrate, $\phidist$.
Note that many of the expressions obtained in this section do not depend explicitly on $\cE$ and $\cI$ when they are written in terms of $R$ and/or $\theta$ (although they do, of course, depend implicitly on both $\cE$ and $\cI$ when they are written in terms of $t$). As a consequence, although they differ when expressed as functions of $t$, a number of the expressions given below coincide with those recently obtained by D’Ambrosio \etal \cite{DAmbrosio2025movingcontactline} in their work on the effect of contact-line motion on the deposition of particles from a droplet undergoing pure evaporation.

\subsection{CR Mode}
\label{sec:particles_CR}

For a droplet evolving in the CR mode, the characteristic equations \eqref{eq:characteristic_form} may be expressed as
\begin{subequations}
\label{eq:characteristic_form_CR}
\begin{equation}
\frac{\textrm{d} \phi}{\textrm{d} \theta} = \frac{\phi \left(\cE \JE + \cI \JI \right)}{h \ \textrm{d} \theta/ \textrm{d}t}
\quad \textrm{and} \quad
\frac{\textrm{d}r}{\textrm{d}\theta} = \frac{\bar{u}}{\textrm{d}\theta/\textrm{d}t}.
\tag*{(\ref{eq:characteristic_form_CR}a,b)}
\end{equation}
\end{subequations}
Substituting the expressions for $h$, $\JE$, $\JI$, $R$, $\theta$, and $\bar{u}$ given by equations \eqref{eq:h_solution}, \eqref{eq:JE}, \eqref{eq:JI}, \eqref{eq:CR_evolution}, and \eqref{eq:ubar_CR}, respectively, into (\ref{eq:characteristic_form_CR}b) and solving for $r_0$ yields 
\begin{equation}
\label{eq:r0_CR}
r_0 = \left[1-\theta^{1/2}\left\{\theta^{-3/4} - 1 + \left(1-r^2\right)^{3/2}\right\}^{2/3}\right]^{1/2}.
\end{equation} 
Setting $r=1$ in \eqref{eq:r0_CR} shows that the initial position of a particle that has travelled to the pinned contact line is given by
\begin{equation}
\label{eq:r0at1_CR}
r_0(1,t) = \left[1-\left(1-\theta^{3/4}\right)^{2/3}\right]^{1/2}.
\end{equation}

\begin{figure}[tp]
\centering
\begin{subfigure}{0.45\textwidth}
\centering
\includegraphics[keepaspectratio,width=1\textwidth,height=1\textheight]{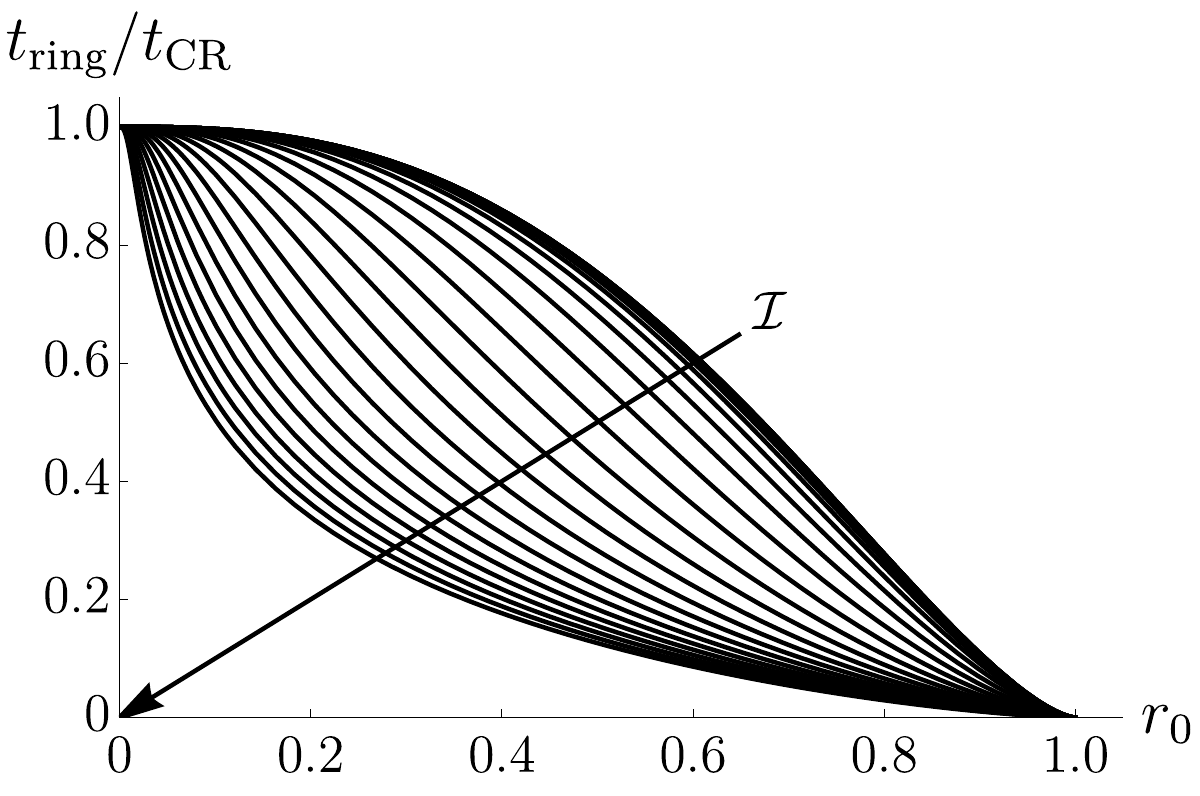}
\caption{}
\end{subfigure}	
\hfill
\begin{subfigure}{0.45\textwidth}
\centering
\includegraphics[keepaspectratio,width=1\textwidth,height=1\textheight]{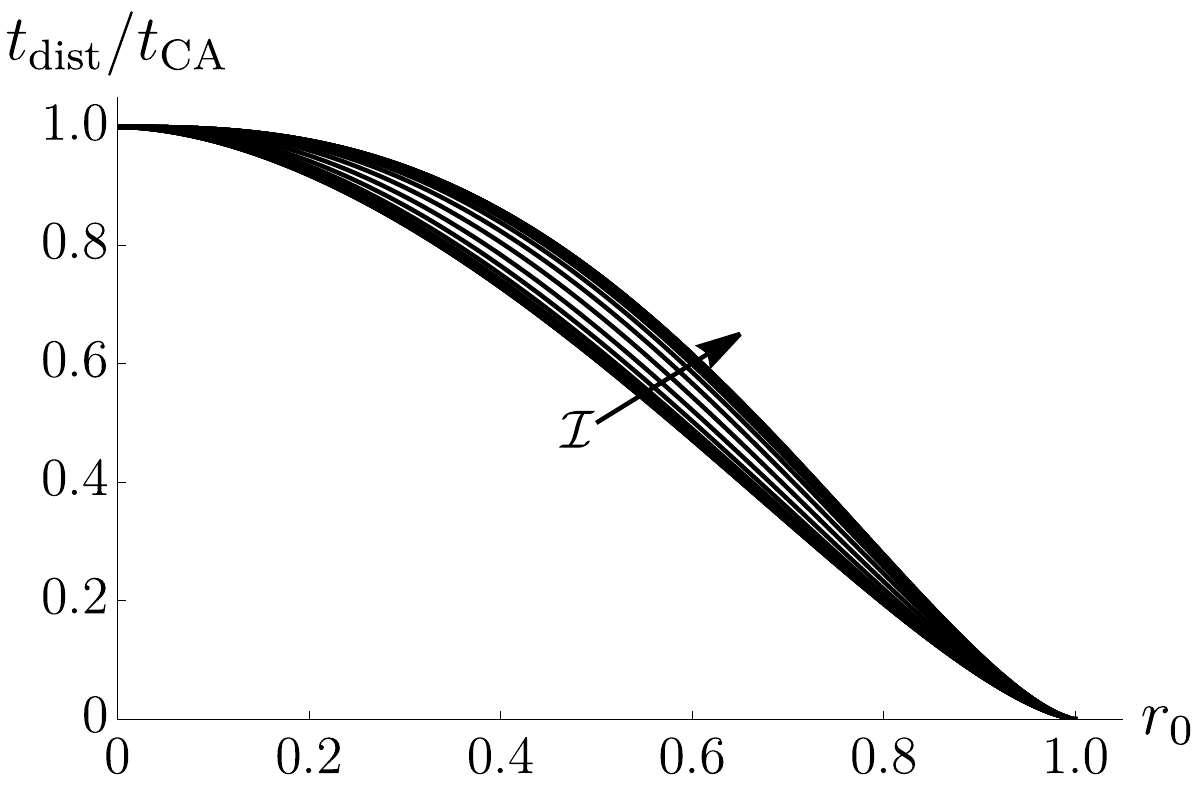}
\caption{}
\end{subfigure}	
\caption{
(a) $\tring/\tCR$ given by \eqref{eq:tring_CR}
and 
(b) $\tdist/\tCA$ obtained substituting \eqref{eq:Rdist} into \eqref{eq:CA_evolution}
plotted as functions of $r_0$ 
for $\cE=1$ and $\cI=2^n$ ($n=-15,-14,\ldots,15$).
The arrows indicate the directions of increasing $\cI$.
}
\label{fig:figure10}
\end{figure}

Substituting the expression for the evolution of $\theta$ in the CR mode given by \eqref{eq:CR_evolution} into \eqref{eq:r0at1_CR} we find that the time for a particle that is at the initial position $r=r_0$ to reach the pinned contact line, denoted by $t=\tring=\tring(r_0)$, is given by
\begin{equation}
\label{eq:tring_CR}
\tring = \frac{\pi}{32 \cI} \log\left[\frac{\cE+2\cI}{\cE+2\cI\left\{1-\left(1-r_0^2\right)^{3/2}\right\}^{4/3}}\right].
\end{equation}
In particular, for a droplet undergoing pure evaporation
\begin{equation}
\label{eq:tring_CR_evap}
\tring = \frac{\pi}{16} \left[1-\left\{1-\left(1-r_0^2\right)^{3/2}\right\}^{4/3}\right],
\end{equation}
while for a droplet undergoing pure imbibition
\begin{equation}
\label{eq:tring_CR_imbib}
\tring = \frac{\pi}{32} \log\left[\left\{1-\left(1-r_0^2\right)^{3/2}\right\}^{-4/3}\right].
\end{equation}
Figure \ref{fig:figure10}(a) shows $\tring/\tCR$ plotted as a function of $r_0$ for a range of values of $\cI$. In particular, Figure \ref{fig:figure10}(a) shows that $\tring$ decreases monotonically from $\tring=\tCR$ at $r_0=0$ to $\tring=0$ at $r_0=1$, i.e., particles that are initially near the contact line are deposited (in the ring deposit) before those that are initially near the centre of the droplet.

\begin{figure}[tp]	
\centering
\begin{subfigure}{0.45\textwidth}
\centering
\includegraphics[keepaspectratio,width=1\textwidth,height=1\textheight]{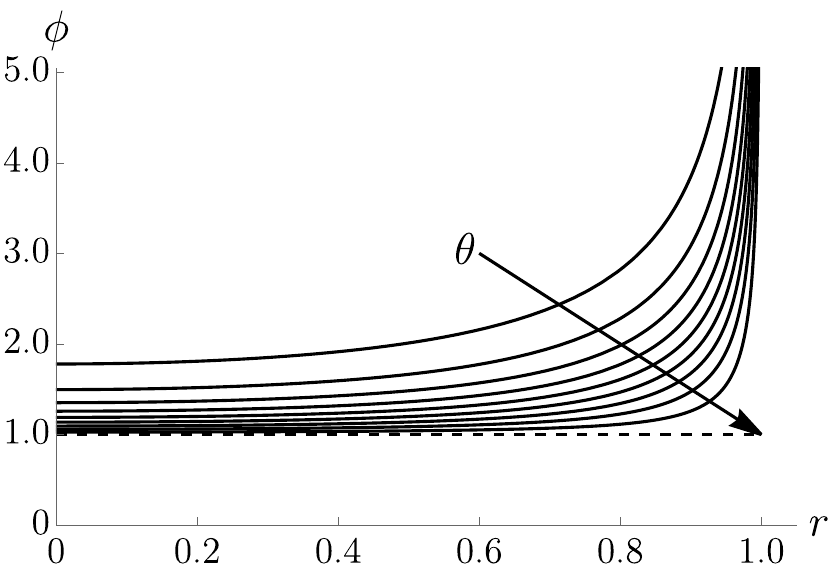}
\caption{}
\end{subfigure}	
\hfill
\begin{subfigure}{0.45\textwidth}
\centering
\includegraphics[keepaspectratio,width=1\textwidth,height=1\textheight]{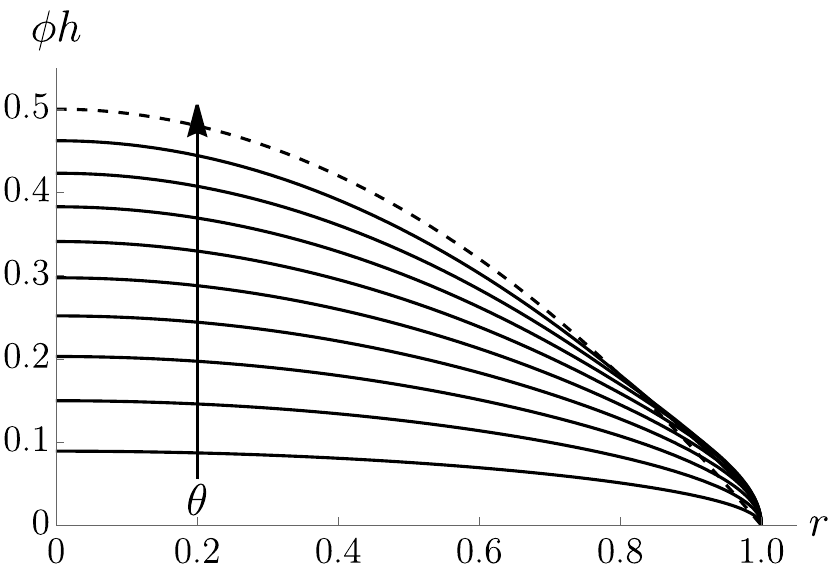}
\caption{}
\end{subfigure}	
\caption{
(a) $\phi$ given by \eqref{eq:phi_CR} 
and 
(b) $\phi h$ given by \eqref{eq:phih_CR} 
in the CR mode
plotted as functions of $r$ 
for $\theta=1/10,1/5,\ldots,9/10$.
The dashed lines indicate the initial values $\phi_0(r)\equiv1$ and $\phi_0(r)h(r,0)=\left(1-r^2\right)/2$ corresponding to $\theta=1$, and
the arrows indicate the directions of increasing $\theta$.
}
\label{fig:figure11}
\end{figure}

Substituting the expression for $r$ given implicitly by \eqref{eq:r0_CR} into (\ref{eq:characteristic_form_CR}a) and solving for $\phi$ shows that
\begin{equation}
\label{eq:phi_CR_r0}
\phi = \left[1 + \left(\theta^{3/4} - 1\right)\left(1-r_0^2\right)^{-3/2}\right]^{-1/3}.
\end{equation}
Using \eqref{eq:r0_CR} to eliminate $r_0$ yields 
\begin{equation}
\label{eq:phi_CR}
\phi = \left[1 + \left(\theta^{-3/4} - 1\right)\left(1-r^2\right)^{-3/2}\right]^{1/3},
\end{equation}
and hence
\begin{equation}
\label{eq:phih_CR}
\phi h = \frac{\theta\left(1-r^2\right)}{2}\left[ 1 + \left(\theta^{-3/4} - 1\right)\left(1-r^2\right)^{-3/2}\right]^{1/3}
\end{equation}
(see, for example, Zheng \cite{zheng2009study} and D'Ambrosio \etal \cite{DAmbrosio2023effect}).
Figure \ref{fig:figure11} shows (a) $\phi$ and (b) $\phi h$ plotted as functions of $r$ for a range of values of $\theta$. In particular, Figure \ref{fig:figure11}(a) illustrates that $\phi$ is a monotonically increasing function of $r$ that takes its minimum value at the centre of the droplet and is singular at the contact-line according to $\phi = O((1-r)^{-1/2}) \to \infty$ as $r \to 1^-$, and a monotonically decreasing function of $\theta$ (i.e., a monotonically increasing function of $t$). On the other hand, Figure \ref{fig:figure11}(b) illustrates that $\phi h$ is a monotonically decreasing function of $r$ that takes its maximum value at the centre of the droplet and goes to zero at the contact-line according to $\phi h = O((1-r)^{1/2}) \to 0^+$ as $r \to 1^-$, and is a monotonically increasing function of $\theta$ (i.e., a monotonically decreasing function of $t$) near the centre of the droplet but a non-monotonic function of $\theta$ (i.e., a non-monotonic function of $t$) near the contact line.
In particular, $\phi h \to 0^+$ everywhere within the droplet as $\theta \to 0^+$ (i.e., as $t \to \tCR^-$), showing that all of the particles eventually leave the droplet, and in this mode are eventually transferred to the ring deposit at $r=1$ (i.e., at the pinned contact line).

\begin{figure}[tp]
\centering
\begin{subfigure}{0.45\textwidth}
\centering
\includegraphics[keepaspectratio,width=1\textwidth,height=1\textheight]{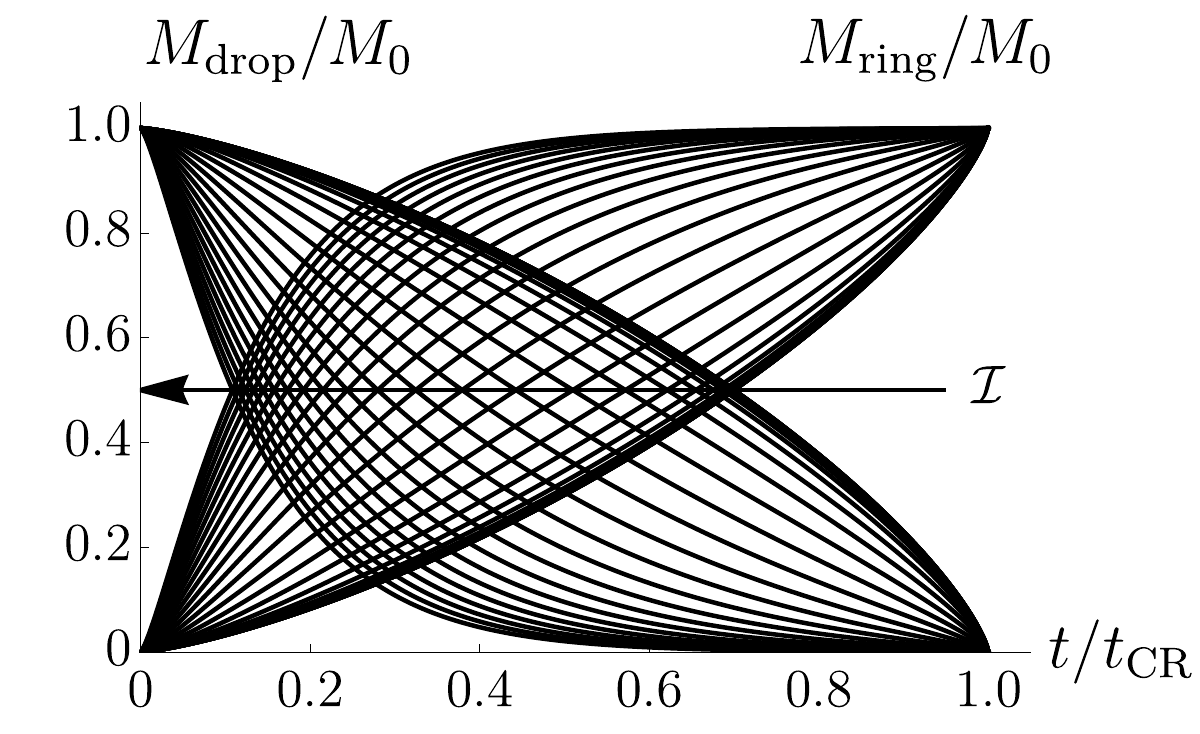}
\caption{}
\end{subfigure}
\hfill
\begin{subfigure}{0.45\textwidth}
\centering
\includegraphics[keepaspectratio,width=1\textwidth,height=1\textheight]{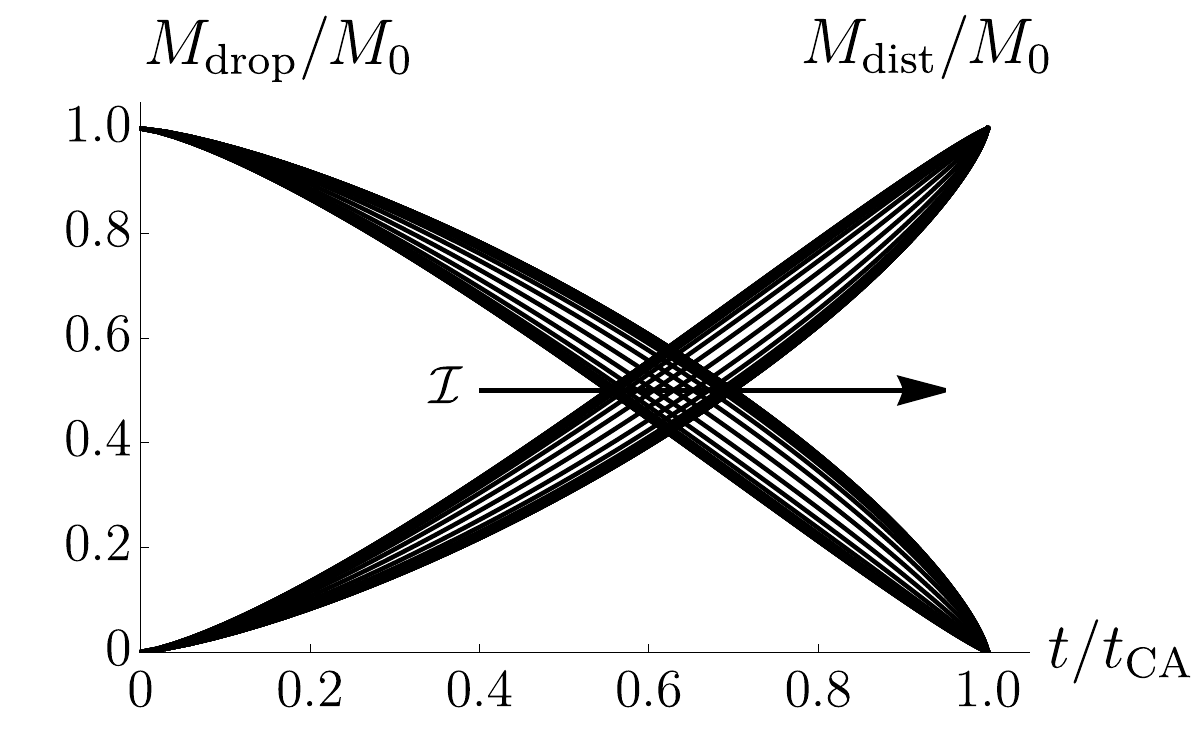}
\caption{}
\end{subfigure}
\caption{
The evolutions of
(a) $\Mdrop/M_0$ and $\Mring/M_0$ 
plotted as functions of $t/\tCR$ in the CR mode 
and 
(b) $\Mdrop/M_0$ and $\Mdist/M_0$ 
plotted as functions of $t/\tCA$ in the CA mode
for $\cE=1$ and $\cI=2^n$ ($n=-15,-14,\ldots,15$).
The arrows indicate the directions of increasing $\theta$.
}
\label{fig:figure12}
\end{figure}

Substituting the expression for $r_0(1,t)$ given by \eqref{eq:r0at1_CR} into \eqref{eq:Mdrop_2} and \eqref{eq:Mring_2} shows that $\Mdrop$ and $\Mring$ are given by 
\begin{equation}
\label{eq:Mdrop_CR}
\Mdrop = M_0\left[1-\left(1-\theta^{3/4}\right)^{4/3}\right]
\end{equation}
and 
\begin{equation}
\label{eq:Mring_CR}
\Mring = M_0\left(1-\theta^{3/4}\right)^{4/3},
\end{equation}
respectively (see, for example, Deegan \etal \cite{deegan2000contact}, Boulogne \etal \cite{boulogne2017coffee}, and D'Ambrosio \etal \cite{DAmbrosio2023effect}), 
together with $\Mdist \equiv 0$.
Figure \ref{fig:figure12}(a) shows the evolutions of $\Mdrop/M_0$ and $\Mring/M_0$ plotted as functions of $t/\tCR$ for a range of values of $\cI$. In particular, Figure \ref{fig:figure12}(a) illustrates how, as in the extensively studied situation of a droplet undergoing pure evaporation in the CR mode, all of the particles initially within the droplet are eventually transferred to the ring deposit at $r=1$.

\subsection{CA Mode}
\label{sec:particles_CA}

For a droplet evolving in the CA mode, the characteristic equations \eqref{eq:characteristic_form} may be expressed as
\begin{subequations}
\label{eq:characteristic_form_CA}
\begin{equation}
\frac{\textrm{d} \phi}{\textrm{d} R} = \frac{\phi \left( \cE \JE + \cI \JI \right)}{h \ \textrm{d} R / \textrm{d}t}
\quad \textrm{and} \quad 
\frac{\textrm{d}r}{\textrm{d}R} = \frac{\bar{u}}{\textrm{d} R/\textrm{d}t}.
\tag*{(\ref{eq:characteristic_form_CA}a,b)}
\end{equation}
\end{subequations}
Substituting the expressions for $h$, $\JE$, $\JI$, $R$, $\theta$, and $\bar{u}$ given by equations \eqref{eq:h_solution}, \eqref{eq:JE}, \eqref{eq:JI}, \eqref{eq:CA_evolution}, and \eqref{eq:ubar_CA}, respectively, into (\ref{eq:characteristic_form_CA}b) and solving for $r_0$ yields
\begin{equation}
\label{eq:r0_CA}
r_0 = \left[1 - R^{3/2}\left\{R^{-9/4} - 1 + \left(1 - \frac{r^2}{R^2}\right)^{3/2}\right\}^{2/3}\right]^{1/2}.
\end{equation}
Setting $r=R$ in \eqref{eq:r0_CA} shows that the initial position of a particle that has travelled to the unpinned contact line is given by
\begin{equation}
\label{eq:r0atR_CA}
r_0(R,t) = \left[1 - \left(1 - R^{9/4}\right)^{2/3}\right]^{1/2}.
\end{equation}

The expression for the evolution of $R$ in the CA mode given by \eqref{eq:CA_evolution} is implicit, but an explicit expression for the time for a particle that is at the initial position $r=r_0$ to reach the unpinned contact line, denoted by $t=\tdist=\tdist(r_0)$, can be obtained by rearranging \eqref{eq:r0atR_CA} for $R$ in terms of $r_0$ to yield
\begin{equation}
\label{eq:Rdist}
R=\left[1-\left(1-r_0^2\right)^{3/2}\right]^{4/9}
\end{equation}
and substituting \eqref{eq:Rdist} into \eqref{eq:CA_evolution} to yield $\tdist$ (omitted here for brevity).
In particular, for a droplet undergoing pure evaporation
\begin{equation}
\label{eq:tdist_CA_evap}
\tdist = \frac{3 \pi}{32} \left[1-\left(1-\left(1-r_0^2\right)^{3/2}\right)^{8/9}\right],
\end{equation}
while for a droplet undergoing pure imbibition
\begin{equation}
\label{eq:tdist_CA_imbib}
\tdist = \frac{\pi}{32} \left[1-\left(1-\left(1-r_0^2\right)^{3/2}\right)^{4/3}\right].
\end{equation}
Figure \ref{fig:figure10}(b) shows $\tdist/\tCA$ plotted as a function of $r_0$ for a range of values of $\cI$. In particular, Figure \ref{fig:figure10}(b) shows that $\tdist$ decreases monotonically from $\tdist=\tCA$ at $r_0=0$ to $\tdist=0$ at $r_0=1$, i.e., particles that are initially near the contact line are deposited (in the distributed deposit) before those that are initially near the centre of the droplet.

\begin{figure}[tp]
\centering
\begin{subfigure}{0.45\textwidth}
\centering
\includegraphics[keepaspectratio,width=1\textwidth,height=1\textheight]{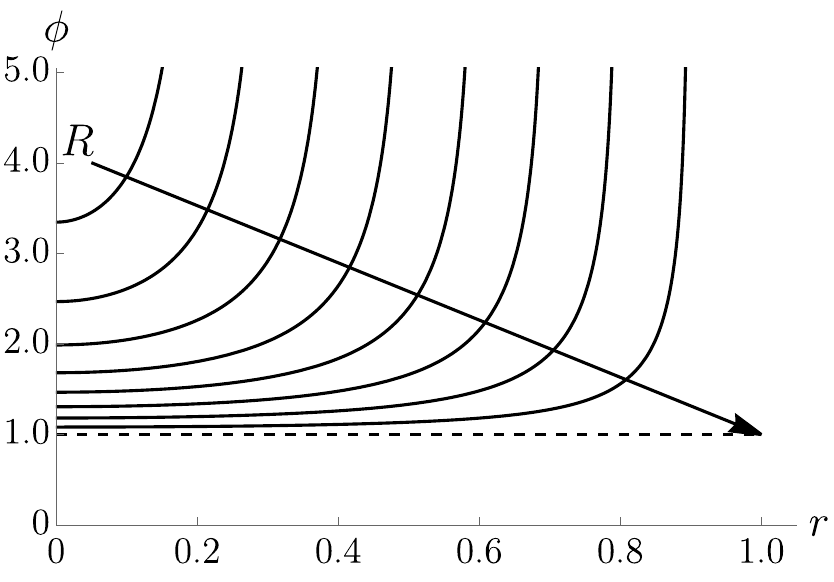}
\caption{}
\end{subfigure}	
\hfill
\begin{subfigure}{0.45\textwidth}
\centering
\includegraphics[keepaspectratio,width=1\textwidth,height=1\textheight]{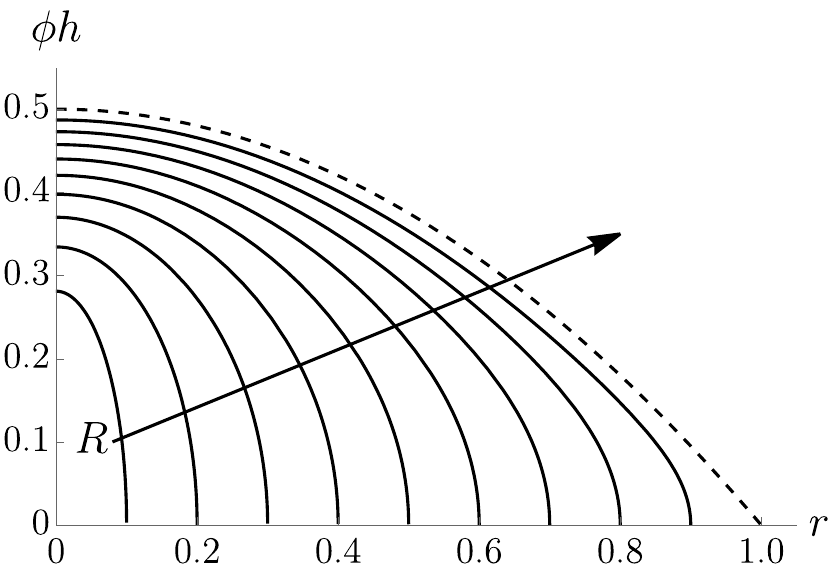}
\caption{}
\end{subfigure}	
\caption{
(a) $\phi$ given by \eqref{eq:phi_CA} 
and 
(b) $\phi h$ given by \eqref{eq:phih_CA} 
in the CA mode
plotted as functions of $r$
for $R=1/10,1/5,\ldots,9/10$.
The dashed lines indicate the initial values $\phi_0(r)\equiv1$ and $\phi_0(r)h(r,0)=\left(1-r^2\right)/2$ corresponding to $R=1$, and
the arrows indicate the directions of increasing $R$.
}
\label{fig:figure13}
\end{figure}

Substituting the expression for $r$ given implicitly by \eqref{eq:r0_CA} into (\ref{eq:characteristic_form_CA}a) and solving for $\phi$ shows that
\begin{equation}
\label{eq:phi_CA_r0}
\phi = \left[1 + \left(R^{9/4} - 1\right)\left(1-r_0^2\right)^{-3/2}\right]^{-1/3}.
\end{equation}
Using \eqref{eq:r0atR_CA} to eliminate $r_0$ yields
\begin{equation}
\label{eq:phi_CA}
\phi = \left[1 + \left(R^{-9/4}-1\right)\left(1 - \frac{r^2}{R^2}\right)^{-3/2}\right]^{1/3},
\end{equation}
and hence
\begin{equation}
\label{eq:phih_CA}
\phi h = \frac{R}{2}\left(1-\frac{r^2}{R^2}\right)\left[1 + \left(R^{-9/4}-1\right)\left(1-\frac{r^2}{R^2}\right)^{-3/2}\right]^{1/3}.
\end{equation}
Figure \ref{fig:figure13} shows (a) $\phi$ and (b) $\phi h$ plotted as functions of $r$ for a range of values of $R$.  In particular, Figure \ref{fig:figure13}(a) illustrates that $\phi$ is a monotonically increasing function of $r$ that takes its minimum value at the centre of the droplet and is singular at the contact line according to $\phi=O((R-r)^{-1/2})$ as $r \to R^-$, and a monotonically decreasing function of $R$ (i.e., a monotonically increasing function of $t$). On the other hand, Figure \ref{fig:figure13}(b) illustrates that $\phi h$ is monotonically decreasing function of $r$ that takes its maximum value at the centre of the droplet and goes to zero at the contact-line according to $\phi h=O((R-r)^{1/2}) \to 0^+$ at $R \to R^-$, and is a monotonically increasing function of $R$ (i.e., a monotonically decreasing function of $t$).
In particular, $\phi h \to 0^+$ everywhere within the droplet as $R \to 0^+$, i.e., as $t \to \tCA^-$, showing that all of the particles eventually leave the droplet, and in this mode are eventually transferred to the distributed deposit in $0 \le r < 1$.

Substituting the expression for $r_0(R,t)$ given by \eqref{eq:r0atR_CA} into \eqref{eq:Mdrop_2} and \eqref{eq:Mdist_2} shows that $\Mdrop$ and $\Mdist$ are given by
\begin{equation}
\label{eq:Mdrop_CA}
\Mdrop = M_0\left[ 1-\left(1-R^{9/4}\right)^{4/3}\right]
\end{equation}
and 
\begin{equation}
\label{eq:Mdist_CA}
\Mdist = M_0\left(1-R^{9/4}\right)^{4/3},
\end{equation}
respectively, together with $\Mring \equiv 0$.
Figure \ref{fig:figure12}(b) shows the evolutions of $\Mdrop/M_0$ and $\Mdist/M_0$ plotted as functions of $t/\tCA$ for a range of values of $\cI$. In particular, Figure \ref{fig:figure12}(b) illustrates how, as in the situation of a droplet undergoing pure evaporation in the CA mode studied by D'Ambrosio \etal \cite{DAmbrosio2025movingcontactline}, all of the particles initially within the droplet are eventually transferred to the distributed deposit in $0 \le r < 1$.

Substituting the expressions for $h$, $r_0$, and $\phi$ given by equations \eqref{eq:h_solution}, \eqref{eq:r0_CA}, and \eqref{eq:phi_CA}, respectively, into \eqref{eq:phidist_CA_equation} shows that the final density per unit area of the distributed deposit of particles on the substrate, $\phidist$, is given by
\begin{equation}
\label{eq:phidist_CA}
\phidist = \frac{3r^{1/4}}{8}\left(1-r^{9/4}\right)^{1/3}
\end{equation}
(see D'Ambrosio \etal \cite{DAmbrosio2025movingcontactline}), 
i.e., the final density of the distributed deposit in the CA mode is zero at both the centre of the droplet and at the initial position of the contact line, and takes its maximum value of $\phidist=3^{4/3}2^{-35/9} \simeq 0.2921$ at $r=2^{-8/9} \simeq 0.5400$.
Note that, perhaps rather unexpectedly, the final density of the distributed deposit $\phidist$ (but not, of course, the temporal evolution towards it) is independent of the values of $\cE$ and $\cI$, i.e., is independent of both the nature and the strength of the physical mechanism(s) driving the mass loss from the droplet.

\subsection{SS Mode}
\label{sec:particles_SS}

For a droplet evolving in the SS mode, the characteristic equations are given by \eqref{eq:characteristic_form_CR} in the CR phase for $0 \le t \le t^*$ and \eqref{eq:characteristic_form_CA} in the CA phase for $t^* < t \le \tSS$. 
Hence the solution for $r_0$ is given by \eqref{eq:r0_CR} in the CR phase and \eqref{eq:r0_CA} in the CA phase, and the corresponding expressions for $\tring$ and $\tdist$ (both omitted here for brevity) can be calculated in the same way as described previously for the CR and CA modes, respectively.

\begin{figure}[tp]
\centering
\begin{subfigure}{0.45\textwidth}
\centering
\includegraphics[keepaspectratio,width=1\textwidth,height=1\textheight]{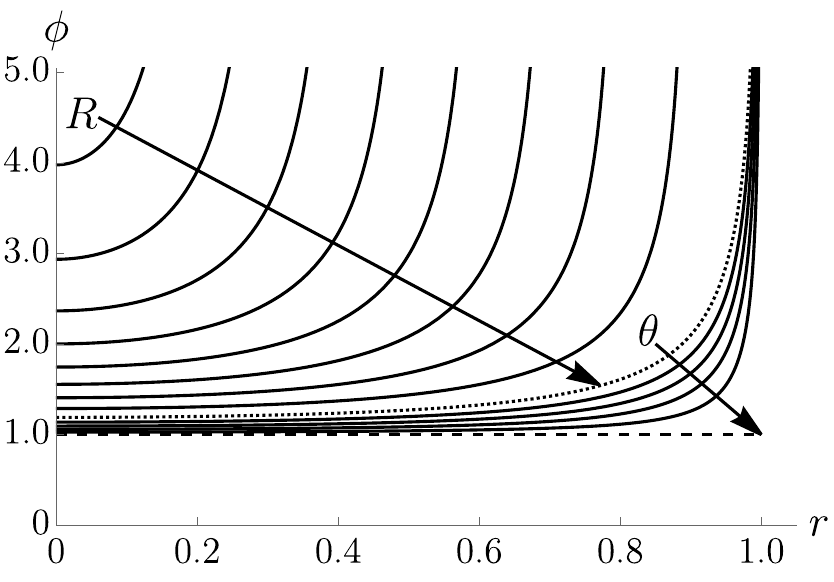}
\caption{}
\end{subfigure}	
\hfill
\begin{subfigure}{0.45\textwidth}
\centering
\includegraphics[keepaspectratio,width=1\textwidth,height=1\textheight]{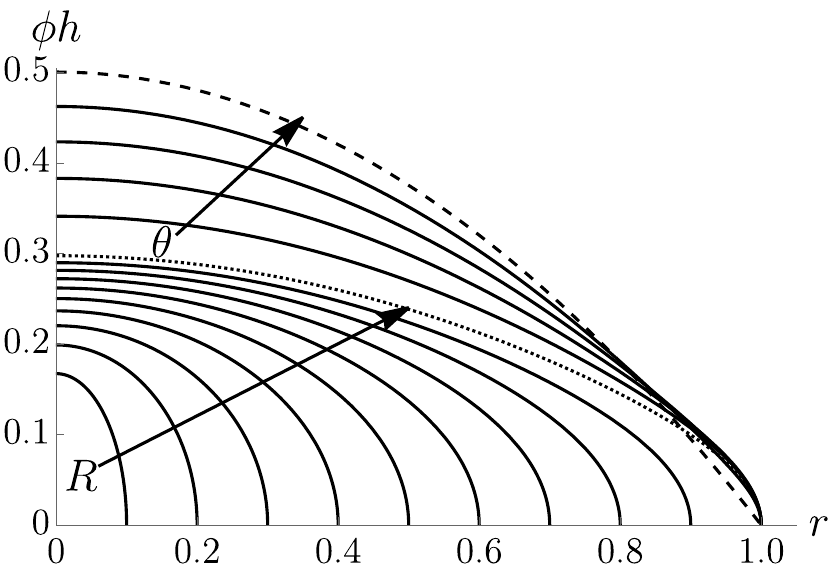}
\caption{}
\end{subfigure}	
\caption{
(a) $\phi$ given by \eqref{eq:phi_CR} and \eqref{eq:phi_CA} 
and 
(b) $\phi h$ given by \eqref{eq:phih_CR} and \eqref{eq:phih_CA}
in the SS mode
plotted as functions of $r$ 
for $\theta=3/5$, $7/10$, $4/5$, $9/10$ in the CR phase and $R=1/10,1/5,\ldots,9/10$ in the CA phase
when $\theta^*=1/2$.
The dashed lines indicate the initial values $\phi_0(r)\equiv1$ and $\phi_0(r)h(r,0)=\left(1-r^2\right)/2$ corresponding to $\theta=1$, 
the dotted lines indicate the solution at $t=t^*$ corresponding to $\theta=\theta^*=1/2$, and
the arrows indicate the directions of increasing $\theta$ and $R$.
}
\label{fig:figure14}
\end{figure}

Solving for $\phi$ yields \eqref{eq:phi_CR} in the CR phase and
\begin{equation}
\label{eq:phi_SS_r0}
\phi = \left[\frac{1+\left(\theta^{*-3/4}-1\right)\left(1-r_0^2\right)^{-3/2}}{1+\left(R^{9/4}-1\right)\left(1-r_0^2\right)^{-3/2}}\right]^{1/3}
\end{equation}
in the CA phase.
Hence $\phi$ is given by \eqref{eq:phi_CR} in the CR phase and
\begin{equation}
\label{eq:phi_SS}
\phi = \left[1 + \left(\theta^{*-3/4}R^{-9/4} - 1\right)\left(1-\frac{r^2}{R^2}\right)^{-3/2}\right]^{1/3}
\end{equation}
in the CA phase, and $\phi h$ is given by \eqref{eq:phih_CR} in the CR phase and
\begin{equation}
\label{eq:phih_SS}
\phi h = \frac{\theta^*R}{2}\left(1-\frac{r^2}{R^2}\right)\left[1 +  \left(\theta^{*-3/4}R^{-9/4}-1\right)\left(1-\frac{r^2}{R^2}\right)^{-3/2}\right]^{1/3}
\end{equation}
in the CA phase.
Figure \ref{fig:figure14} shows (a) $\phi$ and (b) $\phi h$ plotted as functions of $r$ for a range of values of $\theta$ in the CR phase and $R$ in CA phase. In particular, Figure \ref{fig:figure14} illustrates that both quantities inherit their corresponding behaviours in the CR and CA modes in the corresponding phases.
In particular, $\phi h \to 0^+$ everywhere within the droplet as $R \to 0^+$, i.e., as $t \to \tSS^-$, showing that all of the particles eventually leave the droplet, and in this mode are eventually transferred to either the ring deposit at $r=1$ or the distributed deposit in $0 \le r < 1$.

\begin{figure}[tp]
\begin{subfigure}{0.45\textwidth}
\centering
\includegraphics[keepaspectratio,width=1\textwidth,height=1\textheight]{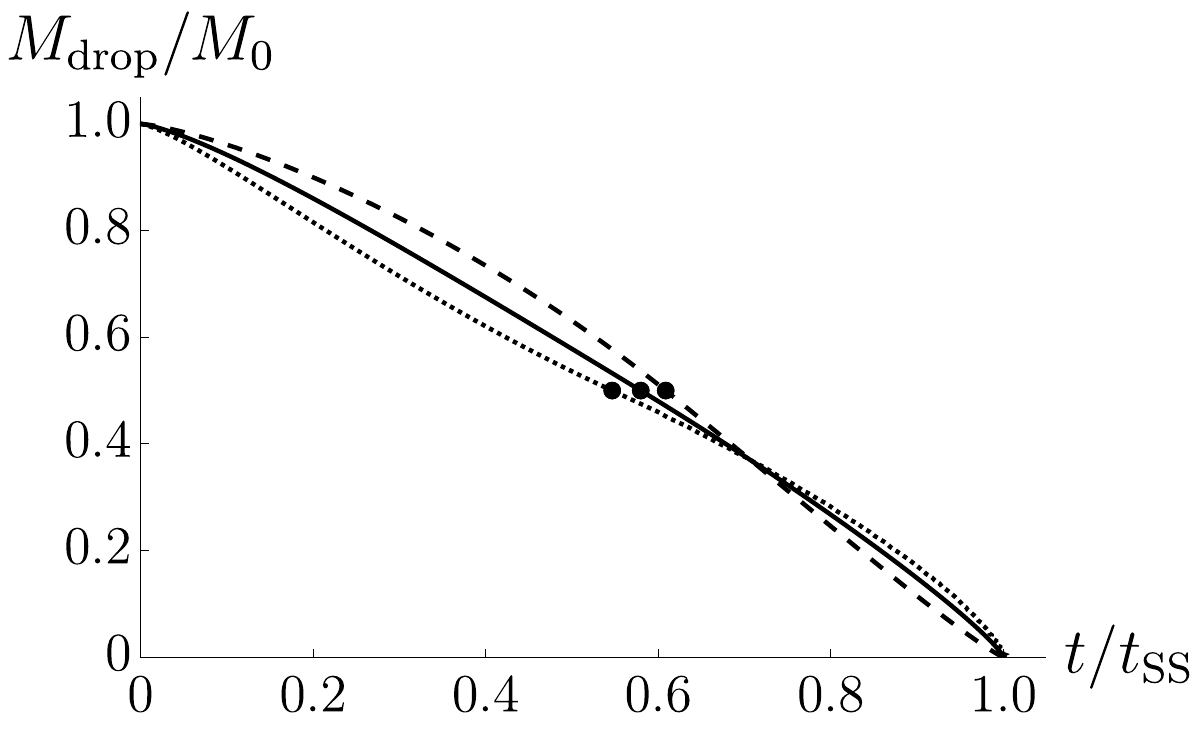}
\caption{}
\end{subfigure}
\hfill
\begin{subfigure}{0.45\textwidth}
\centering
\includegraphics[keepaspectratio,width=1\textwidth,height=1\textheight]{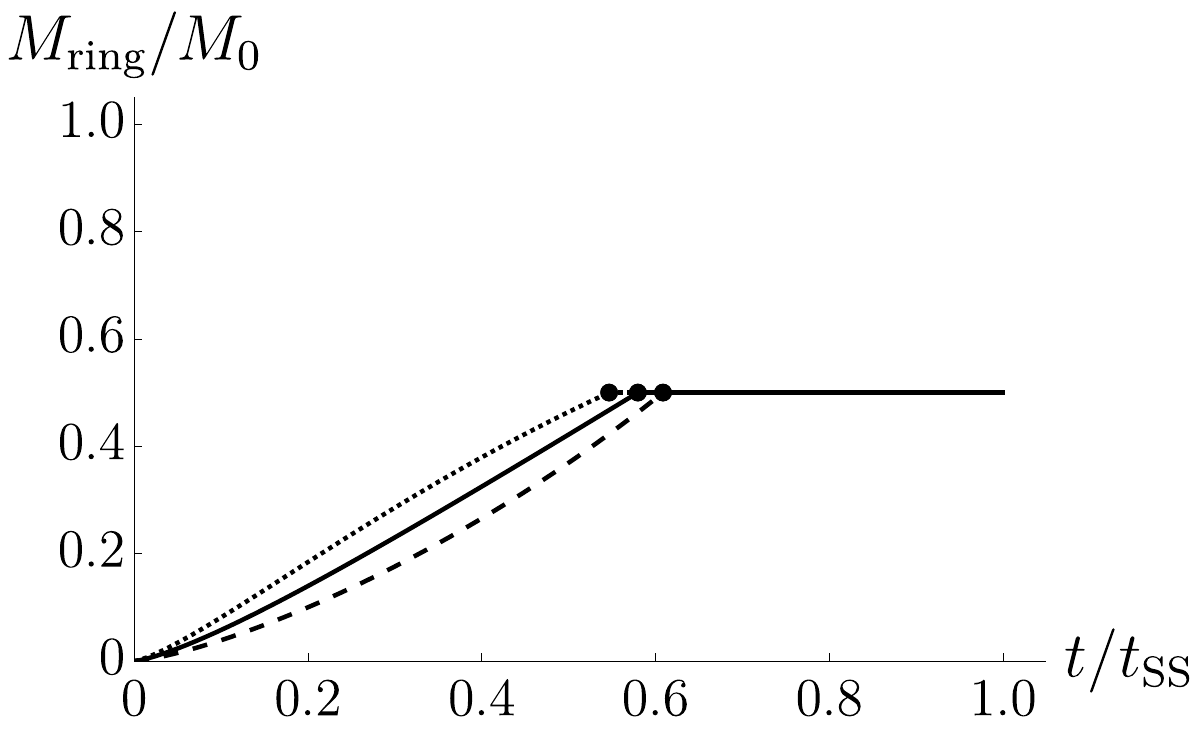}
\caption{}
\end{subfigure}
\hfill
\begin{subfigure}{0.45\textwidth}
\centering
\includegraphics[keepaspectratio,width=1\textwidth,height=1\textheight]{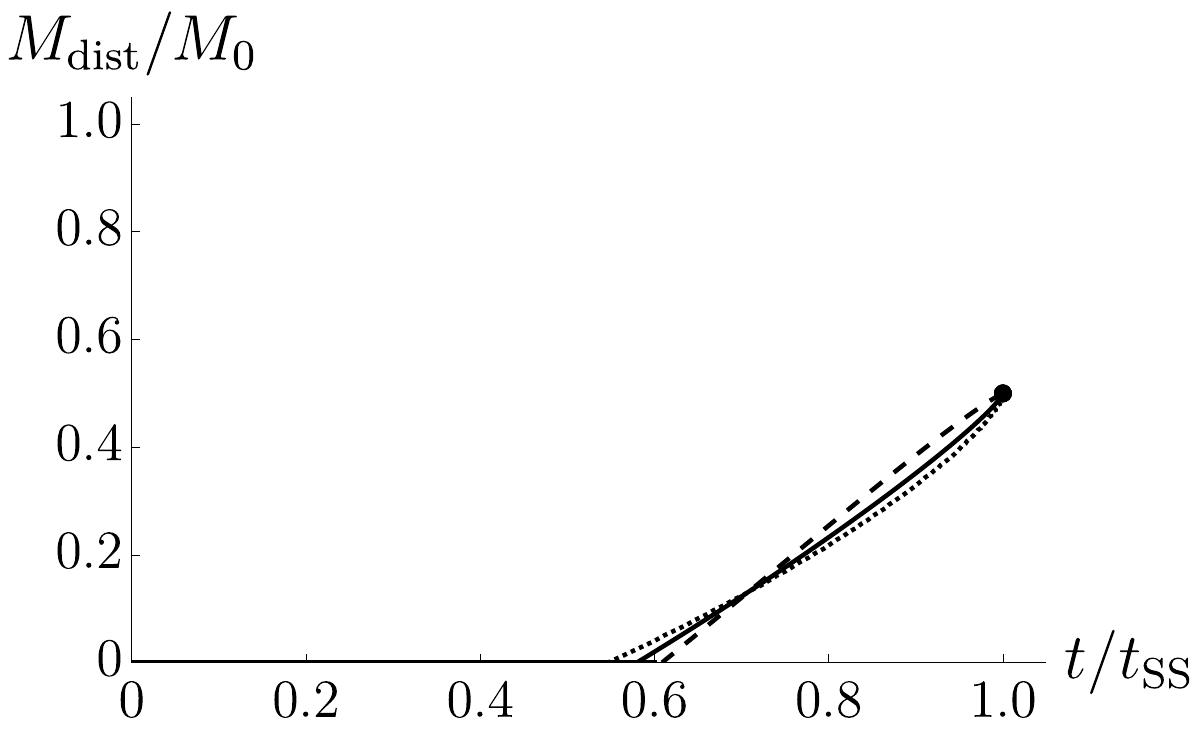}
\caption{}
\end{subfigure}
\hfill
\centering
\begin{subfigure}{0.45\textwidth}
\centering
\includegraphics[keepaspectratio,width=1\textwidth,height=1\textheight]{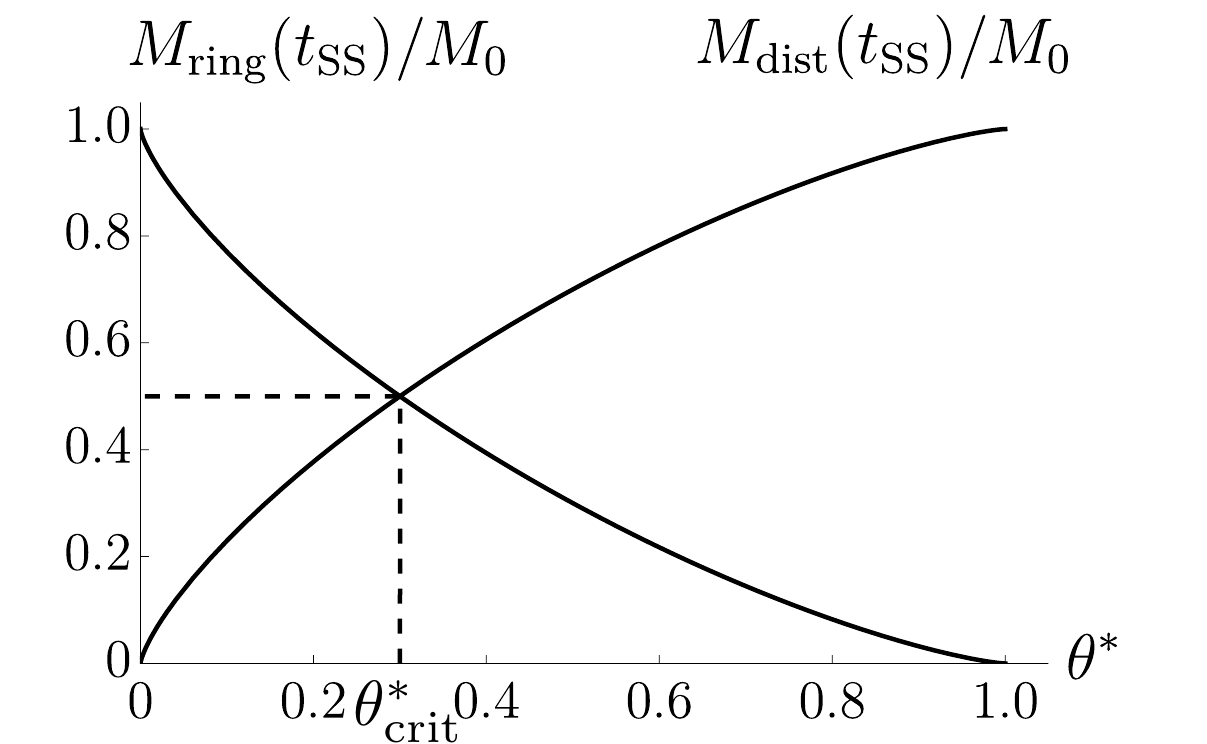}
\caption{}
\end{subfigure}
\caption{
The evolutions of
(a) $\Mdrop/M_0$, (b) $\Mring/M_0$, and (c) $\Mdist/M_0$
in the SS mode
plotted as functions of $t/\tSS$ 
for pure evaporation (dashed line), 
simultaneous evaporation and imbibition with $\cE=\cI=1$ (solid line), and 
pure imbibition (dotted line) 
when $\theta^*=3/10$.
The dots in (a) denote the time at which the CR phase ends, i.e., $t=t^*$.
(d) $\Mring(\tSS)/M_0 \, (=\Mring(t^*)/M_0)$ and $\Mdist(\tSS)/M_0$
in the SS mode
plotted as functions of $\theta^*$
in the SS mode.
The dashed lines in (d) indicate the critical value of $\theta^*=\thetastarcrit=(1-2^{-3/4})^{4/3} \simeq 0.3000$ for which $\Mring(\tSS)/M_0=\Mdist(\tSS)/M_0=1/2$.
}
\label{fig:figure15}
\end{figure}

$\Mdrop$, $\Mring$ and $\Mdist$ are given by \eqref{eq:Mdrop_CR}, \eqref{eq:Mring_CR} and $\Mdist \equiv 0$ in the CR phase, and by
\begin{equation}
\label{eq:Mdrop_SS}
\Mdrop = M_0\left[1-\left(1-\theta^{*3/4} R^{9/4}\right)^{4/3}\right],
\end{equation}
\begin{equation}
\label{eq:Mring_SS}
\Mring \equiv \Mring(t^*) = M_0\left(1-\theta^{*3/4}\right)^{4/3},
\end{equation}
and
\begin{equation}
\label{eq:Mdist_SS}
\Mdist = M_0\left[\left(1-\theta^{*3/4}R^{9/4}\right)^{4/3}-\left(1-\theta^{*3/4}\right)^{4/3}\right]
\end{equation}
in the CA phase.
Figures \ref{fig:figure15}(a,b,c) show the evolutions of (a) $\Mdrop/M_0$, (b) $\Mring/M_0$, and (c) $\Mdist/M_0$ plotted as functions of $t/\tSS$ for pure evaporation (dashed line), simultaneous evaporation and imbibition with $\cE=\cI=1$ (solid line), and pure imbibition (dotted line) when $\theta^*=3/10$, while
Figure \ref{fig:figure15}(d) shows the final (scaled) masses in the deposit ring, $\Mring(\tSS)/M_0 \, (=\Mring(t^*)/M_0)$, and in the distributed deposit, $\Mdist(\tSS)/M_0$, plotted as functions of $\theta^*$. 
In particular, Figure \ref{fig:figure15}(d) illustrates that there is a greater mass of particles in the final deposit ring than in the final distributed deposit (i.e., $\Mring(\tSS)>\Mdist(\tSS)$) when $0 \le \theta^* < \thetastarcrit$, and vice versa (i.e., $\Mring(\tSS)<\Mdist(\tSS)$) for $\thetastarcrit < \theta^* \le 1$, where $\thetastarcrit=(1-2^{-3/4})^{4/3} \simeq 0.3000$ is the critical value of $\theta^*$ for which $\Mring(\tSS)/M_0=\Mdist(\tSS)/M_0=1/2$.

\begin{figure}[tp]
\centering
\begin{subfigure}{0.45\textwidth}
\centering
\includegraphics[keepaspectratio,width=1\textwidth,height=1\textheight]{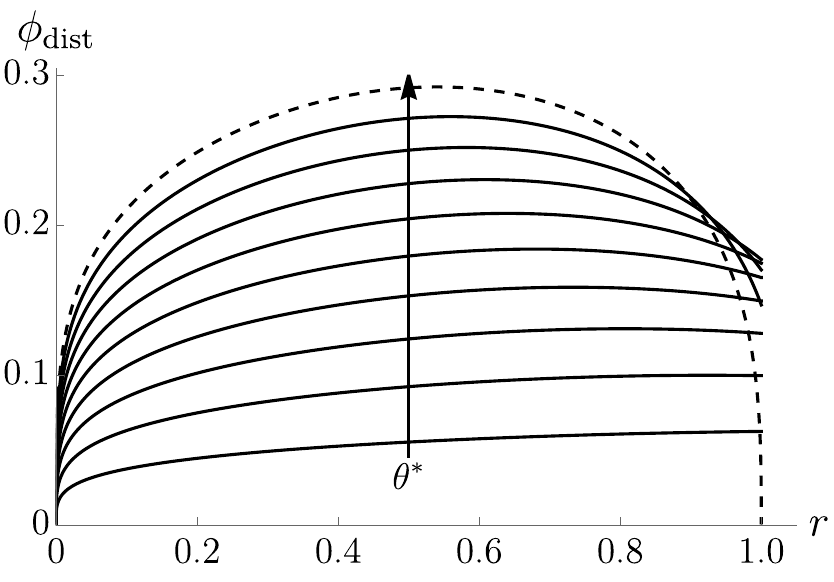}
\end{subfigure}
\caption{
$\phidist$ given by \eqref{eq:phidist_SS}
in the CA phase of the SS mode
plotted as a function of $r$
for $\theta^*=1/10,1/5,\ldots,9/10$.
The dashed line indicates the solution in the CA mode given by \eqref{eq:phidist_CA} corresponding to $\theta^* = 1$, and
the arrow indicates the direction of increasing $\theta^*$.
}
\label{fig:figure16}
\end{figure}

Substituting the expressions for $h$, $r_0$, and $\phi$ given by equations \eqref{eq:h_solution}, \eqref{eq:r0_CA}, and \eqref{eq:phi_SS}, respectively, into \eqref{eq:phidist_SS_equation} shows that in the CA phase of the SS mode $\phidist$ is given by
\begin{equation}
\label{eq:phidist_SS}
\phidist = \frac{3\theta^{*3/4}r^{1/4}}{8}\left(1-\theta^{*3/4}r^{9/4}\right)^{1/3}
\end{equation}
(see D'Ambrosio \etal \cite{DAmbrosio2025movingcontactline}).
Figure \ref{fig:figure16} shows $\phidist$ given by \eqref{eq:phidist_SS} plotted as a function of $r$ for a range of values $\theta^*$, including the solution in the CA mode given by \eqref{eq:phidist_CA} corresponding to $\theta^*=1$. In particular, Figure \ref{fig:figure16} illustrates that the final density of the distributed deposit in the SS mode is zero at the centre of the droplet, and takes its maximum value of $\phidist=3\theta^{*3/4}(1-\theta^{*3/4})^{1/3}/8$ at $r=1$ (i.e., at the initial position of the contact line) for $0 < \theta^* \le \thetastarmin = 2^{-8/3} \simeq 0.1575$ and $\phidist=3^{4/3}2^{-35/9}\theta^{*2/3} \simeq 0.2921\,\theta^{*2/3}$ at $r=2^{-8/9}\theta^{*-1/3} \simeq 0.5400\,\theta^{*-1/3}$ for $\thetastarmin < \theta^* \le 1$.
Note that, as in the CA mode, the final density of the distributed deposit $\phidist$ is independent of the values of $\cE$ and $\cI$, i.e., is independent of both the nature and the strength of the physical mechanism(s) driving the mass loss from the droplet.

\subsection{SJ Mode}
\label{sec:particles_SJ}

For a droplet evolving in the SJ mode, the characteristic equations are given by \eqref{eq:characteristic_form_CR} in the $n$th CR phase for $t_{n-1} < t < t_n$ ($n=1,2,3,\ldots$). 
Hence the solution for $r_{0,n}$ is
\begin{equation}
\label{eq:r0_SJ}
r_{0,n} = R_n \left[ 1 - \left( \frac{\theta}{\thetamax}\right)^{1/2}\Biggl\{\left( \frac{\theta}{\thetamax}\right)^{-3/4} - 1 + \left( 1-\frac{r^2}{R_n^2}\right)^{3/2} \Biggr\}^{2/3} \right]^{1/2}
\end{equation}
for $t_{n-1} < t < t_n$ ($n=1,2,3,\ldots$), and the corresponding expression for the time for a particle that is at the initial position $r = r_{0,n}$ to reach the $n$th pinned contact line, denoted by $t=\tringn=\tringn(r_{0,n})$, (omitted here for brevity) can be calculated in the same way as described previously for $\tring$ in the CR mode (see D'Ambrosio \etal \cite{DAmbrosio2025movingcontactline}).

\begin{figure}[tp]
\centering
\begin{subfigure}{0.45\textwidth}
\centering
\includegraphics[keepaspectratio,width=1\textwidth,height=1\textheight]{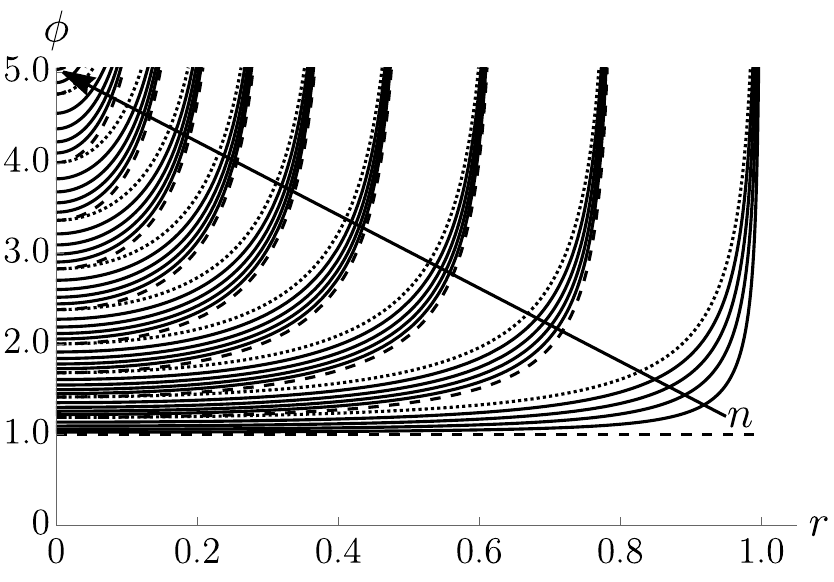}
\caption{}
\end{subfigure}	
\hfill
\begin{subfigure}{0.45\textwidth}
\centering
\includegraphics[keepaspectratio,width=1\textwidth,height=1\textheight]{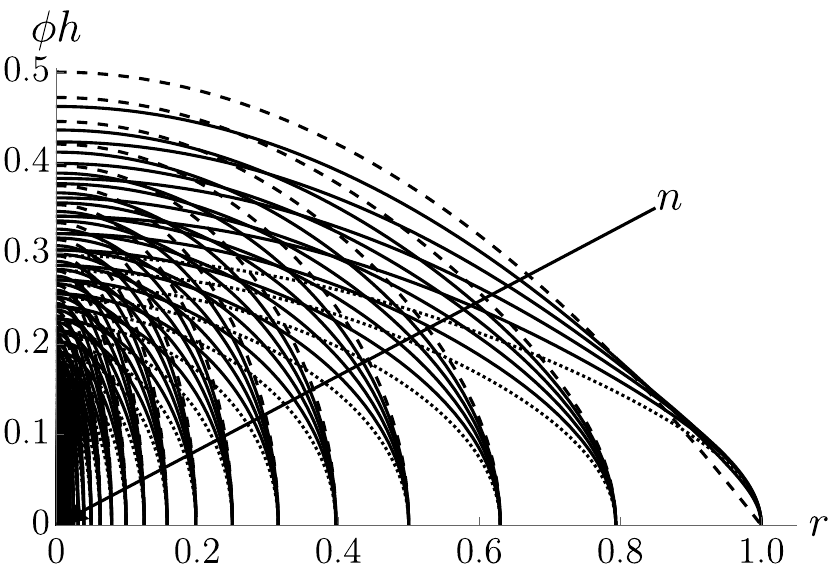}
\caption{}
\end{subfigure}	
\caption{
(a) $\phi$ given by \eqref{eq:phi_SJ_r0} 
and 
(b) $\phi h$ given by \eqref{eq:phih_SJ}
in the SJ mode
plotted as functions of $r$ at $\theta=3/5$, $7/10$, $4/5$, $9/10$ in each CR phase
when $\thetamin=1/2$.
The dashed and the dotted lines indicate the values at $t=t_{n-1}$ and $t=t_n$ ($n=1,2,3,\ldots$) (i.e., at the start and the end of each CR phase) corresponding to $\theta=\thetamax=1$ and $\theta=\thetamin=1/2$, respectively, and
the arrows indicate the directions of increasing $n$.
}
\label{fig:figure17}
\end{figure}

Solving for $\phi$ in the $n$th CR phase yields
\begin{equation}
\label{eq:phi_SJ_r0}
\phi = \phi_{n,\textrm{max}}(r_{0,n}) \left[1+\left\{\left(\frac{\theta}{\thetamax}\right)^{3/4}-1\right\}\left(1-\frac{r_{0,n}^2}{R_n^2}\right)^{-3/2}\right]^{-1/3}
\quad \textrm{for} \quad t_{n-1} < t < t_n \quad (n=1,2,3,\ldots),
\end{equation}
where $\phi_{n,\textrm{max}}(r_{0,n})=\phi(r_{0,n-1},t_{n-1})$ denotes the concentration of particles within the droplet at the end of the $(n-1)$th CR phase. 
Hence $\phi$ is given by
\begin{equation}
\label{eq:phi_SJ}
\phi = \left[1 + \left(\theta^{-3/4}R_n^{-9/4} - 1\right)\left(1-\frac{r^2}{R_n^2}\right)^{-1/2}\right]^{1/3}
\quad \textrm{for} \quad t_{n-1} < t < t_n \quad (n=1,2,3,\ldots)
\end{equation}
and $\phi h$ is given by
\begin{equation}
\label{eq:phih_SJ}
\phi h = \frac{\theta R_n}{2}\left(1-\frac{r^2}{R_n^2}\right)\left[1 + \left(\theta^{-3/4}R_n^{-9/4} - 1\right)\left(1-\frac{r^2}{R_n^2}\right)^{-3/2}\right]^{1/3}
\quad \textrm{for} \quad t_{n-1} < t < t_n \quad (n=1,2,3,\ldots).
\end{equation}
Figure \ref{fig:figure17} shows (a) $\phi$ and (b) $\phi h$ plotted as functions of $r$ for a range of values of $\theta$ in each CR phase. In particular, Figure \ref{fig:figure17} illustrates that both quantities inherit their corresponding behaviour in the CR mode in each CR phase.
In particular, $\phi h \to 0^+$ everywhere within the droplet as $t \to \tSJ^-$, showing that all of the particles eventually leave the droplet, and in this mode are eventually transferred to the (theoretically infinite number of) ring deposits at $r=R_n$ ($n=1,2,3,\ldots$).

\begin{figure}[tp]
\centering
\begin{subfigure}{0.45\textwidth}
\centering
\includegraphics[keepaspectratio,width=1\textwidth,height=1\textheight]{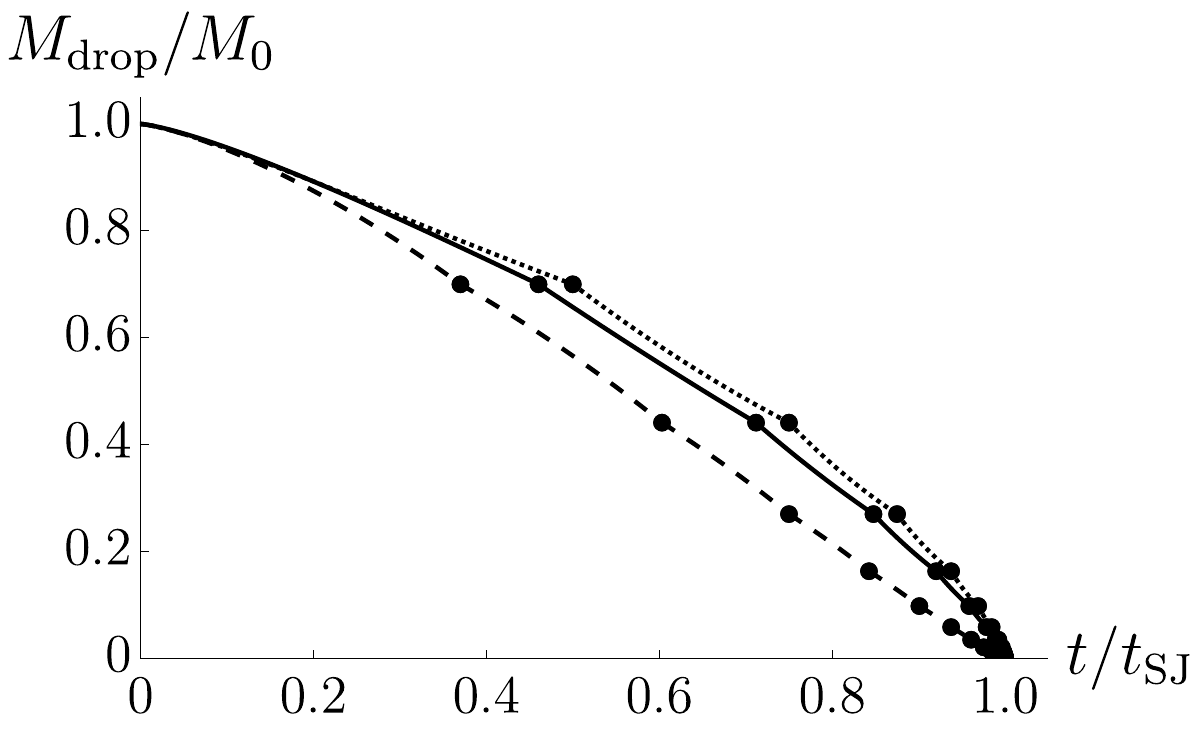}
\caption{}
\end{subfigure}
\hfill
\begin{subfigure}{0.45\textwidth}
\centering
\includegraphics[keepaspectratio,width=1\textwidth,height=1\textheight]{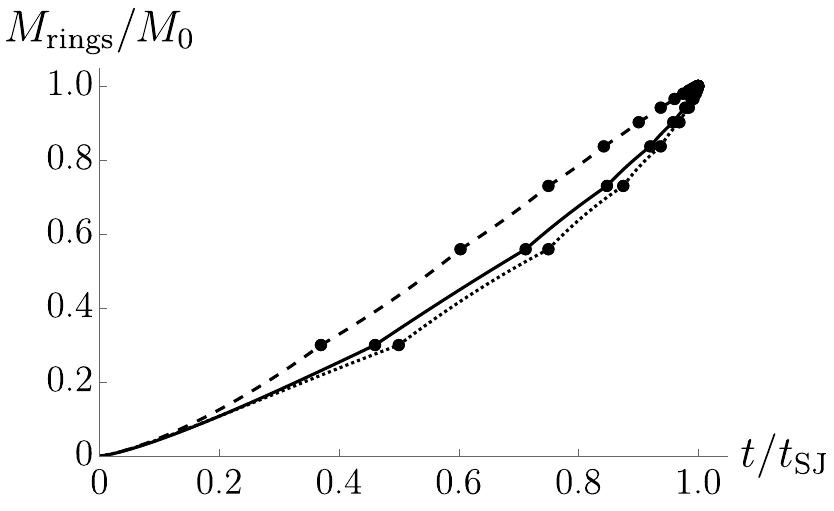}
\caption{}
\end{subfigure}
\hfill
\begin{subfigure}{0.45\textwidth}
\centering		
\includegraphics[keepaspectratio,width=1\textwidth,height=1\textheight]{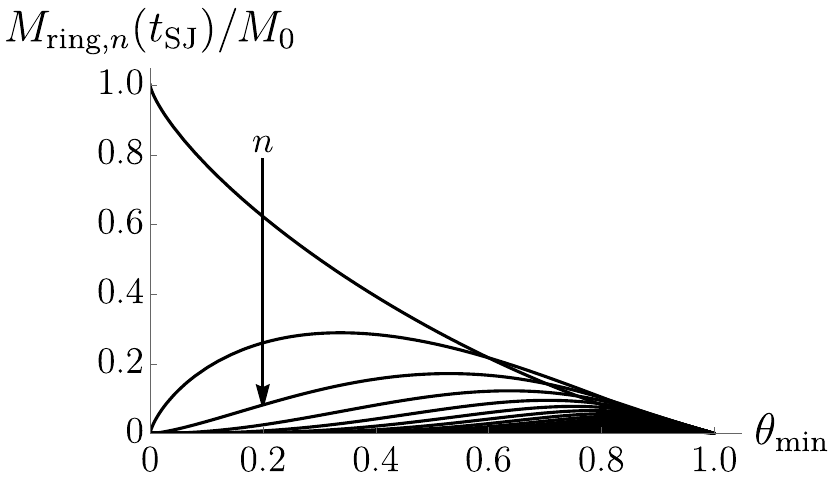}
\caption{}
\end{subfigure}
\caption{
The evolutions of
(a) $\Mdrop/M_0$ and (b) $\Mrings/M_0$
in the SJ mode
plotted as functions of $t/\tSJ$
for pure evaporation (dashed line), 
simultaneous evaporation and imbibition with $\cE=\cI=1$ (solid line), and 
pure imbibition (dotted line)
when $\thetamin=1/2$.
The dots in (a) and (b) denote the times at which the $n$th CR phase ends, i.e., $t=t_n$ ($n=1,2,3,\ldots)$.
(c) 
$\Mringn(\tSJ)/M_0 \, (=\Mringn(t_n)/M_0)$
in the SJ mode
plotted as a function of $\thetamin$
for $n=1,2,3,\ldots$, and
the arrow indicates the direction of increasing $n$.
}
\label{fig:figure18}
\end{figure} 

$\Mdrop$ and $\Mringn$ are given by
\begin{equation}
\label{eq:Mdrop_SJ}
\Mdrop = M_0\left[1-\left(1-\theta^{3/4}R_n^{9/4}\right)^{4/3}\right] 
\quad \textrm{for} \quad t_{n-1} < t < t_n
\end{equation}
and
\begin{equation}
\label{eq:Mrings_SJ}
\Mringn = M_0\left[\left(1-\theta^{3/4}R_n^{9/4}\right)^{4/3}-\left(1-\thetamax^{3/4}R_n^{9/4}\right)^{4/3}\right]
\quad \textrm{for} \quad t_{n-1} < t < t_n,
\end{equation}
together with
$\Mrings$, which is determined by summing the non-zero values of $\Mringn$,
and
$\Mdist \equiv 0$.
Figures \ref{fig:figure18}(a,b) show the evolutions of (a) $\Mdrop/M_0$ and (b) $\Mrings/M_0$ plotted as functions of $t/\tSJ$ for pure evaporation (dashed line), simultaneous evaporation and imbibition with $\cE=\cI=1$ (solid line), and pure imbibition (dotted line) when $\thetamin=1/2$, while Figure \ref{fig:figure18}(c) shows the final (scaled) mass of particles in the $n$th deposit ring, $\Mringn(\tSJ) \, (=\Mringn(t_n)/M_0)$, plotted as function of $\thetamin$ for $n=1,2,3,\ldots$. In particular, Figure \ref{fig:figure18}(c) illustrates that, as D'Ambrosio \etal \cite{DAmbrosio2025movingcontactline} pointed out, $\Mringn(\tSJ)$ decreases monotonically with $n$ for $0 \le \thetamin \le \thetamincrit=(2^{3/4}-1)^{4/3} \simeq 0.6001$, but is non-monotonic for $\thetamincrit \le \thetamin \le 1$.

\section{Particle Paths}
\label{sec:paths}

In Sections \ref{sec:model_transport} and \ref{sec:particles} we considered the transport and deposition of particles in the regime $\hat{\theta}_0^2 \ll \textrm{Pe}^* \ll 1$ in which diffusion of particles is faster than axial advection but slower than radial advection of particles, and hence at leading order axial diffusion causes the concentration of particles to be independent of $z$.
In this section, we calculate the paths of the particles within the droplet in the alternative regime in which diffusion of particles is slower than both radial and axial advection of particles, and hence at leading order diffusion of particles plays no role and the transport of particles is purely advective.
Specifically, we consider the regime in which the reduced P\'{e}clet number satisfies $\textrm{Pe}^* \gg 1$, and hence at leading order the path taken by any particle, denoted by $(r,z)=(r(t),z(t))$, is determined by solving
\begin{equation}
\label{eq:particle_paths_equation_1}
\frac{\textrm{d}r}{\textrm{d}t} = u \quad \textrm{and} \quad \frac{\textrm{d}z}{\textrm{d}t} = w 
\quad \textrm{for} \quad 0 \le z \le h \quad \textrm{and} \quad 0 \le r \le R,
\end{equation}
subject to initial conditions of the form $(r(0),z(0))=(r_0,z_0)$, where $(r_0,z_0)$ is the initial position of the particle. For brevity, only the CR and CA modes will be considered here, but, in principle, the analysis could be extended to other modes (including the SS and SJ modes).

Following the same approach as Kang \etal \cite{kang2016alternative}, D'Ambrosio \etal \cite{DAmbrosio2023effect}, and Coombs \etal \cite{coombs2024colloidal1} we assume that if a particle reaches the (descending) free surface of the droplet then it remains on it until the droplet has completely evaporated and/or imbibed. We refer to this as free-surface capture. Once a particle has been captured, it moves along the free surface according to
\begin{equation}
\label{eq:particle_paths_equation_2}
\frac{\textrm{d}r}{\textrm{d}t} = u \quad \textrm{and} \quad \frac{\textrm{d}z}{\textrm{d}t} = w-\JE = \frac{\partial h}{\partial t} + u \frac{\partial h}{\partial r} 
\quad \textrm{on} \quad z=h \quad \textrm{for} \quad 0 \le r \le R,
\end{equation}
and is eventually transported to the contact line. 
We also assume that once a particle reaches the surface of the substrate it remains there (i.e., it does not enter the substrate or move along its surface) until the droplet has completely evaporated and/or imbibed. We refer to this as substrate capture.
In addition, we assume that the presence of particles on the free surface does not disrupt the shape of the droplet or the flow within it, and that the presence of particles on the surface of substrate does not inhibit the flow of liquid through it (this latter assumption could readily be relaxed in future work if required).
Equations \eqref{eq:particle_paths_equation_1} and \eqref{eq:particle_paths_equation_2} must, in general, be solved numerically, which we implemented using Mathematica \cite{Mathematica} routine {\tt NDSolve[]}. Specifically, we used the method of lines and confirmed convergence under refinement.

\subsection{Particle Paths in the CR Mode}
\label{sec:paths_CR}

\begin{figure}[tp]
\centering
\begin{subfigure}{0.45\textwidth}
\centering
\includegraphics[keepaspectratio,width=1\textwidth,height=1\textheight]{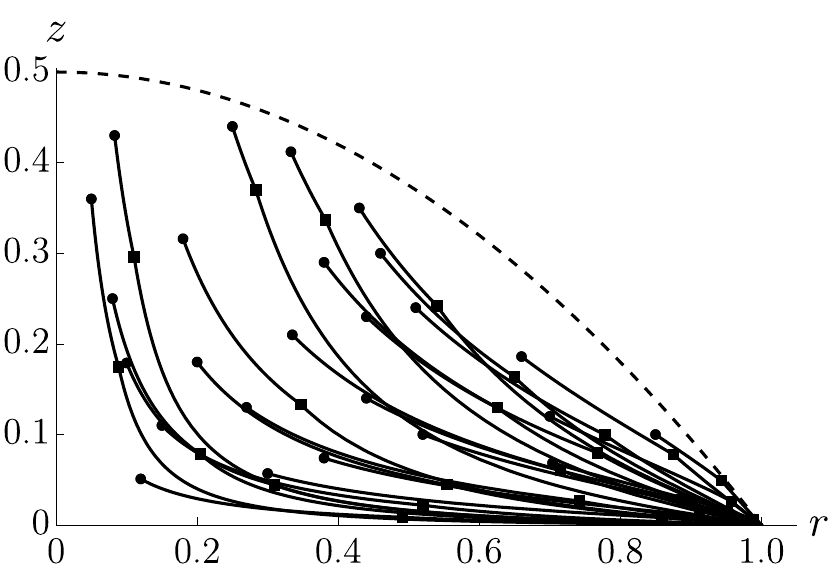}
\caption{}
\end{subfigure}
\hfill
\begin{subfigure}{0.45\textwidth}
\centering
\includegraphics[keepaspectratio,width=1\textwidth,height=1\textheight]{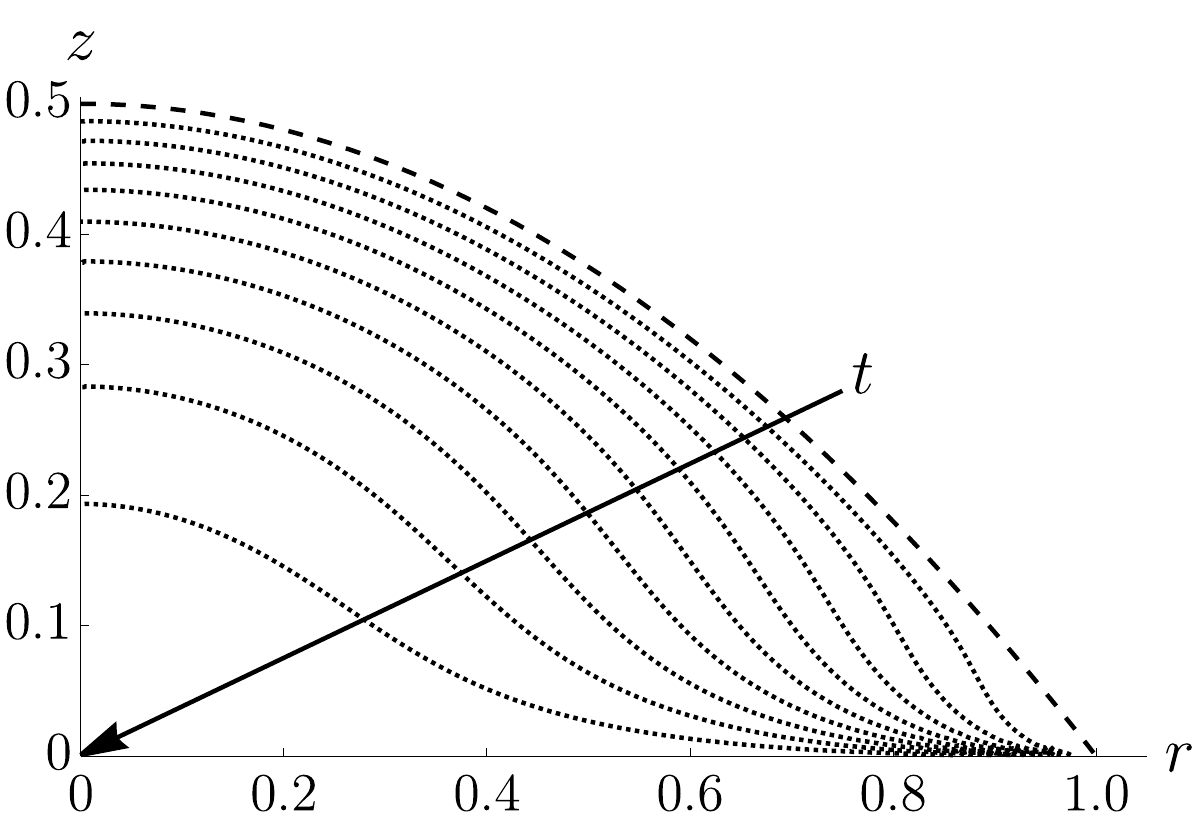}
\caption{}
\end{subfigure}
\centering
\begin{subfigure}{0.45\textwidth}
\centering
\includegraphics[keepaspectratio,width=1\textwidth,height=1\textheight]{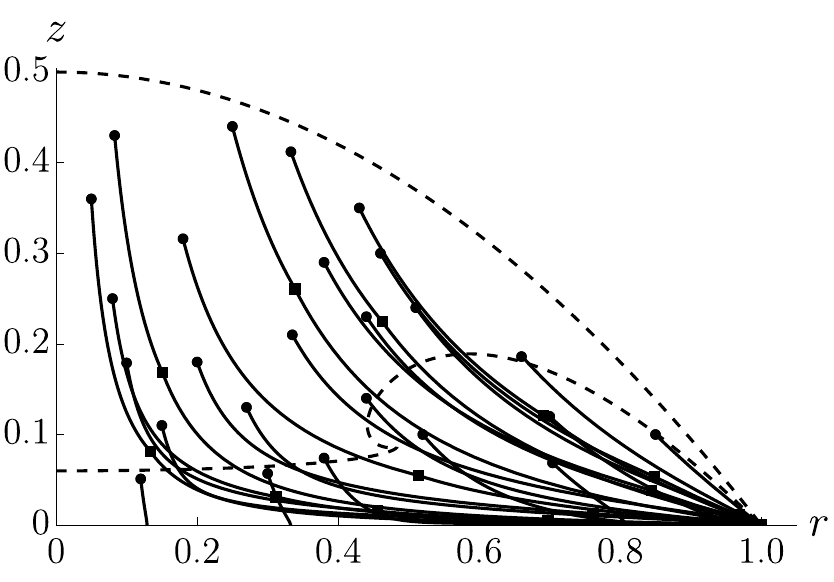}
\caption{}
\end{subfigure}
\hfill
\begin{subfigure}{0.45\textwidth}
\centering
\includegraphics[keepaspectratio,width=1\textwidth,height=1\textheight]{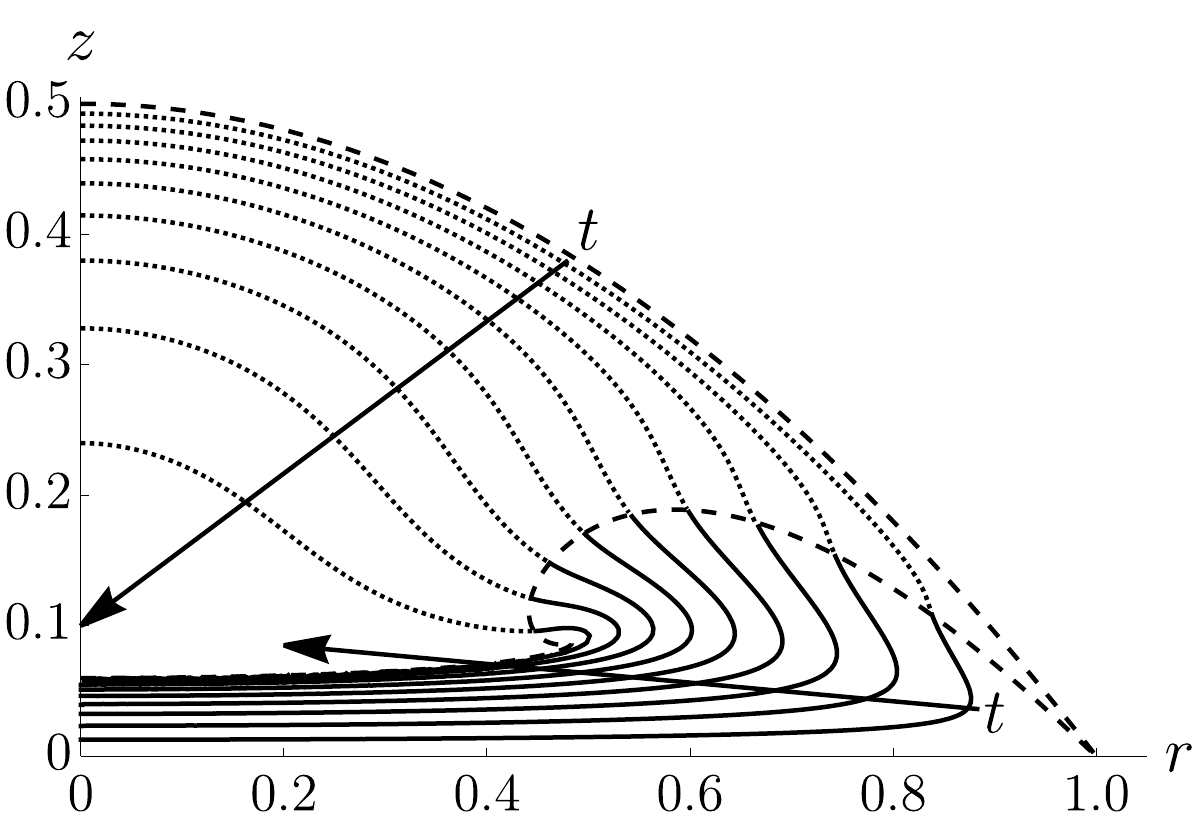}
\caption{}
\end{subfigure}
\centering
\begin{subfigure}{0.45\textwidth}
\centering
\includegraphics[keepaspectratio,width=1\textwidth,height=1\textheight]{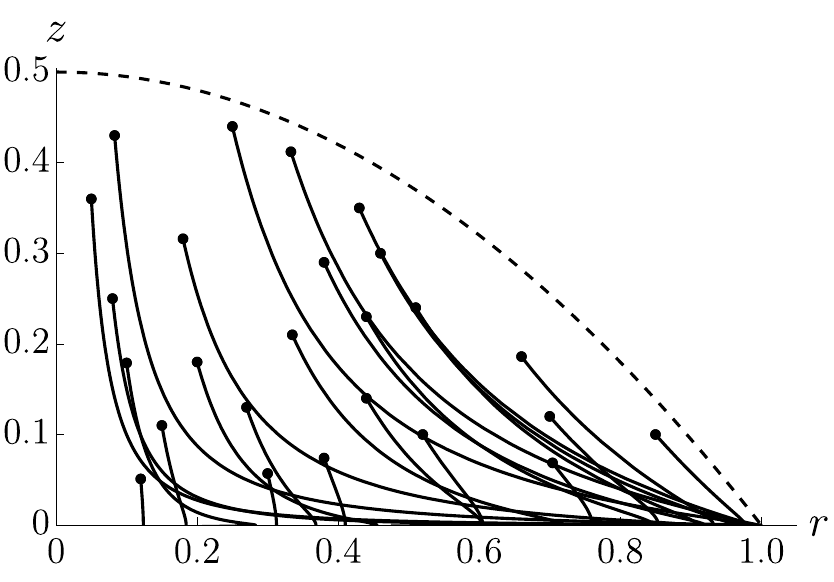}
\caption{}
\end{subfigure}
\hfill
\begin{subfigure}{0.45\textwidth}
\centering
\includegraphics[keepaspectratio,width=1\textwidth,height=1\textheight]{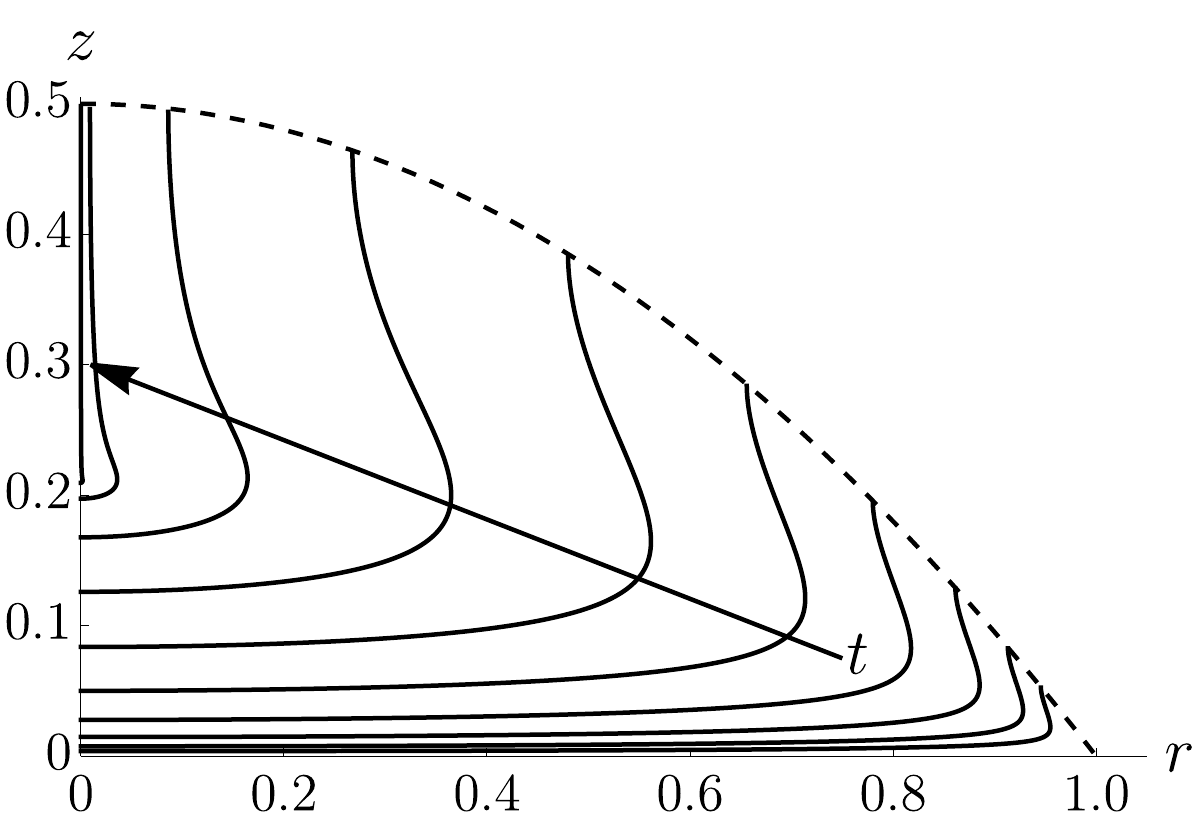}
\caption{}
\end{subfigure}
\caption{
Plots of (a,c,e) the paths of a selection of representative particles and (b,d,f) the locus of the initial positions of particles that are captured by either the free surface (dotted lines) or the substrate (solid lines) for a droplet evolving in the CR mode at $t=(1/10,1/5,\ldots,9/10) \times \tCR$ for (a,b) pure evaporation and (c,d) simultaneous evaporation and imbibition with $\cE = \cI = 1$, and at $t=2^n \times \pi$ ($n=-10,-9,\ldots,0$) for (e,f) pure imbibition.
In (a,c,e) the dots ($\bullet$) denote the initial positions of the particles and the squares ($\blacksquare$) denote the positions at which free-surface capture occurs.
The parabolic dashed line indicates the initial free-surface profile of the droplet, $h(r,0)=\left(1-r^2\right)/2$, 
the dashed line in (c,d) indicates the critical ``watershed'' curve on which free-surface and substrate capture occur simultaneously, and
the arrows in (b,d,f) indicate the directions of increasing $t$.
}
\label{fig:figure19}
\end{figure}

The equations governing the paths of the particles within a droplet evolving in the CR mode are obtained by substituting the expression for $u$ given by \eqref{eq:u_CR} and the expression for $w$ given by \eqref{eq:w_CR} into \eqref{eq:particle_paths_equation_1} and \eqref{eq:particle_paths_equation_2}.
Figures \ref{fig:figure19}(a,c,e) show the paths of a selection of representative particles and Figures \ref{fig:figure19}(b,d,f) show the locus of the initial positions of particles that are captured by either the free surface (dotted lines) or the substrate (solid lines) at various times for (a,b) pure evaporation, (c,d) simultaneous evaporation and imbibition with $\cE = \cI = 1$, and (e,f) pure imbibition.
In particular, Figure \ref{fig:figure19} shows that in the regime $\textrm{Pe}^* \gg 1$ the particles are always advected by the flow within the droplet radially outwards (i.e., towards the contact line) and downwards (i.e., towards the substrate) and always captured by either the free surface or the substrate.
Specifically, Figures \ref{fig:figure19}(a,b) show that for pure evaporation all of the particles are eventually captured by the free surface and then transported to the contact line, forming a ring deposit at the (pinned) contact line, as sketched in Figure \ref{fig:figure3}(a) (see, for example, D'Ambrosio \etal \cite{DAmbrosio2023effect}). This behaviour is qualitatively the same as that in the regime $\hat{\theta}_0^2 \ll \textrm{Pe}^* \ll 1$ discussed in Sections \ref{sec:model_transport} and \ref{sec:particles}.
On the other hand, Figures \ref{fig:figure19}(e,f) show that for pure imbibition all of the particles are eventually captured by the substrate, forming a distributed deposit within the initial footprint of the droplet, as sketched in Figure \ref{fig:figure3}(b). This behaviour is qualitatively different to that in the regime $\hat{\theta}_0^2 \ll \textrm{Pe}^* \ll 1$.
Figures \ref{fig:figure19}(c,d) show that for simultaneous evaporation and imbibition both kinds of capture occur and there is a critical ``watershed'' curve (i.e., a locus of initial positions of the particles) on which free-surface and substrate capture occur simultaneously and above which the particles are captured by the free surface and below which they are captured by the substrate, forming a combined ring and distributed deposit as sketched in Figure \ref{fig:figure3}(c). Again, this behaviour is qualitatively different to that in the regime $\hat{\theta}_0^2 \ll \textrm{Pe}^* \ll 1$.

\subsection{Particle Paths in the CA Mode}
\label{sec:paths_CA}

\begin{figure}[tp]
\centering
\begin{subfigure}{0.45\textwidth}
\centering
\includegraphics[keepaspectratio,width=1\textwidth,height=1\textheight]{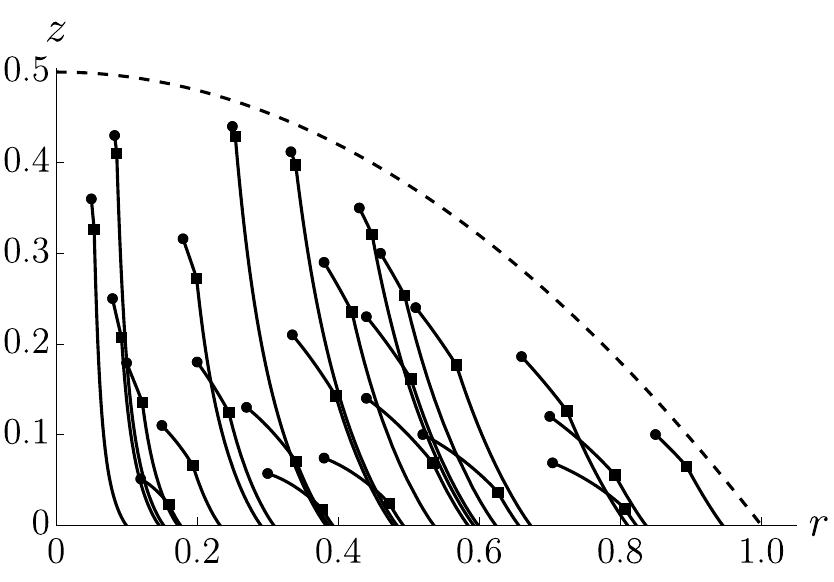}
\caption{}
\end{subfigure}
\hfill
\begin{subfigure}{0.45\textwidth}
\centering
\includegraphics[keepaspectratio,width=1\textwidth,height=1\textheight]{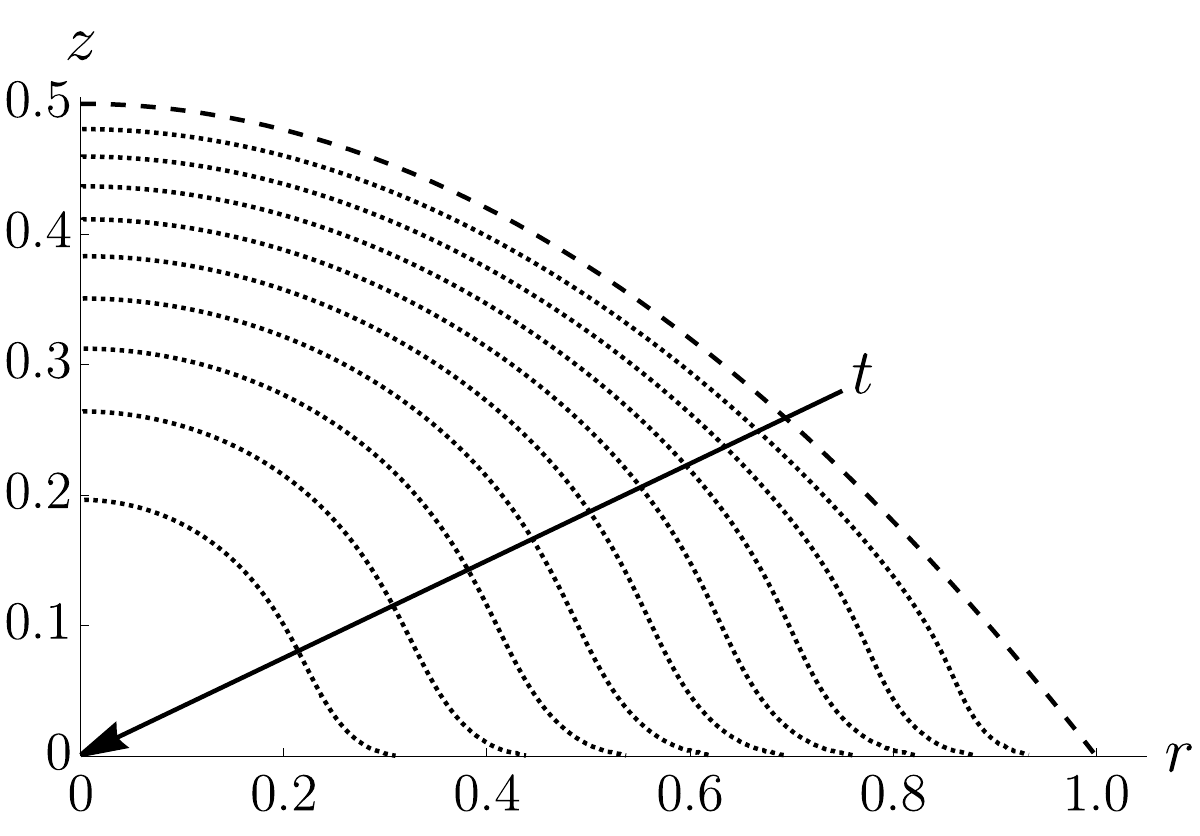}
\caption{}
\end{subfigure}
\centering
\begin{subfigure}{0.45\textwidth}
\centering
\includegraphics[keepaspectratio,width=1\textwidth,height=1\textheight]{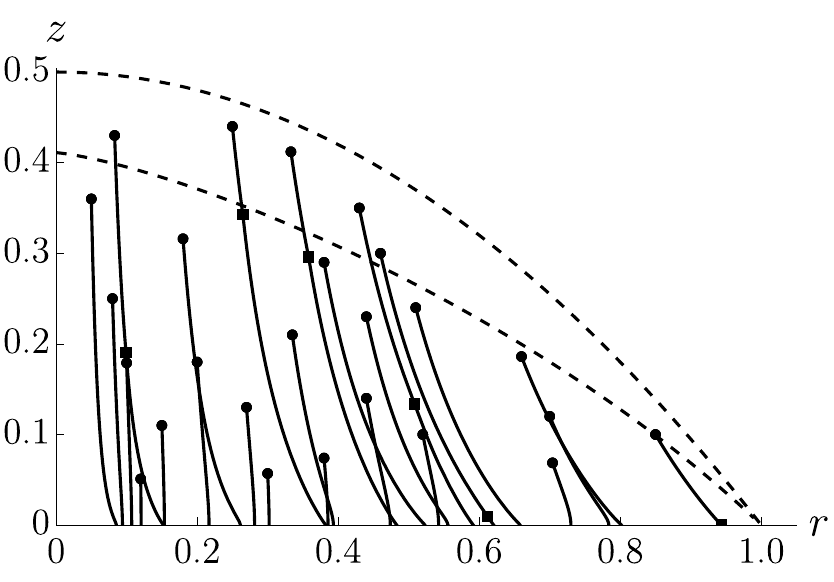}
\caption{}
\end{subfigure}
\hfill
\begin{subfigure}{0.45\textwidth}
\centering
\includegraphics[keepaspectratio,width=1\textwidth,height=1\textheight]{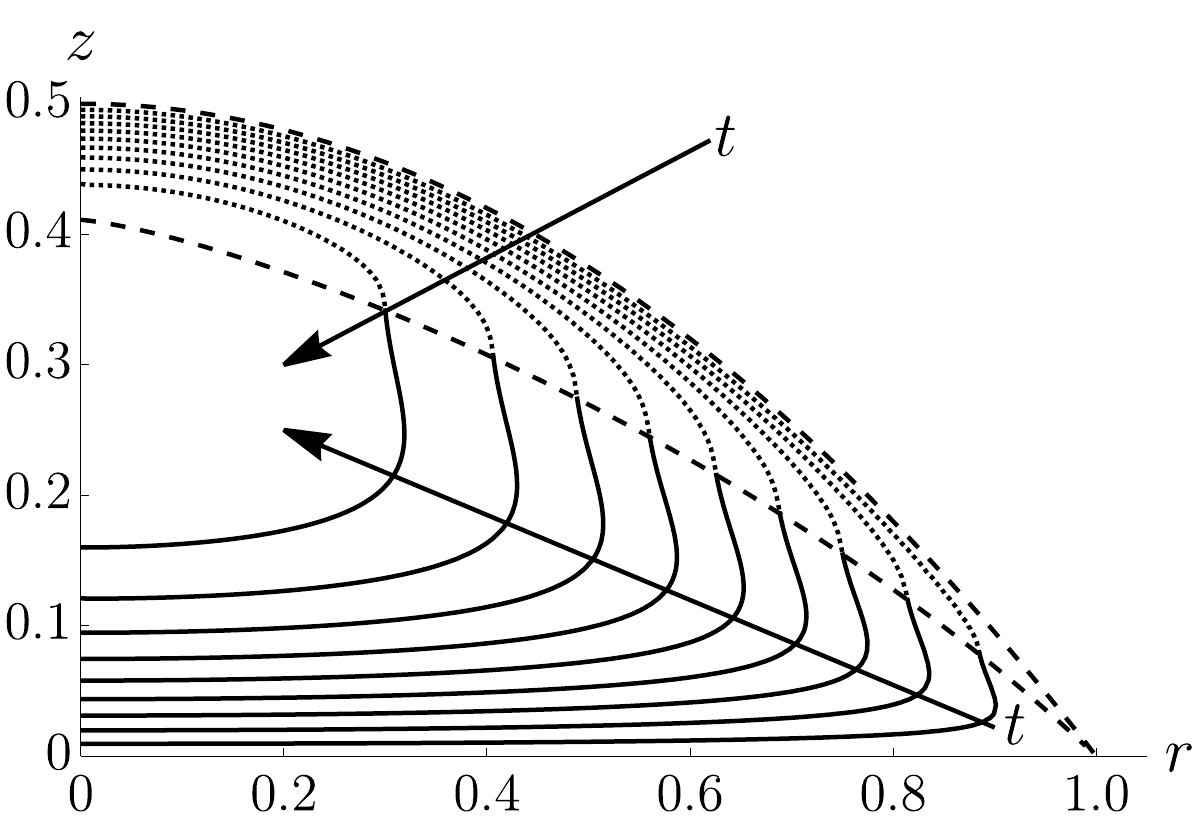}
\caption{}
\end{subfigure}
\centering
\begin{subfigure}{0.45\textwidth}
\centering
\includegraphics[keepaspectratio,width=1\textwidth,height=1\textheight]{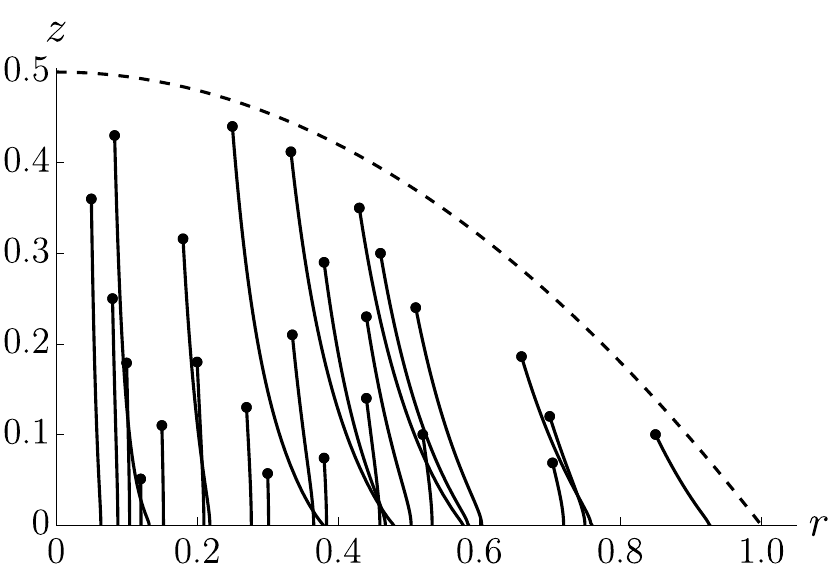}
\caption{}
\end{subfigure}
\hfill
\begin{subfigure}{0.45\textwidth}
\centering
\includegraphics[keepaspectratio,width=1\textwidth,height=1\textheight]{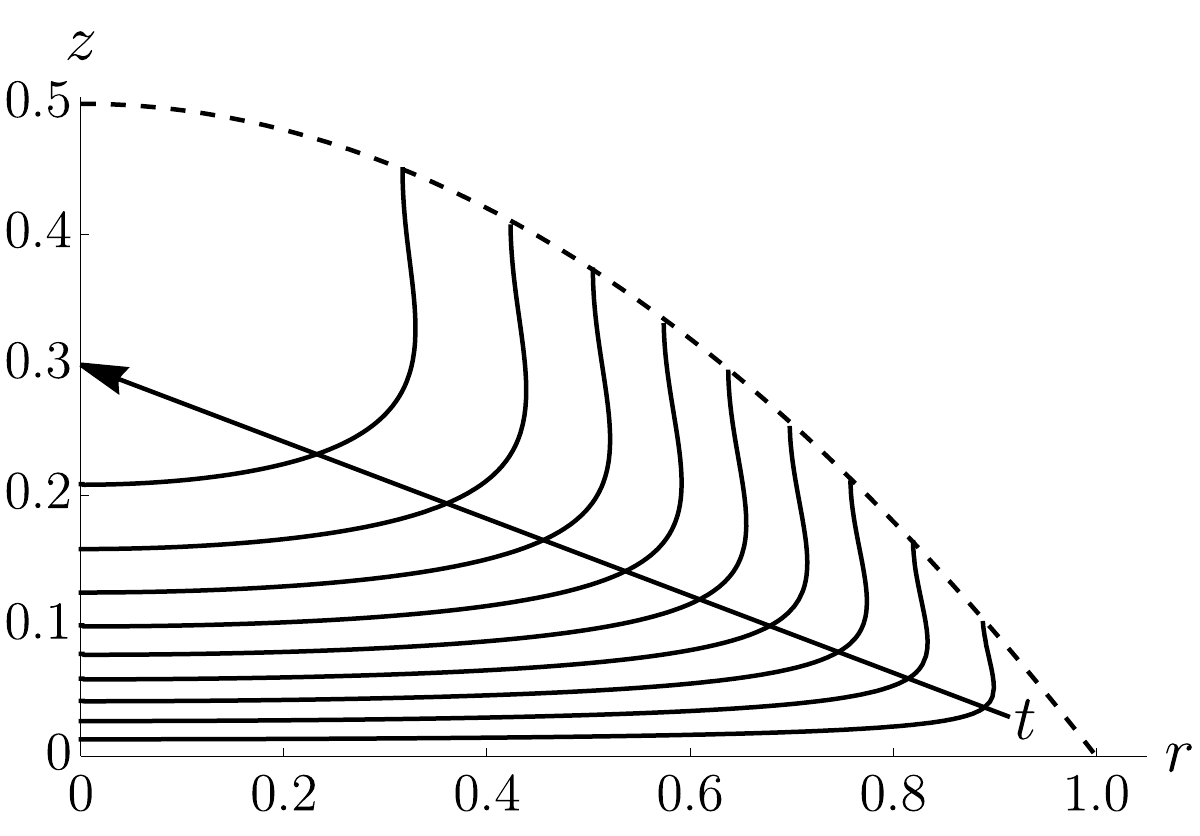}
\caption{}
\end{subfigure}
\caption{
Plots of (a,c,e) the paths of a selection of representative particles and (b,d,f) the locus of the initial positions of particles that are captured by either the free surface (dotted lines) or the substrate (solid lines) for a droplet evolving in the CR mode at $t=(1/10,1/5,\ldots,9/10) \times \tCA$ for (a,b) pure evaporation, (c,d) simultaneous evaporation and imbibition with $\cE = \cI = 1$, and (e,f) pure imbibition.
In (a,c,e) the dots ($\bullet$) denote the initial positions of the particles, and in (a,c) the squares ($\blacksquare$) denote the positions at which free-surface capture occurs.
The parabolic dashed line indicates the initial free-surface profile of the droplet, $h(r,0)=\left(1-r^2\right)/2$,
the dashed line in (c,d) indicates the critical ``watershed'' curve on which free-surface and substrate capture occur simultaneously, and
the arrows in (b,d,f) indicate the directions of increasing $t$.
}
\label{fig:figure20}
\end{figure}

The equations governing the paths of the particles within a droplet evolving in the CA mode are obtained by substituting the expression for $u$ given by \eqref{eq:u_CA} and the expression for $w$ given by \eqref{eq:w_CA} into \eqref{eq:particle_paths_equation_1} and \eqref{eq:particle_paths_equation_2}.
Figure \ref{fig:figure20}, which is analogous to Figure \ref{fig:figure19} but for a droplet evolving in the CA mode, shows that the particle paths are qualitatively the same as those for a droplet evolving in the CR mode. However, the motion of the (unpinned) contact line means that the particles are always deposited as a distributed deposit within the initial footprint of the droplet, as sketched in Figure \ref{fig:figure3}(b). This behaviour is qualitatively the same as that in the regime $\hat{\theta}_0^2 \ll \textrm{Pe}^* \ll 1$ discussed in Sections \ref{sec:model_transport} and \ref{sec:particles}.

\section{Conclusions}
\label{sec:conclusions}

In the present work we formulated and analysed a mathematical model for the evolution of, and deposition from, a thin particle-laden droplet on an infinitely thick, isotropic, flooded, porous substrate with interconnected pores undergoing simultaneous evaporation and imbibition.
In particular, we obtained analytical expressions for the evolution of the droplet, as well as for the flow within the droplet and the substrate, and for the transport and deposition onto the substrate of the particles for droplets evolving in the CR, CA, SS and SJ modes.
While the physical mechanisms driving evaporation (namely, the diffusion-driven flow of vapour in the atmosphere away from the free surface of the droplet) and imbibition (namely, the pressure-driven flow of liquid in the substrate away from the base of the droplet) are rather different, perhaps rather unexpectedly, we found that there are a number of qualitative and quantitative similarities as well as differences in the resulting behaviour of the droplet as it loses mass to its environment.
For example, we showed that a droplet undergoing pure imbibition in the CR mode never completely imbibes (i.e., it has an infinitely long lifetime), but if it evaporates (either with or without imbibition also occurring) then it has a finite lifetime, and increasing the strength of evaporation and/or imbibition shortens its lifetime.
We also showed that in the regime in which diffusion of particles is faster than axial advection but slower than radial advection of particles, the final deposit from a droplet evolving in the CR, CA, SS and SJ modes (namely a ring deposit, a distributed deposit, a combined ring and distributed deposit, and multiple ring deposits, respectively) is independent of both the nature and the strength of the physical mechanism(s) driving the mass loss from the droplet.
However, we also showed that in the alternative regime in which diffusion of particles is slower than both radial and axial advection of particles, the nature of the final deposit from a droplet evolving in the CR mode depends on the mechanism driving the mass loss. In particular, while a droplet undergoing pure evaporation again leads to a ring deposit, a droplet undergoing pure imbibition leads to a distributed deposit, and for a droplet undergoing simultaneous evaporation and imbibition there is a critical ``watershed'' curve (i.e., a locus of initial positions of the particles) above which the particles are captured by the free surface and below which they are captured by the substrate, leading to a combined ring and distributed deposit.
Not only are these results of theoretical interest, but they are also relevant to a wide variety of practical applications, including ink-jet printing, DNA chip manufacturing, and disease diagnostics, that would benefit from an improved ability to predict and/or control the final deposit pattern from a droplet undergoing simultaneous evaporation and imbibition.

Of course, despite the progress made in the present work, there is still much to be learned about the behaviour of a droplet undergoing simultaneous evaporation and imbibition. In addition to relaxing one or more of the assumptions made in the present work (e.g., the assumptions that the droplet is thin, that there is no evaporation from the ``unwetted'' surface of the substrate, that particles do not pass into the substrate, and that the accumulation of particles on the surface of the substrate does not inhibit the flow of liquid into the substrate), perhaps the most exciting direction for future work is to investigate other physical scenarios for the flow within the substrate, such as an anisotropic substrate or a partially flooded substrate, which require the formation and analysis of new mathematical models.

\subsection*{Acknowledgments}

DC gratefully acknowledges the support of the United Kingdom Engineering and Physical Sciences Research Council (EPSRC) via a Mathematical Sciences Doctoral Training Partnership (DTP) Studentship held at the University of Strathclyde in Glasgow (EP/W52394X/1).
All of the authors wish to thank Dr Hannah-May D'Ambrosio (University of Glasgow) for sharing a preliminary version of her work on deposition from an evaporating droplet with a moving contact line, and for her insightful comments on a draft version of the present manuscript.

\subsection*{Data Availability Statement}

All of the data used in the present work are either given within it or can be generated from the mathematical results presented within it.

\bibliographystyle{unsrt}
\bibliography{referencesNov24} 

\end{document}